\documentclass[a4paper,11pt]{article}
\usepackage{jheppub}
\usepackage{upgreek}
\usepackage{subcaption}
\usepackage[compat=1.1.0]{tikz-feynman}
\tikzfeynmanset{warn luatex=false}
\usepackage{amsmath}
\usepackage{slashed}
\usepackage{cleveref}
\usepackage{xcolor}
\usepackage{colortbl}
\usepackage{multirow}
\usepackage{tcolorbox}
\usepackage{adjustbox}
\newcommand{\mut}{\mu^{}_{3}}
\newcommand{\qu}{q_{\rm u}^{}}
\newcommand{\qd}{q_{\rm d}^{}}
\newcommand{\yt}{\mathtt{y}_{\rm t}^{}}
\newcommand{\yu}{\mathtt{y}_{\rm u}^{}}
\newcommand{\yc}{\mathtt{y}_{\rm c}^{}}
\newcommand{\mh}{m_{\rm h}^{}}
\newcommand{\mt}{m_{\rm t}^{}}

\newcommand{\mdm}{m_{\rm DM}^{}}
\newcommand{\TeV}{{\rm TeV}}
\newcommand{\GeV}{{\rm GeV}}
\newcommand{\MeV}{{\rm MeV}}
\newcommand{\SM}{{\rm SM}}
\newcommand{\mpsi}{m_{\psi}^{}}
\newcommand{\mphi}{m_{\Phi}^{}}
\newcommand{\deltam}{\delta m}
\newcommand{\lPhi}{\lambda_{\Phi}^{}}
\newcommand{\lH}{\lambda_{\rm H}^{}}
\newcommand{\lphiH}{\lambda_{\Phi\rm H}^{}}
\newcommand{\Eq}{Eq\,.~}
\newcommand{\Fig}{Fig\,.~}
\newcommand{\Sec}{Sec\,.~}

\newcommand{\ie}{{\it i.e., }}
\newcommand{\hc}{{\rm h.c.}}
\newcommand{\tc}{{t\bar{c}}}
\newcommand{\tcZ}{t\to c Z}
\newcommand{\tuZ}{t\to u Z}
\newcommand{\tch}{t\to c h}
\newcommand{\tuh}{t\to u h}
\newcommand{\tcg}{t\to c g}
\newcommand{\tug}{t\to u g}
\newcommand{\ddbar}{D^{0} \!-\!\bar{D^{0}}}
\newcommand{\llbar}{\ell\bar{\ell}}

\newcommand{\sphiN}{\sigma_{\rm \Phi N}^{\rm eff}}

\definecolor{deepmagenta}{rgb}{0.8, 0.0, 0.8}
\definecolor{mediumtealblue}{rgb}{0.0, 0.33, 0.71}
\definecolor{warmblack}{rgb}{0.0, 0.26, 0.26}
\definecolor{bostonuniversityred}{rgb}{0.8, 0.0, 0.0}
\definecolor{junglegreen}{rgb}{0.16, 0.67, 0.53}
\definecolor{lightcornflowerblue}{rgb}{0.6, 0.81, 0.93}
\definecolor{mypink1}{rgb}{0.858, 0.188, 0.478}
\definecolor{mypink2}{RGB}{219, 48, 122}
\definecolor{mypink3}{cmyk}{0, 0.7808, 0.4429, 0.1412}
\definecolor{mygray}{gray}{0.2}
\definecolor{ForestGreen}{RGB}{34,139,34}

\usepackage[font=small]{caption}
\usepackage[font={small},labelfont={bf}]{caption}
\usepackage{url}
\definecolor{MyDarkBlue}{rgb}{0.1, 0.1, 0.8}
\definecolor{SBlue}{rgb}{0.2, 0.4, 0.7} 
\definecolor{MyLightBlue}{rgb}{0.22,0.51,0.9}
\definecolor{MyGreen}{rgb}{0.0, 0.5, 0.0}
\definecolor{BrickRed}{rgb}{0.8, 0.25, 0.33}
\usepackage{hyperref}
\hypersetup{colorlinks, citecolor=BrickRed,linkcolor=MyDarkBlue, urlcolor=MyGreen}
\title{Up-type FCNC in presence of Dark Matter}
\author[a]{Subhaditya Bhattacharya, }
\author[a]{Lipika Kolay, }
\author[a]{Dipankar Pradhan}
\author[a]{and Abhik Sarkar}
\affiliation[a]{Department of Physics, Indian Institute of Technology Guwahati, North Guwahati, 781039, India}
\emailAdd{subhab@iitg.ac.in}
\emailAdd{klipika@iitg.ac.in}
\emailAdd{d.pradhan@iitg.ac.in}
\emailAdd{sarkar.abhik@iitg.ac.in}
\abstract{Dark Matter (DM) is a known unknown. Apart, current experimental constraints on flavor-changing neutral current (FCNC) processes involving up-type quarks also provide scope to explore physics beyond the Standard Model (SM). In this article, we establish a connection between the flavor sector and the DM sector with minimal extension of the SM. 
Here a singlet complex scalar field, stable under $\mathcal{Z}_3$ symmetry, acts as DM and couples to SM up-type quarks through a heavy Dirac vector-like quark (VLQ), which shares the same $\mathcal{Z}_3$ charge as of the DM. The model thus addresses the observed $\ddbar$ mixing, top-FCNC interactions, and $D^0$ meson decays, together with DM relic density, while evading the direct and indirect DM search bounds. The model can be probed at the future high-energy muon collider, through distinctive signatures of VLQ production, where the VLQ decays into DM and SM particles,  abiding by the existing bounds.}
\begin{document}
\maketitle
\flushbottom
\section{Introduction}
\label{sec:intro}
The Standard Model (SM) of particle physics has been proved to be an appropriate theory to describe fundamental particles and interactions to a great extent. 
Amongst its shortcomings, processes involving flavor-changing neutral current (FCNC) indicate towards physics beyond the Standard Model (BSM) \cite{Castro:2022qkg, Altmannshofer:2024jyv, Sher:2022aaa}. Presence of non-luminous dark matter (DM) \cite{Bertone:2016nfn}, which contributes significantly ($\sim 26.8\%$) \cite{Planck:2018vyg} to the energy budget of the universe, is another challenge, as SM doesn't contain any such particle. Despite astrophysical and cosmological evidences \cite{2010arXiv1001.0316R, Begeman:1989kf, Massey:2010hh, Barrena:2002dp, Schmidt:2001tb, Hoekstra:2008db}, DM is still undiscovered. Some characteristics to associate DM are its non-zero mass, lack of electromagnetic interactions, and dynamical stability over cosmological timescales, consistent with the predictions of the Lambda Cold Dark Matter ($\Lambda$CDM) paradigm \cite{Efstathiou:1990xe, Guth:1980zm, Bullock:2017xww}. Among the leading CDM candidates, classified by their production mechanisms and interaction strengths, are Weakly Interacting Massive Particles (WIMPs) \cite{Gondolo:1990dk}, Feebly Interacting Massive Particles (FIMPs) \cite{Hall:2009bx}, and Strongly Interacting Massive Particles (SIMPs) \cite{Hochberg:2014dra}, amongst others.
DM searches are going on through direct detection (DD) \cite{Liu:2017drf, Misiaszek:2023sxe, Cebrian:2021mvb}, indirect detection (ID) \cite{Gaskins:2016cha, PerezdelosHeros:2020qyt, Strigari:2018utn}, and collider experiments \cite{Boveia:2018yeb, Buchmueller:2017qhf, Birkedal:2004xn}. Null results have consequently resulted in exclusion bounds on pertinent observables, which serve as the dominant constraints in exploring DM phenomenology.

FCNC processes are powerful indirect probes of BSM physics due to their strong suppression in the SM. While flavor physics has primarily focused on the down-type quark sector, the up sector remains less explored, yet offers a theoretically clean environment to test BSM effects. Charm meson processes, such as those involving the $D^0$ meson, 
have their short-distance contributions suppressed by loop-level dynamics and CKM factors, making them largely dominated by long-distance effects \cite{Burdman:2003rs}. If $\ddbar$ mixing originates entirely from BSM contributions, it imposes stringent constraints on New Physics (NP) scenarios \cite{Golowich:2007ka}. Leptonic decays of $D^0$, where both long- and short-distance effects are suppressed, have been studied model-independently \cite{Bharucha:2020eup, Fajfer:2015mia, deBoer:2015boa, Bause:2019vpr} and can be significantly enhanced in NP models with extended Higgs sectors or leptoquarks \cite{Fajfer:2008tm, Kowalska:2018ulj, Ishiwata:2015cga}. Similarly, top-quark FCNC decays like $t \to u_j X$ ($X = \gamma, g, Z, h$) are extremely rare in the SM, offering a clean channel to search for BSM signals. Recent experimental advances have notably improved our sensitivity to such rare processes, making the up-type quark sector an increasingly important frontier for NP searches \cite{Chen:2023eof, Chen:2022dzc, Balaji:2020qjg, Balaji:2021lpr}.

In this work, a potential correlation between DM and FCNC observables is investigated. Our model involves a complex scalar dark matter (CSDM) stabilized by \(\mathcal{Z}_3\) symmetry \cite{Belanger:2012zr, Bhattacharya:2024nla}, which helps in semi annihilation \cite{Wu:2016mbe}, as opposed to a real scalar stabilized by a \(\mathcal{Z}_2\) symmetry \cite{Guo:2010hq}. Such processes provide additional depletion channels to satisfy relic density, evading DD bounds, 
and ID bounds. After strong constraints on minimal scalar and fermion DM models, such extensions are phenomenologically motivated. Apart, we have 
a Dirac vector-like quark (VLQ) that interacts exclusively with the right-handed up-type quarks\footnote{For a vector-like lepton extension of the complex scalar DM model, see \cite{Lahiri:2024rxc}. For scenarios involving a Majorana fermion \(\psi\), refer to \cite{Bigaran:2023ris, Chen:2022dzc, Acaroglu:2021qae}. In such cases, the \(\mathcal{Z}_3\) symmetry cannot be employed for DM stabilization; instead, a \(\mathcal{Z}_2\) symmetry must be utilized.}, due to its specific $\rm U(1)_{\mathtt{Y}}$ hypercharge assignment, \(\mathtt{Y} = 2/3\). This provides further annihilation and co-annihilation channels of DM. 

Due to strong interaction, the vector-like quark (VLQ) \cite{Alves:2023ufm} is expected to yield a significant cross-section at the LHC. The absence of a signal excess thus constrains the VLQ to a 
high-mass regime. The VLQ decays to up-type quark and DM, providing $t\bar{c} + \slashed{E}$ signal as a potential probe for the model. Multi-TeV muon colliders offer a promising platform to explore such heavy VLQs, while extensive studies have been done at both hadron \cite{ATLAS:2016scx, ATLAS:2022tla, ATLAS:2022ozf, CMS:2024xbc, CMS:2024qdd,Carvalho:2018jkq,Aguilar-Saavedra:2013qpa,Freitas:2022cno} and lepton colliders \cite{HajiRaissi:2018yub, Chala:2017xgc}. 

This paper is organized as follows: \Cref{sec:model} introduces the VLQ model that links the dark sector to the Standard Model. In \Cref{sec:constraints}, we discuss various experimental constraints on the model. \Cref{sec:result} presents our analysis of FCNC effects, dark matter phenomenology, and search strategies for future colliders. Finally, \Cref{sec:summary} provides a summary and conclusion.

\section{Model framework}
\label{sec:model}
In this article, we present a minimal extension of a complex scalar singlet DM model by incorporating a Dirac VLQ, which carries $\rm U(1)_{\mathtt{Y}}$ hypercharge \(\mathtt{Y} = 2/3\)\footnote{Alternatively, the scalar can acquire hypercharge, enabling the singlet fermion to serve as DM \cite{Jueid:2024cge}.} and couples only to the right-handed up-type quarks of the SM.
The SM extended Lagrangian is given by\footnote{The source of origin of low energy quark portal renormalizable interaction term, $\overline{\psi}u^{}_R\Phi$, could be a dimension-5 effective operator $\left(C_{u}/\Lambda \right) \overline{\uppsi}H \Phi u^{}_R$, where $\uppsi=(\psi^0~\psi)^{\rm T}$ is a vector like Quark doublet, and also transform similarly under $\mathcal{Z}_3$ like $\Phi$. After EWSB, obtain a term like $\left(C_{u}v/\sqrt{2}\Lambda\right)\overline{\psi} u^{}_R \Phi\equiv \mathtt{y}_{u} \overline{\psi} u^{}_R \Phi$.},
\begin{gather}
\begin{split}
\mathcal{L}_{}\,\supset\,&-\mu_H^2|H|^2-\lambda_{\rm H}|H|^4+|\partial_{\mu}\Phi |^2-m_{\Phi}^2|\Phi|^2-\lambda_{\Phi}|\Phi|^4-\frac{\mu_{3}}{2}\left[\Phi^3+\left(\Phi^{*}\right)^3\right] \\& -\lambda_{\Phi \rm H}|\Phi|^2\left(H^{\dagger}H-v^2/2\right)
+ \overline{\psi}[i\gamma^{\mu}\left(\partial_{\mu}+ig_{s} T^{a} G^{a}_{\mu} +ig^{\prime}{\tt Y} B_{\mu}\right)]\psi \\& -\mpsi\overline{\psi} \psi\, -\,\left(\yu \overline{\psi} u^{}_R\Phi+\yc \overline{\psi} c^{}_R\Phi +\yt \overline{\psi} t^{}_R\Phi+\hc\right)\,.
\end{split}
\label{eq:model}
\end{gather}
Here, the parameters associated with the kinetic term of $\psi$ carry their usual meanings, and $\mu_H^2 < 0$. The model is perturbative if $\lPhi \leq \pi$, $\lH \leq 4\pi$ \cite{Lerner:2009xg} and $\{\mathtt{y}_{c}^{},~\mathtt{y}_{t}^{},~\mathtt{y}_{u}^{}\}<\sqrt{4\pi}$, while the unitarity bounds are always weaker than this. To keep the potential bounded from below, we choose the quartic interaction in such a way that it satisfies the vacuum stability conditions, $\lambda_H>0,~\lambda_{\Phi}>0,~\lphiH+2\sqrt{\lPhi\lH}>0$. Moreover, the maximum allowed value of the cubic parameter, $\mu_3$ is approximately ${\rm max}~\mu_3\approx 2\sqrt{2}\sqrt{\frac{\lambda_{\Phi}}{\delta}}\;\mphi$, while $\delta=2$ gives the absolute stability bound \cite{Belanger:2012zr}. The SM and BSM fields relevant to our analysis, along with their respective quantum numbers, are detailed in \Cref{tab:tab1}.
\begin{table}[htb!]
\begin{center}
\begin{tabular}{|c|cccc|}\hline
\rowcolor{gray!25}{\bf Fields}& $\rm SU(3)_{c}$ & $\rm SU(2)_{\tt{L}}$ & $\rm U(1)_{\tt{Y}}$ &$\mathcal{Z}_3$ \\\hline
\rowcolor{green!10}  $H$  & 1 & 2 & 1/2 & $1$ \\
\rowcolor{cyan!10}  $u_{R}^{},c_{R}^{},t_{R}^{}$  & 3 & 1 & 2/3 & $1$ \\
\rowcolor{lime!10}  $\Phi$  & 1 & 1 & 0 & $\omega/\omega^2$ \\
\rowcolor{magenta!10}  $\psi$  & 3 & 1 & 2/3 & $\omega/\omega^2$ \\\hline
\end{tabular}
\end{center}
\caption{The model particle contents and their corresponding quantum numbers.}
\label{tab:tab1}
\end{table} 

The free parameters of this model are
\begin{align}
\{\mphi,~\mpsi,~\yu,~\yc,~\yt,~\lphiH,~\mut\}\,.
\label{eq:free}
\end{align}
These serve as the basis for analyzing the relic density of DM and its detection prospects, as well as for exploring various flavor physics phenomena involving up-type quarks.
\section{Experimental constraints}
\label{sec:constraints}
\subsection{Flavor Changing Processes}
\label{sec:flavor-constraints}
In this work, we have considered a fermionic mediator $\chi$ interacting with the complex scalar DM $\Phi$ as well as the SM fermions via the Yukawa portal, as can be seen from \Eq\eqref{eq:model}. The hypercharge of the VLQ $\psi$ allows it to only interact with the up-type quarks $u,c,t$. Constraints from the low-energy process related to the charm and up-quark and high-energy processes related to the top quark can be derived, which can put bounds on the allowed couplings and masses of the NP fields. The important processes of low-energy are meson mixing: $\ddbar$, rare decay: $D^{0} \to \ell^{+} \ell^{-}$, semi-leptonic decays: $D^{0} \to (\pi^{0}, \rho^{0}, \omega) \ell \bar{\ell}$ etc. The important high-energy processes are the rare decays of top quark like: $ t \to c(u) X $ with $X\in \gamma, ~g,~ Z$ and $h$. Following, we will discuss our model's impact on these processes. 

\subsubsection{Meson mixing}
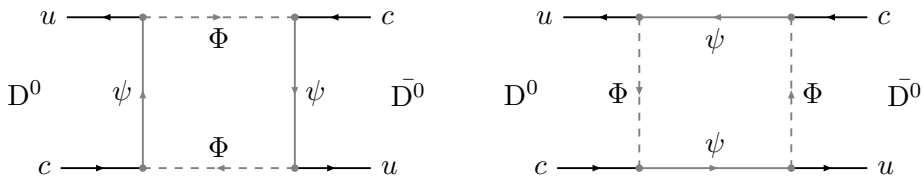
\begin{figure}[htb!]
\centering
\begin{tikzpicture}
\begin{feynman}
\vertex (a1){\(c\)};
\vertex[right=1.3cm of a1](a2);
\vertex[above=2cm of a2](a3);
\vertex[left=1cm of a3](a4){\(u\)};
\vertex[right=2cm of a2](a5);
\vertex[right=2cm of a3](a6);
\vertex[above left=1cm and 0.25cm  of a1] (c){\(\rm D^0\)};
\vertex[right=1cm of a6](a7){\(c\)};
\vertex[right=1cm of a5](a8){\(u\)};
\vertex[above right=1cm and 0.25cm  of a8] (d){\(\rm \bar{D^0}\)};
\diagram* { 
(a1) --[fermion,line width=0.25mm, arrow size=0.7pt](a2),
(a2) --[fermion, line width=0.25mm,arrow size=0.7pt,style=black!50,edge label=\(\color{black}\psi\)](a3),
(a3)--[fermion, line width=0.25mm,arrow size=0.7pt](a4),
(a5) --[charged scalar,line width=0.25mm,arrow size=0.7pt,style=black!50,edge label'=\(\color{black}\Phi\)](a2),
(a3) --[charged scalar,line width=0.25mm,style=black!50,arrow size=0.7pt,edge label'={\(\color{black}\Phi\)}](a6),
(a7) --[fermion,line width=0.25mm, arrow size=0.7pt](a6),
(a6) --[fermion,line width=0.25mm, arrow size=0.7pt,style=black!50,edge label=\(\color{black}\psi\)](a5),
(a5) --[fermion,line width=0.25mm, arrow size=0.7pt](a8)};
\node at (a2)[circle,fill,style=gray,inner sep=1pt]{};
\node at (a3)[circle,fill,style=gray,inner sep=1pt]{};
\node at (a5)[circle,fill,style=gray,inner sep=1pt]{};
\node at (a6)[circle,fill,style=gray,inner sep=1pt]{};
\end{feynman}
\end{tikzpicture}\qquad
\begin{tikzpicture}
\begin{feynman}
\vertex (a1){\(c\)};
\vertex[right=1.3cm of a1](a2);
\vertex[above=2cm of a2](a3);
\vertex[left=1cm of a3](a4){\(u\)};
\vertex[right=2cm of a2](a5);
\vertex[right=2cm of a3](a6);
\vertex[right=1cm of a6](a7){\(c\)};
\vertex[right=1cm of a5](a8){\(u\)};
\vertex[above left=1cm and 0.25cm  of a1] (c){\(\rm D^0\)};
\vertex[above right=1cm and 0.25cm  of a8] (d){\(\rm \bar{D^0}\)};
\diagram* { 
(a1) --[fermion,line width=0.25mm, arrow size=0.7pt](a2),
(a3) --[charged scalar,style=black!50, arrow size=0.7pt,line width=0.25mm,edge label'=\(\color{black}\Phi\)](a2),
(a3) --[fermion, line width=0.25mm,arrow size=0.7pt](a4),
(a2) --[fermion,line width=0.25mm,arrow size=0.7pt,style=black!50,edge label=\(\color{black}\psi\)](a5),
(a6) --[fermion,line width=0.25mm,arrow size=0.7pt,style=black!50,edge label={\(\color{black}\psi\)}](a3),
(a7) --[fermion,line width=0.25mm, arrow size=0.7pt](a6),
(a5) --[charged scalar,line width=0.25mm, arrow size=0.7pt,style=black!50,edge label'=\(\color{black}\Phi\)](a6),
(a5) --[fermion,line width=0.25mm, arrow size=0.7pt](a8)};
\node at (a2)[circle,fill,style=gray,inner sep=1pt]{};
\node at (a3)[circle,fill,style=gray,inner sep=1pt]{};
\node at (a5)[circle,fill,style=gray,inner sep=1pt]{};
\node at (a6)[circle,fill,style=gray,inner sep=1pt]{};
\end{feynman}
\end{tikzpicture}
\caption{The relevant Feynman diagrams contributing to the process $\ddbar$ mixing.}
\label{fig:meson_mixing}
\end{figure}

The $\ddbar$ mixing is an important probe to study the NP, since it is highly suppressed in the SM via CKM and loop contributions. The meson mixing observable $\Delta m $ is defined as 
\begin{equation} \label{eq:mixing_exp_val}
\Delta m = \frac{|\mathcal{M}|}{m_{D}}\,.
\end{equation}
Here, $\mathcal{M}$ is the amplitude of the meson mixing process. The experimental value is measured by LHCb \cite{LHCb:2022cak}, BaBar \cite{BaBar:2016kvp}, Belle \cite{Belle:2014yoi}, CDF \cite{CDF:2013gvz}. In our analysis, we have used the value averaged by HFLAV \cite{HeavyFlavorAveragingGroupHFLAV:2024ctg}, which is given as:
\begin{equation}
\Delta m =\left( 0.997 \pm 0.116 \right)\times 10^{10} \, \rm s^{-1}.
\end{equation}
This mixing is primarily influenced by long-distance amplitudes, which are challenging to compute. The SM short-distance effect is very suppressed of the order of $\sim \mathcal{O}(10^{6})\, \rm s^{-1}$ \cite{Curtin2009}. Assuming that the observed mixing arises solely from non-SM contributions, one can derive strong constraints on potential NP models \cite{Golowich:2007ka, Isidori:2011qw}. 
In our work, we have also considered that the effect is entirely dominated by the NP effect. In our case, we get the mixing via one-loop box diagrams as shown in the \Fig\ref{fig:meson_mixing}. 

The loop contributions can be written as:
\begin{equation}
\mathcal{L}_{\rm mixing} = \mathcal{C}_{RR} [\bar{u} \gamma_{\mu} P_{R} c ]\,[\bar{u} \gamma^{\mu} P_{R} c ] \,,
\end{equation}
with $\mathcal{C}_{RR}$ being the loop contribution. Here, we only get a non-negligible contribution to the right-handed vector current. As mentioned, since the contribution from the SM short-distance effect is highly suppressed, we assume the entire effect is coming via NP.   
The expression for the mass difference $\Delta m_{\footnotesize \rm NP} $ is given by:
\begin{equation}
\Delta m_{\footnotesize \rm NP} = \frac{2}{3} \, B_{D} \,  m_{D} \, f_{D}^2 \, \mathcal{C}_{RR}\,,
\end{equation}
$B_{2} = 0.757 $ \cite{Carrasco:2015pra} is the bag factor coming from:
\begin{equation}
\langle D^{0} | (\bar{u} \gamma_{\mu} P_{R} c )\, ( \bar{u}  \gamma^{\mu} P_{R} c ) | \bar{D^0}\rangle = \frac{2}{3}\,  B_{D} \,  f_{D}^2 m_{D}^2 \,,
\end{equation}
with $f_{D} = 212.0$ MeV, being the decay constant of the $D^{0}$ meson \cite{FlavourLatticeAveragingGroupFLAG:2024oxs}.
Another way to express the mixing observable is through the ratio of the NP contribution to the SM contribution, as shown in \cite{Kolay:2024wns, Kolay:2025jip}. This is typically done to eliminate uncertainties from decay constants or bag parameters. In our case, since the SM contribution to the process is negligible, we simply take $\Delta m$ as the observable.
The allowed parameter space from $\ddbar$ mixing will be discussed in the next section.

\subsubsection{Rare decays}\label{sec:rare_decay}
Another significant avenue for probing NP is through rare decay processes, as they are highly suppressed in the SM due to loop suppression and the GIM mechanism. For charm mesons, this suppression is even stronger. Most charm processes are dominated by long-distance effects. Similar to the case of meson mixing, the SM contributions to rare decays are several orders of magnitude smaller than experimental results, leading to the assumption that the entire contribution comes from BSM physics. In SM, the branching ratio to muon channel is of the order of $\sim \mathcal{O}(10^{-18})$ \cite{Burdman:2001tf} and the long-distance effect $\sim \mathcal{O}(10^{-13})$. The mass of $D^{0}$ meson allows it to decay to muon and electron pairs. The corresponding experimental branching ratios are given by:
\begin{eqnarray}
\mathcal{B}({D^0\to e^+~e^-}) &< &7.9\times 10^{-8} \,~\text{ \cite{Belle:2010ouj}}, \\
\mathcal{B}({D^0\to \mu^+~\mu^-}) &<& 3.1\times 10^{-9}\,~\text{ \cite{LHCb:2022jaa}}.
\label{eq:rare_exp_val}
\end{eqnarray}
\begin{figure}[htb!]
\centering
\begin{tikzpicture}
\begin{feynman}
\vertex(a);
\vertex[above left = 1cm and 1cm  of a] (a1);
\vertex[below left = 1cm and 1cm  of a] (a2);
\vertex[left = 1cm  of a1] (a3){\(c\)};
\vertex[left = 1cm  of a2] (a4){\(u\)};
\vertex[right =1.5cm  of a] (b);
\vertex[above right = 0.75cm and 0.75cm  of b] (b1){\(\ell\)};
\vertex[below right = 0.75cm and 0.75cm  of b] (b2){\(\ell\)};
\diagram*{
(a3) -- [line width=0.25mm,fermion, arrow size=0.7pt,style=black] (a1),
(a1) -- [line width=0.25mm,fermion,quarter left, arrow size=0.7pt,style=black!50, edge label={\({\color{black}\psi} \)}] (a),
(a) -- [line width=0.25mm,fermion, quarter left, arrow size=0.7pt,style=black!50, edge label={\({\color{black}\psi} \)}] (a2),
(a2) -- [line width=0.25mm,fermion, arrow size=0.7pt,style=black, edge label'={\({\color{black}} \)}] (a4),
(a2) -- [line width=0.25mm,charged scalar, arrow size=0.7pt,style=black!50, edge label={\({\color{black}\Phi} \)}] (a1),
(a) -- [line width=0.25mm,boson, arrow size=0.7pt,style=black!75, edge label'={\({\color{black}\gamma,~Z} \)}] (b),
(b2) -- [line width=0.25mm,fermion, arrow size=0.7pt,style=black] (b),
(b) -- [line width=0.25mm,fermion, arrow size=0.7pt,style=black] (b1)};
\node at (a)[circle,fill,style=gray,inner sep=1pt]{};
\node at (b)[circle,fill,style=black,inner sep=1pt]{};
\node at (a1)[circle,fill,style=gray,inner sep=1pt]{};
\node at (a2)[circle,fill,style=gray,inner sep=1pt]{};
\end{feynman}
\end{tikzpicture}\hspace{0.5cm}
\begin{tikzpicture}
\begin{feynman}\label{fig:Feynman_D02mumu2}
\vertex(a);
\vertex[above left = 1cm and 1cm  of a] (a1);
\vertex[below left = 1cm and 1cm  of a] (a2);
\vertex[left = 1cm  of a1] (a3){\(c\)};
\vertex[left = 1cm  of a2] (a4){\(u\)};
\vertex[right =1.5cm  of a] (b);
\vertex[above right = 0.75cm and 0.75cm  of b] (b1){\(\ell\)};
\vertex[below right = 0.75cm and 0.75cm  of b] (b2){\(\ell\)};
\diagram*{
(a3) -- [line width=0.25mm,fermion, arrow size=0.7pt,style=black] (a1),
(a) -- [line width=0.25mm,charged scalar,quarter right, arrow size=0.7pt,style=black!50, edge label'={\({\color{black}\Phi} \)}] (a1),
(a2) -- [line width=0.25mm,charged scalar,quarter right, arrow size=0.7pt,style=black!50, edge label'={\({\color{black}\Phi} \)}] (a),
(a2) -- [line width=0.25mm,fermion, arrow size=0.7pt,style=black, edge label'={\({\color{black}} \)}] (a4),
(a1) -- [line width=0.25mm,fermion, arrow size=0.7pt,style=black!50, edge label'={\({\color{black}\psi} \)}] (a2),
(a) -- [line width=0.25mm,scalar, arrow size=0.7pt,style=black!75, edge label'={\({\color{black}h} \)}] (b),
(b2) -- [line width=0.25mm,fermion, arrow size=0.7pt,style=black] (b),
(b) -- [line width=0.25mm,fermion, arrow size=0.7pt,style=black] (b1)};
\node at (a)[circle,fill,style=gray,inner sep=1pt]{};
\node at (b)[circle,fill,style=black,inner sep=1pt]{};
\node at (a1)[circle,fill,style=gray,inner sep=1pt]{};
\node at (a2)[circle,fill,style=gray,inner sep=1pt]{};
\end{feynman}
\end{tikzpicture}
\caption{Feynman diagrams contributing to the rare decay of charm meson $D^0\to \llbar $.}
\label{fig:Feynman_D02mumu}
\end{figure}
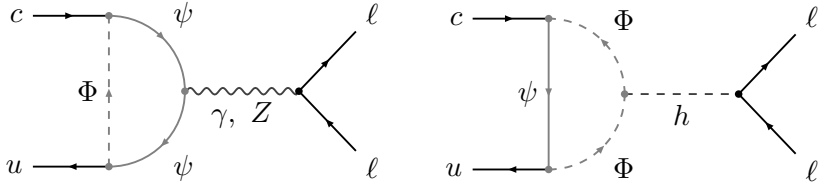

The effective Hamiltonian describing the decay is given by: 
\begin{eqnarray}
\mathcal{H}_{\rm eff}^{ c \to u \ell \ell} = - \frac{4 G_{F}}{\sqrt{2}} \lambda_{q} \sum_{ i = 1,...,10,S^{(\prime)},   P^{(\prime)},T,T5}  \mathcal{C}_{i} \mathcal{O}_{i} \,,
\end{eqnarray}
$\lambda_{q}$ being the CKM element $V_{cq}^{\ast} V_{uq}$. $q= (b,s,d)$,depending on the particle in the loop.
The contributing operators have a structure \cite{Bharucha:2020eup}:
\begin{eqnarray}
\mathcal{O}_{S} = \frac{g_{e}^2}{16 \pi^2} \left( \bar{c} P_{R} u \right) (\bar{\ell }~ \ell ), & \quad &  \mathcal{O}_{S}^{\prime} = \frac{g_{e}^2}{16 \pi^2} \left( \bar{c} P_{L} u \right) (\bar{\ell } ~\ell )\,, \\
\mathcal{O}_{P} = \frac{g_{e}^2}{16 \pi^2} \left( \bar{c} P_{R} u \right) (\bar{\ell } \gamma_{5} \ell ) ,& \quad &  \mathcal{O}_{S}^{\prime} = \frac{g_{e}^2}{16 \pi^2} \left( \bar{c} P_{L} u \right) (\bar{\ell } \gamma_{5} \ell )\,,\\
\mathcal{O}_{10} = - \frac{\alpha_{e}}{4\pi} (\bar{u}_{L} \gamma^{\mu} c_{L})(\bar{\ell} \gamma_{\mu} \gamma_{5} \ell ) ,  &\quad& \mathcal{O}_{10}^{\prime} = - \frac{\alpha_{e}}{4\pi} (\bar{u}_{R} \gamma^{\mu} c_{R})(\bar{\ell} \gamma_{\mu} \gamma_{5} \ell )\,.
\end{eqnarray}
The CKM elements are embedded within the Wilson coefficients, which for short-distance effects are detailed in \cite{Bharucha:2020eup}. Here, $\ddbar$ mixing arises from one-loop box diagrams, while the decay $D^{0} \to \llbar$ occurs via one-loop penguin diagrams, as shown in \Fig\ref{fig:Feynman_D02mumu}.
The branching ratio can be written as: 
\begin{eqnarray}\label{eq:rare_decay_formula}
\mathcal{B}(D^{0} \to \ell \bar{\ell} ) = \tau_{D} \frac{G_{F}^2 \alpha_{e}^2 } {64 \pi^3 } f_{D}^2 m_{D}^{5} \beta(m_{D}^2) \left( |P|^2 + \beta(m_{D}^2) |S|^2\right)\,, 
\end{eqnarray}
with,
\begin{subequations}
\begin{gather}
S =  \frac{1}{m_{c}} \left( C_{S} - C_{S}^{\prime}\right) \,,\\
\beta(q^2)  =  \sqrt{1 - 4 m_{\ell}^2/ q^2 }  \,, \\
P  =  \frac{1}{m_{c}} (C_{P} - C_{P}^{\prime}) + \frac{2m_{\ell}}{m_{D}^2} (C_{10 } - C_{10}^{\prime}) \,. 
\end{gather}
\end{subequations}
In our work, we obtain contributions to the Wilson operators $C_{10}^{(\prime)}$ and $C_{S}^{(\prime)}$. 
The same loop of \Fig \ref{fig:Feynman_D02mumu} will also contribute to other semi-leptonic processes of the $D$ meson. 
Since the loop contributions to the $c \to u \ell \bar{\ell}$ process are extremely small, we do not get any effective constraints on the couplings from this process, and thus it will be disregarded in the combined analysis. Similarly, we also do not obtain significant contributions from other semi-leptonic $D$-meson processes, such as $D^{0} \to (\pi^{0}, \rho^{0}, \omega) \ell \bar{\ell}$, $D_{s}^{+} \to K^{+} \ell \bar{\ell}$, $D^{0} \to \pi^{0} \nu \bar{\nu}$, and other observables related to $c \to u \ell \bar{\ell}$ and $c \to u \nu \bar{\nu}$ transitions.

Furthermore, CP-violating decays of the \( D \)-meson are also expected to receive NP contributions in our model. One such observable is the difference in CP asymmetries in \( D^{0} \to K^{+} K^{-} \) and \( D^{0} \to \pi^{+} \pi^{-} \) decays. The experimental value of the observable \( \Delta A_{CP} \) is given by LHCb \cite{LHCb:2019hro} as:
\begin{gather}
\Delta A_{\rm CP}^{\rm exp}\,=\,(-15.4\pm 2.9)\times 10^{-4}\,.
\end{gather}
In contrast, the SM contribution is significantly smaller, estimated as \( \Delta A_{\rm CP}^{\rm SM}\,=\,(2.0\pm 0.3)\times 10^{-5} \).
The decays \( D^{0} \to K^{+} K^{-} \) and \( D^{0} \to \pi^{+} \pi^{-} \) in our model are governed by gluon-mediated penguin processes. Furthermore, since all the couplings in our models are assumed to be real, the resulting contributions, even when accounting for hadronic enhancement, remain insufficient to satisfy the experimental value of $\Delta A_{\rm CP}$ within the viable parameter space.
For more details on the contribution to \( \Delta A_{\ rmCP} \) while considering a Dirac fermion in a t-channel DM model can be found in \cite{Altmannshofer:2012ur}, while analysis involving a Majorana fermion is discussed in ref. \cite{Acaroglu:2021qae}.
Further studies involving a \( Z^{\prime} \) mediator are discussed in \cite{Wang:2011uu}.
However, the invisible decay of the $D^0$ meson, $\mathcal{B}(D^0 \to \text{invisible}) < 9.4 \times 10^{-5}$~\cite{Belle:2016qek, Belle-II:2018jsg}, could be relevant for low-mass DM phenomenology where $(\sum_im_{{\rm DM}^{}_i}^{}<m_{D^0}^{}\approx 1.864~\GeV)$ \cite{Li:2023sjf}.

\subsubsection{Top-FCNC decays}
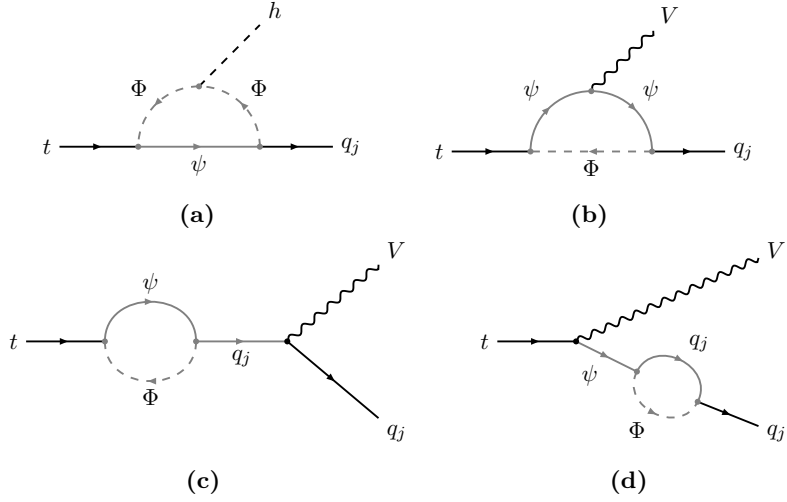
\begin{figure}[htb!]
\centering
\subfloat[]{\begin{tikzpicture}[baseline={(current bounding box.center)},style={scale=0.80, transform shape}]
\begin{feynman}
\vertex(a){\(t\)};
\vertex[right =1.5cm  of a] (a1);
\vertex[right =3.5cm  of a] (a2);
\vertex[right =5cm of a] (a3){\(q_j\)};
\vertex[above right =1cm and 1cm of a1] (b);
\vertex[above right=1cm and 1cm of b] (b1){\(h\)};
\diagram*{
(a) -- [line width=0.25mm, fermion, arrow size=0.7pt,style=black] (a1),
(a2) -- [line width=0.25mm, quarter right, charged scalar, arrow size=0.7pt,style=black!50,edge label'={\({\color{black}\rm\Phi} \)}] (b),
(b)-- [line width=0.25mm, quarter right, charged scalar, arrow size=0.7pt,style=black!50,edge label'={\({\color{black}\rm\Phi} \)}]  (a1), 
(a2) -- [line width=0.25mm, fermion,arrow size=0.7pt,edge label={\(\rm \)},style=black] (a3),
(a1) -- [line width=0.25mm, fermion, arrow size=0.7pt,style=black!50, edge label'={\({\color{black}\rm\psi} \)}] (a2),
(b)-- [line width=0.25mm, scalar,arrow size=0.7pt,edge label={\({\color{black}\rm} \)},style=black] (b1)};
\node at (a1)[circle,fill,style=gray,inner sep=1pt]{};
\node at (b)[circle,fill,style=gray,inner sep=1pt]{};
\node at (a2)[circle,fill,style=gray,inner sep=1pt]{};
\end{feynman}
\end{tikzpicture}}\qquad
\subfloat[]{\begin{tikzpicture}[baseline={(current bounding box.center)},style={scale=0.80, transform shape}]
\begin{feynman}
\vertex(a){\(t\)};
\vertex[ right =1.5cm  of a] (a1);
\vertex[ right =3.5cm  of a] (a2);
\vertex[ right =5cm of a] (a3){\(q_j\)};
\vertex[ above right =1cm and 1cm of a1] (b);
\vertex[ above right=1cm and 1cm of b] (b1){\(V\)};
\diagram*{
(a) -- [line width=0.25mm, fermion, arrow size=0.7pt,style=black] (a1),
(a2) -- [line width=0.25mm, charged scalar, arrow size=0.7pt,style=black!50,edge label={\({\color{black}\rm\Phi} \)}]  (a1), 
(a2)-- [line width=0.25mm, fermion,arrow size=0.7pt,edge label={\(\rm \)},style=black] (a3),
(a1) -- [line width=0.25mm, fermion,quarter left, arrow size=0.7pt,style=black!50,edge label={\({\color{black}\rm\psi} \)}] (b),
(b) -- [line width=0.25mm, fermion, quarter left,arrow size=0.7pt,style=black!50,edge label={\({\color{black}\rm\psi} \)}] (a2),
(b)-- [line width=0.25mm, boson,arrow size=0.7pt,edge label={\({\color{black}\rm} \)},style=black] (b1)};
\node at (a1)[circle,fill,style=gray,inner sep=1pt]{};
\node at (b)[circle,fill,style=gray,inner sep=1pt]{};
\node at (a2)[circle,fill,style=gray,inner sep=1pt]{};
\end{feynman}
\end{tikzpicture}}

\subfloat[]{\begin{tikzpicture}[baseline={(current bounding box.center)},style={scale=0.80, transform shape}]
\begin{feynman}
\vertex (a){\(t\)};
\vertex[right=1.5cm of a](b);
\vertex[right=1.5cm of b](c);
\vertex[right=1.5cm of c](d);
\vertex[above right=1.25cm and 1.5cm of d](e1){\(V\)};
\vertex[below right=1.25cm and 1.5cm of d](e2){\(q_j\)};
\diagram* { 
(a) --[fermion,line width=0.25mm, arrow size=0.7pt](b),
(b) --[fermion, line width=0.25mm,half left, arrow size=0.7pt,style=black!50, edge label=\(\color{black}\psi\)](c),
(c) --[charged scalar, half left,line width=0.25mm, arrow size=0.7pt,style=black!50, edge label=\(\color{black}\Phi\)] (b),
(c) --[fermion, line width=0.25mm,arrow size=0.7pt,style=black!50,edge label'=\({\color{black}q_j}\)](d),
(d) --[boson, line width=0.25mm,arrow size=0.7pt,style=black, edge label=\(\color{black}\)](e1),
(d) --[fermion, line width=0.25mm,arrow size=0.7pt,style=black,edge label=\(\color{black}\)](e2)
};
\node at (b)[circle,fill,style=gray,inner sep=1pt]{};
\node at (c)[circle,fill,style=gray,inner sep=1pt]{};
\node at (d)[circle,fill,style=black,inner sep=1pt]{};
\end{feynman}
\end{tikzpicture}}\qquad
\subfloat[]{\begin{tikzpicture}[baseline={(current bounding box.center)},style={scale=0.80, transform shape}]
\begin{feynman}
\vertex (a){\(t\)};
\vertex[right=1.5cm of a](b);
\vertex[above right=1.25cm and 3cm of b](e1){\(V\)};
\vertex[below right=1.25cm and 3cm of b](e2){\(q_j\)};
\vertex[below right=0.5cm and 1cm of b](c);
\vertex[below right=0.5cm and 1cm of c](d);
\diagram* { 
(a) --[fermion,line width=0.25mm, arrow size=0.7pt](b),
(b) --[fermion, line width=0.25mm, arrow size=0.7pt,style=black!50,edge label'=\(\color{black}\psi\)](c),
(c) --[fermion, line width=0.25mm, half left,arrow size=0.7pt,style=black!50,edge label=\({\color{black}q_j}\)](d),
(c) --[charged scalar, half right,line width=0.25mm, arrow size=0.7pt,style=black!50,edge label'=\(\color{black}\Phi\)] (d),
(b) --[boson, line width=0.25mm,arrow size=0.7pt,style=black,edge label=\(\color{black}\)](e1),
(d) --[fermion, line width=0.25mm,arrow size=0.7pt,style=black,edge label=\(\color{black}\)](e2)
};
\node at (b)[circle,fill,style=black,inner sep=1pt]{};
\node at (c)[circle,fill,style=gray,inner sep=1pt]{};
\node at (d)[circle,fill,style=gray,inner sep=1pt]{};
\end{feynman}
\end{tikzpicture}}
\caption{Feynman diagrams that are contributing to $t\to q_j~ V~(h)$, where $q_j\in\{c,~u\}$ and $V\in\{\gamma,~Z,~g\}$.}
\label{fig:qfcnc}
\end{figure}
The branching ratios of top-FCNC processes are highly suppressed within the SM, while experimental measurements report values several orders of magnitude larger (see \Cref{tab:top_FCNC_prediction}), indicating a potential manifestation of NP. 
In this model, we have Yukawa interactions with the up-type quarks, as can be seen from \Eq\eqref{eq:model}, which can have an impact on the FCNC decays of top quark such as $t \to c(u) X $, where $X=\{ \gamma, g, Z, h\}$. Therefore, we get top-FCNC processes via one-loop penguin diagrams as can be seen from \Fig\ref{fig:qfcnc}.
\begin{table}[htb!]
\centering
\renewcommand{\arraystretch}{1.3}
\setlength{\tabcolsep}{5pt}
\begin{tabular}{|c|c|cc|}
\hline
\rowcolor{gray!15}\textbf{Branching Ratio} & \textbf{SM \cite{TopQuarkWorkingGroup:2013hxj}} &\textbf{Experimental bound}& \textbf{Ref} \\ \hline
\rowcolor{magenta!10}$\mathcal{B}(\tch)$        & $3 \times 10^{-15}$     & $3.7 \times 10^{-4}$ &\\
\rowcolor{magenta!5}$\mathcal{B}(\tuh)$        & $2 \times 10^{-17}$     & $1.9 \times 10^{-4}$  &  \multirow{-2}{*}{CMS~\cite{CMS:2024ubt}}\\
\rowcolor{lime!15}$\mathcal{B}(\tcg)$        & $5 \times 10^{-12}$    & $3.7 \times 10^{-4}$  &             \\
\rowcolor{lime!5}$\mathcal{B}(\tug)$        & $4 \times 10^{-14}$    & $6.1 \times 10^{-5}$    &  \multirow{-2}{*}{ATLAS~\cite{ATLAS:2021amo}} \\
\rowcolor{green!10} $\mathcal{B}(t \to c\gamma )$ & $5 \times 10^{-14}$  & \shortstack{$4.2~ (4.5) \times 10^{-5}$ LH~(RH)}  & \\
\rowcolor{green!5} $\mathcal{B}(t \to u\gamma )$ &   $4 \times 10^{-16}$   & \shortstack{$8.5~ (12) \times 10^{-6}$  LH~(RH)}  &  \multirow{-2}{*}{ATLAS~\cite{ATLAS:2022per}}  \\ 
\rowcolor{cyan!10}$\mathcal{B}(\tcZ)$        &$1 \times 10^{-14}$     & \shortstack{$ 1.3~(1.2) \times 10^{-4}$  LH~(RH)}& \\
\rowcolor{cyan!5}$\mathcal{B}(\tuZ)$        & $7 \times 10^{-17}$    & \shortstack{$ 6.2 ~(6.6) \times 10^{-5}$  LH~(RH)}& \multirow{-2}{*}{ATLAS~\cite{ATLAS:2023qzr}}\\ \hline
\end{tabular}
\caption{Available SM predictions and experimental bounds of the top-FCNC decays by considering left (right) handed  effective couplings denoted by $\rm LH~(RH)$.}
\label{tab:top_FCNC_prediction}
\end{table}
The general Lagrangian responsible for different top-FCNC decays can be given by:
\begin{eqnarray}\label{eq:top_FCNC_Lag}
\mathcal{L}_{t \to q_{j} X }  = && - \bar{q} \gamma_{\mu} \left(C_{tq_{j}Z}^{VL} P_{L} +  C_{tq_{j}Z}^{VR} P_{R} \right) t Z^{\mu}  - \bar{q} i \sigma_{\mu \nu} \epsilon_{Z}^{\mu *} p^{\nu} \left(C_{tq_{j}Z}^{TL} P_{L} +  C_{tq_{j}Z}^{TR} P_{R} \right) t \nonumber \\
&& - \bar{q} i \sigma_{\mu \nu} \epsilon_{\gamma}^{\mu *} p^{\nu} \left(C_{tq_{j}\gamma}^{L} P_{L} +  C_{tq_{j}\gamma}^{R} P_{R} \right) t -  \bar{q} i\sigma_{\mu\nu}T^a \epsilon^{\mu\,*}_{g}p^{\nu}\left(C_{tq_{j}g}^{L} P_L +C_{tq_{j}g}^{R} P_R\right)t \nonumber \\
&& - \bar{q} \left(C_{tq_{j}h}^{L} P_{L} +  
  C_{tq_{j}h}^{R} P_{R} \right) t h \,.
\end{eqnarray}
with $q_{j} = c, u $. The expression for the branching ratios can be given by:
\begin{eqnarray}\label{eq:Br_top_FCNC}
 \mathcal{B} (t \to q_{j} \gamma )  &=& \frac{\tau_{t}m_{t}^3}{16 \pi} \left( |C_{tq_{j}\gamma}^{L}|^2 + |C_{tq_{j}\gamma}^{R}|^2 \right)\,, \\
 \mathcal{B} (t \to q_{j} \, g ) & =&  C_{F}  \frac{\tau_{t} m_{t}^3}{16 \pi} \left( |C_{tq_{j}g }^{L}|^2 + |C_{tq_{j}g }^{R}|^2 \right)\,, \\
\mathcal{B} (t \to q_{j} Z ) &=&  \frac{\tau_{t}}{32 \pi m_{t}} \frac{(m_{t}^2 -m_{Z}^2)^2}{m_{t}^2 m_{Z}^2}  \left\{ (m_{t}^2 + 2 m_{Z}^2) \left( (C_{tq_{j}Z}^{L})^2 + (C_{tq_{j}Z}^{R})^2 \right) \right. \nonumber \\
&& \left. -6 m_{t}m_{Z}^2 \left(  C_{tq_{j}Z}^{VL} C_{tq_{j}Z}^{TR} + C_{tq_{j}Z}^{VR}  C_{tq_{j}Z}^{TL} \right) \right. \nonumber \\  
&& \left. + m_{Z}^2 (m_{t}^2 + m_{Z}^2) \left( (C_{tq_{j}Z}^{TL} )^2 + ( C_{tq_{j}Z}^{TR} )^2 \right) \right\} \,, \\
\mathcal{B} (t \to q_{j} h ) &=& \frac{\tau_{t} m_{t}}{32 \pi} \left( 1- \frac{m_{h}^2}{m_{t}^2} \right)^2 \left( |C_{tq_{j}h}^{L} |^2 + |C_{tq_{j}h}^{R}|^2 \right) \,. 
\end{eqnarray}
The branching ratios to the photon and gluon are suppressed in our case; therefore, their explicit expressions are not presented. The corresponding experimental measurements and SM predictions for each decay are summarized in \Cref{tab:top_FCNC_prediction}.
\subsection{Dark Matter}
\label{sec:dm-constraints}
\begin{figure}[htb!]
\centering\begin{adjustbox}{width=\textwidth}
\begin{tcolorbox}[
colback=gray!5, colframe=black!10, boxrule=0.8pt, arc=4mm, boxsep=0pt, left=0pt, right=0pt, top=0pt, bottom=0pt, width=0.5\textwidth, halign=center]
\subfloat[]{\begin{tikzpicture}
\begin{feynman}
\vertex (a);
\vertex[above left=0.9cm and 0.9cm  of a] (a1){\(\Phi\)};
\vertex[below left=0.9cm and 0.9cm  of a] (a2){\(u\)}; 
\vertex[right=0.9cm of a] (b); 
\vertex[above right=0.9cm and 0.9cm of b] (c1){\(\Phi\)};
\vertex[below right=0.9cm and 0.9cm of b] (c2){\(u\)};
\diagram*{
(a1) -- [line width=0.25mm, charged scalar, arrow size=0.7pt,style=black] (a),
(a2) -- [line width=0.25mm,fermion, arrow size=0.7pt,style=black] (a),
(a) -- [line width=0.25mm,fermion,arrow size=0.7pt, edge label'={\(\color{black}{\psi}\)}, style=black!50] (b),
(b) -- [line width=0.25mm,charged scalar, arrow size=0.7pt] (c1),
(b) -- [line width=0.25mm,fermion, arrow size=0.7pt] (c2)};
\node at (a)[circle,fill,style=black,inner sep=1pt]{};
\node at (b)[circle,fill,style=black,inner sep=1pt]{};
\end{feynman}
\end{tikzpicture}
\label{feyn:wimp-dd1}}
\subfloat[]{\begin{tikzpicture}
\begin{feynman}
\vertex (a);
\vertex[above left=0.45cm and 0.9cm of a] (a1){\(\Phi\)};
\vertex[above right=0.45cm and 0.9cm  of a] (a2){\( \Phi \)}; 
\vertex[below = 0.9cm of a] (b); 
\vertex[below left=0.45cm and 0.9cm of b] (c1){\(N\)};
\vertex[below right=0.45cm and 0.9cm of b] (c2){\(N\)};
\diagram*{
(a1) -- [line width=0.25mm,charged scalar, arrow size=0.7pt,style=black] (a) -- [line width=0.25mm,charged scalar, arrow size=0.7pt,style=black] (a2),
(b) -- [line width=0.25mm, scalar, arrow size=0.7pt, edge label'={\(\color{black}{h}\)}, style=black!50] (a) ,
(c1) -- [line width=0.25mm,fermion, arrow size=0.7pt] (b),
(b) -- [line width=0.25mm,fermion, arrow size=0.7pt] (c2)};
\node at (a)[circle,fill,style=black,inner sep=1pt]{};
\node at (b)[circle,fill,style=black,inner sep=1pt]{};
\end{feynman}
\end{tikzpicture}
\label{feyn:wimp-dd2}}
\end{tcolorbox}
\begin{tcolorbox}[
colback=gray!5, colframe=black!10, boxrule=0.8pt, arc=4mm, boxsep=0pt, left=0pt, right=0pt, top=0pt, bottom=0pt, width=0.5\textwidth, halign=center]
\subfloat[]{\begin{tikzpicture}
\begin{feynman}
\vertex (a);
\vertex[above left=0.9cm and 0.9cm  of a] (a1){\(\Phi\)};
\vertex[below left=0.9cm and 0.9cm  of a] (a2){\(\Phi\)}; 
\vertex[right=0.9cm of a] (b); 
\vertex[above right=0.9cm and 0.9cm of b] (c1){\(b~(u)\)};
\vertex[below right=0.9cm and 0.9cm of b] (c2){\(b~(u)\)};
\diagram*{
(a1) -- [line width=0.25mm, charged scalar, arrow size=0.7pt,style=black] (a),
(a) -- [line width=0.25mm, charged scalar, arrow size=0.7pt,style=black] (a2),
(a) -- [line width=0.25mm, scalar,arrow size=0.7pt, edge label'={\(\color{black}{\rm h}\)}, style=black!50] (b),
(c1) -- [line width=0.25mm,fermion, arrow size=0.7pt] (b),
(b) -- [line width=0.25mm,fermion, arrow size=0.7pt] (c2)};
\node at (a)[circle,fill,style=black,inner sep=1pt]{};
\node at (b)[circle,fill,style=black,inner sep=1pt]{};
\end{feynman}
\end{tikzpicture}
\label{feyn:wimp-id1}}
\subfloat[]{\begin{tikzpicture}
\begin{feynman}
\vertex (a);
\vertex[above left=0.45cm and 0.9cm of a] (a1){\(\Phi\)};
\vertex[above right=0.45cm and 0.9cm  of a] (a2){\( u \)}; 
\vertex[below = 0.9cm of a] (b); 
\vertex[below left=0.45cm and 0.9cm of b] (c1){\(\Phi\)};
\vertex[below right=0.45cm and 0.9cm of b] (c2){\(u\)};
\diagram*{
(a1) -- [line width=0.25mm, charged scalar, arrow size=0.7pt,style=black] (a),
(a2) -- [line width=0.25mm, fermion, arrow size=0.7pt,style=black] (a),
(a) -- [line width=0.25mm, fermion, arrow size=0.7pt, edge label'={\(\color{black}{\psi}\)}, style=black!50] (b) ,
(b) -- [line width=0.25mm, charged scalar, arrow size=0.7pt] (c1),
(b) -- [line width=0.25mm, fermion, arrow size=0.7pt] (c2)};
\node at (a)[circle,fill,style=black,inner sep=1pt]{};
\node at (b)[circle,fill,style=black,inner sep=1pt]{};
\end{feynman}
\end{tikzpicture}
\label{feyn:wimp-id2}}
\end{tcolorbox}
\end{adjustbox}
\caption{The Feynman diagrams are related to the direct (\ref{feyn:wimp-dd1}, \ref{feyn:wimp-dd2}) and indirect detection (\ref{feyn:wimp-id1}, \ref{feyn:wimp-id2}) of WIMP.}
\label{fig:feynman-dd}
\end{figure}
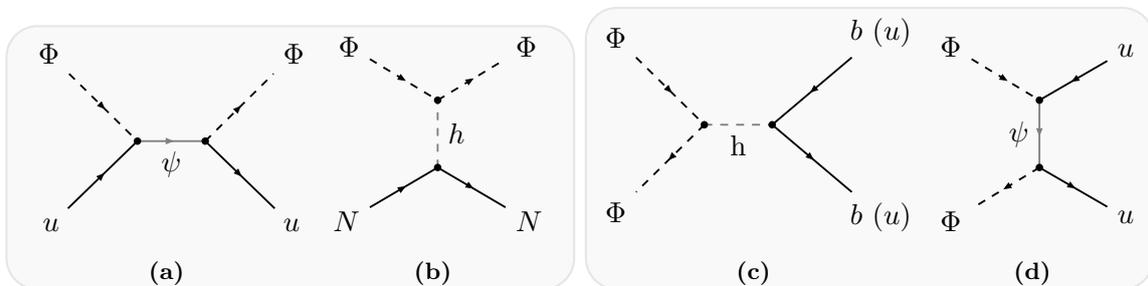

In accordance with \Eq\eqref{eq:model} and \Cref{tab:tab1}, the complex scalar field $\Phi$ is identified as the only viable DM candidate among the fields $\Phi$ and $\psi$. Since both fields transform under the same $\mathcal{Z}_3$ symmetry, the stability of $\Phi$ is kinematically guaranteed by the mass hierarchy $\mphi < \mpsi$. Nevertheless, the heavier dark sector particle $\psi$ can still contribute to the DM relic density through co-annihilation processes, as depicted in \Fig\ref{fig:feynman-relic}.
The mass of thermal DM is constrained by various theoretical and observational studies, providing the following limits \cite{Boyarsky:2008ju, Dutta:2022wdi, Griest:1989wd, Berlin:2017ftj}:
\begin{align}
10~\MeV\lesssim \mdm \lesssim 100~\TeV\,.
\end{align}
The presence of the Higgs portal interaction in the CSDM model, where the Higgs decays into invisible particles, imposes constraints on the coupling $\lphiH$. The 95\% confidence level (CL) upper limit set on the branching fraction of the $125~\GeV$ Higgs boson decay into invisible particles at $\sqrt{s} = 13~\TeV$ by the ATLAS and CMS detectors at the LHC is given by \cite{ATLAS:2023tkt, CMS:2022qva, CMS:2023sdw}:
\begin{align}
\mathcal{B}(h\to\text{inv})
< \begin{cases}
0.107\,,\quad \text{ATLAS} \,,\\
0.150\,,\qquad \text{CMS}  \,.
\end{cases}
\end{align}
Alternatively, the invisible width of the \(Z\) boson, using an integrated luminosity of $\rm 37~(36.3)~fb^{-1}$, produced in proton-proton collisions at a center-of-mass energy of $13~\TeV$ at the ATLAS (CMS) detector, is given by \cite{ATLAS:2023ynf, CMS:2022ett}:
\begin{align}
{\Gamma}(Z\to\rm inv)
< \begin{cases}
506~\MeV\,,\quad \text{ATLAS} \,,\\
523~\MeV\,,\qquad \text{CMS}  \,.
\end{cases}
\end{align}
The WMAP Collaboration constraint on the DM relic density, inferred from $\rm Planck-2018$ measurements, is given by \cite{Planck:2018vyg}:
\begin{align}
\Omega_{\rm DM}h^2=0.1200\pm 0.0012\,.
\label{eq:relic}
\end{align}
The most recent results from the search for nuclear recoil induced by WIMPs using the LUX-ZEPLIN (LZ) two-phase xenon time projection chamber set the strongest spin-independent (SI) exclusion limit on the WIMP-nucleon cross-section, $\sim 2.1 \times 10^{-48}~\text{cm}^2$ at the 90\% CL for a DM mass of $36~\GeV$.
Although several experiments, such as LUX-ZEPLIN \cite{LZ:2024zvo}, XENONnT \cite{XENON:2023cxc}, PandaX-4T \cite{PandaX-4T:2021bab}, and PandaX-xT \cite{PANDA-X:2024dlo}, also place limits on the DM-nucleon scattering cross-section, these limits are less stringent. In this framework, DM-nucleon scattering is governed by inelastic s-channel and t-channel processes, as shown in \Fig\ref{feyn:wimp-dd1} and \Fig\ref{feyn:wimp-dd2}, respectively. Interestingly, the DM-nucleon scattering cross-section depends on $\yu$ and $\lphiH$, along with the $\mpsi$ and $\mphi$ masses, which are also strongly constrained by the DM relic density and FCNC observables.

Gamma-rays might be produced at the Galactic Center through the annihilation of DM, as constrained by observations from Fermi-LAT \cite{Fermi-LAT:2015att} and projected results \cite{Fermi-LAT:2016afa}.
In our model, the presence of the Higgs portal coupling and the direct coupling with up-type quarks predominantly affects the DM annihilation into $b\bar{b}$ (\Fig\ref{feyn:wimp-id1}) and $u\bar{u}$ (\Fig\ref{feyn:wimp-id2}).
Notably, these coupling parameters (\(\yu\) and \(\lphiH\)) are constrained by both the DM relic density (which requires a larger value to obtain the correct relic density) and DD constraints (which require a smaller value to remain consistent with current bounds and be within the reach of future DD experiments) when $\mphi$ and $\mpsi$ are $\mathcal{O}(\GeV-\TeV)$ range. Simultaneously, the magnitude of these couplings is further restricted by ID constraints from Fermi-LAT. The direct and indirect limits on the relic-allowed parameter space are discussed in \Cref{sec:dm-result}.
\subsection{VLQ search at the LHC}
\label{sec:collider-constraints}
VLQs are predicted in many extensions of the Standard Model, but dedicated searches at the LHC remain limited. Most studies in this direction have focused on scenarios where VLQs mix with SM quarks, leading to their decay into SM quarks and gauge bosons \cite{ATLAS:2022hnn,ATLAS:2022tla}. While such scenarios provide explicit signatures of VLQs, their final states bear little resemblance to the case at hand, where the VLQ decays into an SM quark and dark matter. A more comparable scenario arises in supersymmetric quark partner (\textit{squark}) searches, where SUSY particles decay into SM quarks and neutralinos (invisible particles), closely mirroring our scenario. In this section, we recast an existing ATLAS \textit{stop} (supersymmetric top partner) search analysis \cite{ATLAS:2020dsf} from LHC Run II at 13 TeV with an integrated luminosity of 139 ab$^{-1}$. This analysis considers a \textit{stop} pair decaying into a top quark and a neutralino, resulting in a final state with a top-quark pair and missing transverse energy (MET). Using this study, we derive constraints on the masses of the VLQ and dark matter from existing data. Although similar constraints can be obtained from other \textit{squark} or VLQ searches, they are less stringent in comparison.
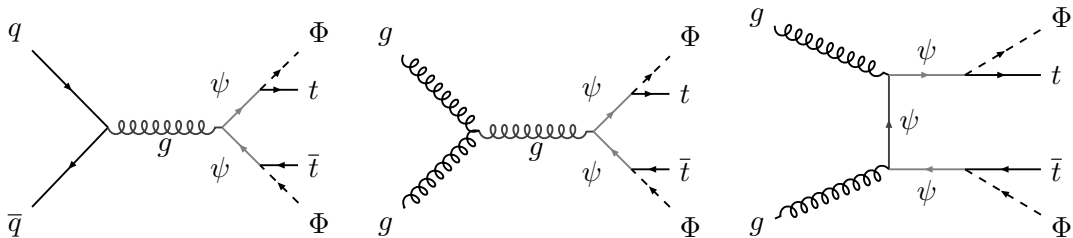
\begin{figure}[htb!]
\centering
\begin{tikzpicture}
\begin{feynman}
\vertex(a);
\vertex[above left = 1.0cm and 1cm  of a] (a1){\(q\)};
\vertex[below left = 1.0cm and 1cm  of a] (a2){\(\overline{q}\)};
\vertex[right = 1.5cm  of a] (b);
\vertex[above right = 0.5cm and 0.5cm  of b] (b1);
\vertex[below right = 0.5cm and 0.5cm  of b] (b2);
\vertex[above right = 0.5cm and 0.5cm  of b1] (b3){\(\Phi\)};
\vertex[below right = 0.5cm and 0.5cm  of b2] (b4){\(\Phi\)};
\vertex[right = 0.5cm  of b1] (b11){\(t\)};
\vertex[right = 0.5cm  of b2] (b22){\(\overline{t}\)};
\diagram*{
(a1) -- [line width=0.25mm,fermion, arrow size=0.7pt,style=black] (a),
(a) -- [line width=0.25mm,fermion, arrow size=0.7pt,style=black] (a2),
(a) -- [line width=0.25mm,gluon, arrow size=0.7pt,style=black!75,edge label'={\({\color{black} g} \)}] (b),
(b) -- [line width=0.25mm,fermion, arrow size=0.7pt,style=black!50,edge label={\({\color{black}\rm\psi} \)}] (b1),
(b1) -- [line width=0.25mm,charged scalar, arrow size=0.7pt,style=black] (b3),
(b1) -- [line width=0.25mm,fermion, arrow size=0.7pt,style=black] (b11),
(b2) -- [line width=0.25mm,fermion, arrow size=0.7pt,style=black!50,edge label={\({\color{black}\rm\psi} \)}] (b),
(b22) -- [line width=0.25mm,fermion, arrow size=0.7pt,style=black] (b2),
(b4) -- [line width=0.25mm,charged scalar, arrow size=0.7pt,style=black] (b2)};
\end{feynman}
\end{tikzpicture}~~
\begin{tikzpicture}
\begin{feynman}
\vertex(a);
\vertex[above left = 1cm and 1cm  of a] (a1){\(g\)};
\vertex[below left = 1cm and 1cm  of a] (a2){\(g\)};
\vertex[right =1.5cm  of a] (b);
\vertex[above right = 0.5cm and 0.5cm  of b] (b1);
\vertex[below right = 0.5cm and 0.5cm  of b] (b2);
\vertex[above right = 0.5cm and 0.5cm  of b1] (b3){\(\Phi\)};
\vertex[below right = 0.5cm and 0.5cm  of b2] (b4){\(\Phi\)};
\vertex[right = 0.5cm  of b1] (b11){\(t\)};
\vertex[right = 0.5cm  of b2] (b22){\(\overline{t}\)};
\diagram*{
(a1) -- [line width=0.25mm,gluon, arrow size=0.7pt,style=black] (a),
(a) -- [line width=0.25mm,gluon, arrow size=0.7pt,style=black] (a2),
(a) -- [line width=0.25mm,gluon, arrow size=0.7pt,style=black!75,edge label'={\({\color{black} g} \)}] (b),
(b) -- [line width=0.25mm,fermion, arrow size=0.7pt,style=black!50,edge label={\({\color{black}\rm\psi} \)}] (b1),
(b1) -- [line width=0.25mm,charged scalar, arrow size=0.7pt,style=black] (b3),
(b1) -- [line width=0.25mm,fermion, arrow size=0.7pt,style=black] (b11),
(b2) -- [line width=0.25mm,fermion, arrow size=0.7pt,style=black!50,edge label={\({\color{black}\rm\psi} \)}] (b),
(b22) -- [line width=0.25mm,fermion, arrow size=0.7pt,style=black] (b2),
(b4) -- [line width=0.25mm,charged scalar, arrow size=0.7pt,style=black] (b2)};
\end{feynman}
\end{tikzpicture}~~
\begin{tikzpicture}
\begin{feynman}
\vertex(a);
\vertex[above left = 0.5cm and 1.5cm  of a] (a01){\(g\)};
\vertex[below = 1.25cm  of a] (b);
\vertex[below left = 0.5cm and 1.5cm  of b] (a02){\(g\)};
\vertex[right = 1cm  of a] (a1);
\vertex[right = 1cm  of b] (b1);
\vertex[ right = 0.5cm and 1cm  of a1] (a10){\(t\)};
\vertex[above right = 0.5cm and 1cm  of a1] (a11){\(\Phi\)};
\vertex[ right = 0.5cm and 1cm  of b1] (b10){\(\overline{t}\)};
\vertex[below right = 0.5cm and 1cm  of b1] (b11){\(\Phi\)};
\diagram*{
(a01) -- [line width=0.25mm, gluon, arrow size=0.7pt,style=black] (a),
(b) -- [line width=0.25mm, fermion, arrow size=0.7pt,style=black!75,edge label'={\({\color{black}\rm\psi} \)}] (a),
(b) -- [line width=0.25mm,gluon, arrow size=0.7pt,style=black] (a02),
(b1) -- [line width=0.25mm,fermion, arrow size=0.7pt,style=black!50,edge label={\({\color{black}\rm\psi} \)}] (b),
(b10) -- [line width=0.25mm,fermion, arrow size=0.7pt,style=black,edge label'={\({\color{black}\rm} \)}] (b1),
(b11) -- [line width=0.25mm,charged scalar, arrow size=0.7pt,style=black,edge label'={\({\color{black}\rm} \)}] (b1),
(a) -- [line width=0.25mm,fermion, arrow size=0.7pt,style=black!50,edge label={\({\color{black}\rm\psi} \)}] (a1),
(a1) -- [line width=0.25mm,fermion, arrow size=0.7pt,style=black,edge label'={\({\color{black}\rm} \)}] (a10),
(a1) -- [line width=0.25mm,charged scalar, arrow size=0.7pt,style=black,edge label'={\({\color{black}\rm} \)}] (a11)};
\end{feynman}
\end{tikzpicture}
\caption{Feynman diagrams corresponding to VLQ production ($t\overline{t}+\slashed{E}_T$ signal) at the LHC.}
\label{fig:pp}
\end{figure}

\begin{figure}[htb!]
\centering
\includegraphics[width=0.6\linewidth]{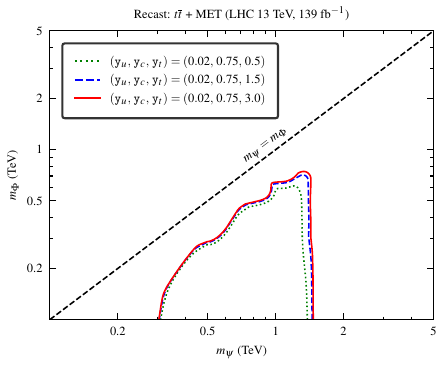}
\caption{95\% C.L. exclusion lines on the $\mpsi-\mphi$ plane from recast of ATLAS top pair + MET analysis \cite{ATLAS:2020dsf}.}
\label{fig:recast}
\end{figure}

The dominant channels of VLQ production (and subsequent decay to top quark and DM) at the LHC are shown in \Fig\ref{fig:pp}. The model implementation is done using \texttt{FeynRules}. The MC events are generated in \texttt{MG5\_aMC} \cite{Alwall:2011uj} and the events are showered using \texttt{Pythia8} \cite{Sjostrand:2007gs}. The showered events are fed into \texttt{CheckMATE2} \cite{Dercks:2016npn} (build upon \texttt{Delphes3} \cite{deFavereau:2013fsa} and \texttt{Fastjet3} \cite{Cacciari:2005hq,Cacciari:2008gp,Cacciari:2011ma}). \texttt{CheckMATE2} uses CL$_{s}$ method \cite{Read:2002hq} to determine the 95\% C.L. exclusion limits. The events are generated at different $\left(\mpsi, \mphi\right)$ benchmarks. The 95\% C.L. exclusion limit is shown in \Fig\ref{fig:recast}. The different signal regions considered for the analysis are detailed in \cite{ATLAS:2020dsf}.
\section{Phenomenological analysis}
\label{sec:result}

\subsection{Flavor observables and model constraints}
\label{sec:flavor-result}
In the previous section, we discussed how a complex scalar DM with a fermionic mediator impacts the flavor observables of up-type quarks, such as $\ddbar$ mixing, rare decays of $D^0$ meson, and top-FCNC decays. In this section, we will show their effect separately and combined on our model parameter space.
\section*{$\boldsymbol{\ddbar}$ mixing:}
\begin{figure}[htb!]
\centering
\subfloat[]{\includegraphics[width=0.47\linewidth]{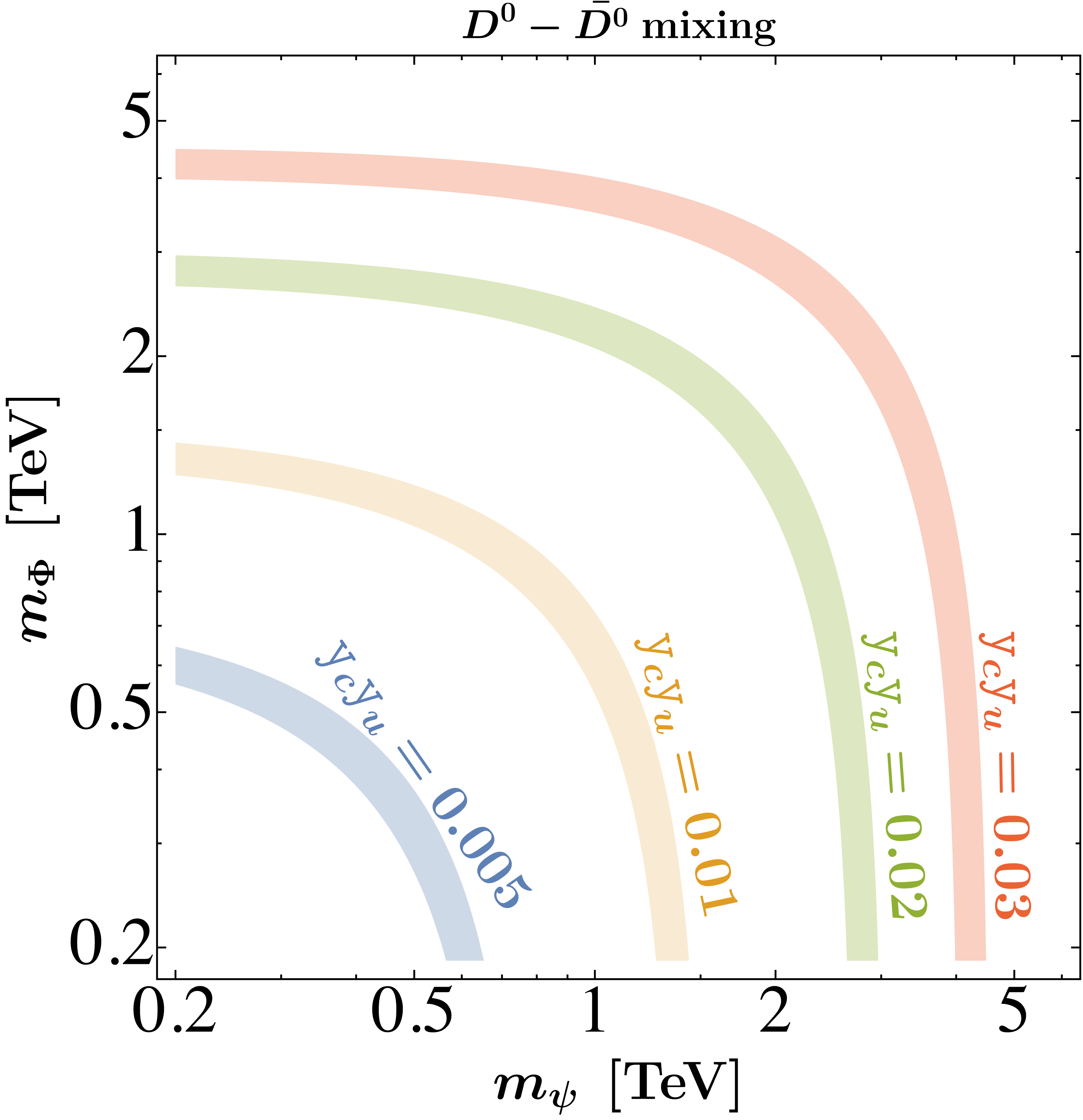}\label{fig:D0mixing}}\hspace{0.5cm}
\subfloat[]{\includegraphics[width=0.47\linewidth]{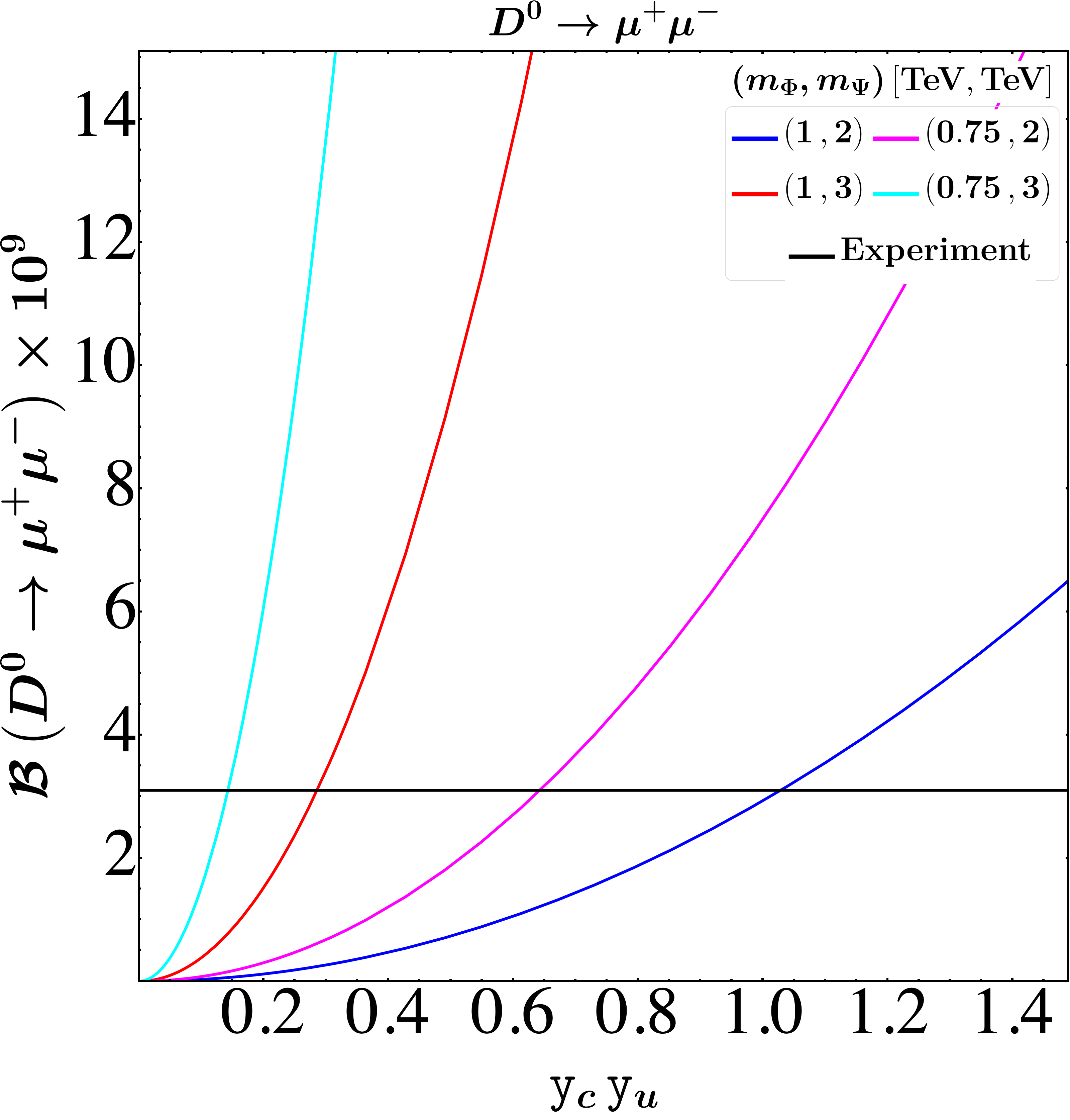}\label{fig:D02mumu}}
\caption{Left: Allowed parameter space from $\ddbar$ mixing for different combinations of the couplings. The bands are allowed region for $1 \sigma$ error band. Right: Variation of the branching ratio $\mathcal{B}(D^0 \to \mu^{+} \mu^{-})$ with the couplings $\yc \yu $, for different combinations of the DM ($\Phi$) and mediator ($\Psi$)  mass. The black line represents the experimental upper limit.}
\label{fig:meson_mixing_region}
\end{figure}

The short-distance effect in $D^0$ meson mixing is very suppressed in SM, so we have assumed the whole contribution is due to beyond SM effects, as discussed. The observable mass difference will have a dependency on the couplings as: $\Delta m  \propto (\yu\, \yc)^2$. 
\Fig \ref{fig:D0mixing} shows the allowed parameter space from $\ddbar$ mixing in $\mpsi-\mphi$ plane. 
The colored region represents the allowed parameter space. 
The different color regions correspond to different coupling combinations of \( \yu \) and \( \yc \).
The bands represent the allowed parameter space, considering the \( 1\sigma \) error of the observable according to \Eq\eqref{eq:mixing_exp_val}. The red, green, orange, and blue color regions correspond to $\yc \yu = 0.03, 0.02, 0.01$  and $0.005$, respectively. The lower mass regions of both DM and mediator are satisfied for smaller coupling values. With increasing the coupling values, we need higher $\mpsi$ and $\mphi$ to satisfy the experimental value. 
\section*{$\boldsymbol{ D^0 \to \mu^{+} \mu^{-} :} $} 
The effect of our model on the rare decay $D^{0} \to \ell \ell $ is discussed in the previous \Cref{sec:rare_decay}. 
For rare decays of charm meson, both the short-distance and long-distance effects in the branching ratio are several orders smaller than that of the experimental bound provided, so here also, we take the whole contribution as coming from NP.
Our model contributes only to the hadronic part of the decay; the lepton vertex does not receive any BSM corrections. The only new physics contributions appear in the vector current operators \( C_{10}^{(\prime)} \). In our case, the decay into an electron pair is further suppressed, as seen from Eq.~\eqref{eq:rare_decay_formula}. The experimental bound on the electron decay mode is also more relaxed than that for the muon mode. For these reasons, we have only imposed the constraint from \( D^0 \to \mu^+ \mu^- \). The parameter space consistent with this decay automatically satisfies the constraints from \( D^0 \to e^+ e^- \) as well.

\Fig\ref{fig:D02mumu} shows the variation of the branching ratio of the decay $D^0 \to \mu^{+} \mu^{-}$ with the coupling combination $\yc \, \yu$ since the branching ratio have a dependency $\propto |\yc \yu |^2$. 
The branching ratio depends on the coupling \( \lambda_{\Phi H} \) through the Feynman diagram shown in the right panel of Fig.~\ref{fig:Feynman_D02mumu}. However, since this is a scalar-mediated diagram, its contribution is negligible compared to the other diagram. So, this decay process cannot be used to place any bound on the coupling \( \lambda_{\Phi H} \).
Different colors of Fig.~\ref{fig:D02mumu} represent various combinations of the DM and mediator masses, while the black line represents the experimental upper limit as given in \Eq\eqref{eq:rare_exp_val}.
For a fixed coupling, the branching ratio increases with increasing \( \mpsi \) and decreases with increasing \( \mphi \). Notably, the cyan line shows the most stringent constraint on coupling, corresponding to \( \mphi = 0.75 \) TeV and \( \mpsi = 2 \) TeV, setting an upper limit \( |\yc \yu| \lesssim 0.12 \).
Unlike the observable \( D^0 \) mixing, where there are measurements, here we only have an upper bound on the branching ratio, allowing any region with couplings below this constraint.
Between the mixing and leptonic decays of the \( D^0 \) meson, the more stringent bound comes from the \( D^0\!-\!\bar{D^0} \) mixing process. The region allowed by mixing is already consistent with the constraint from the \( D^0 \to \mu^+ \mu^- \) decay. Therefore, in the combined analysis, we do not consider the \( D^0 \to \mu^+ \mu^- \) constraint separately.

\begin{figure}[htb!]
\centering
\includegraphics[width=0.475\linewidth]{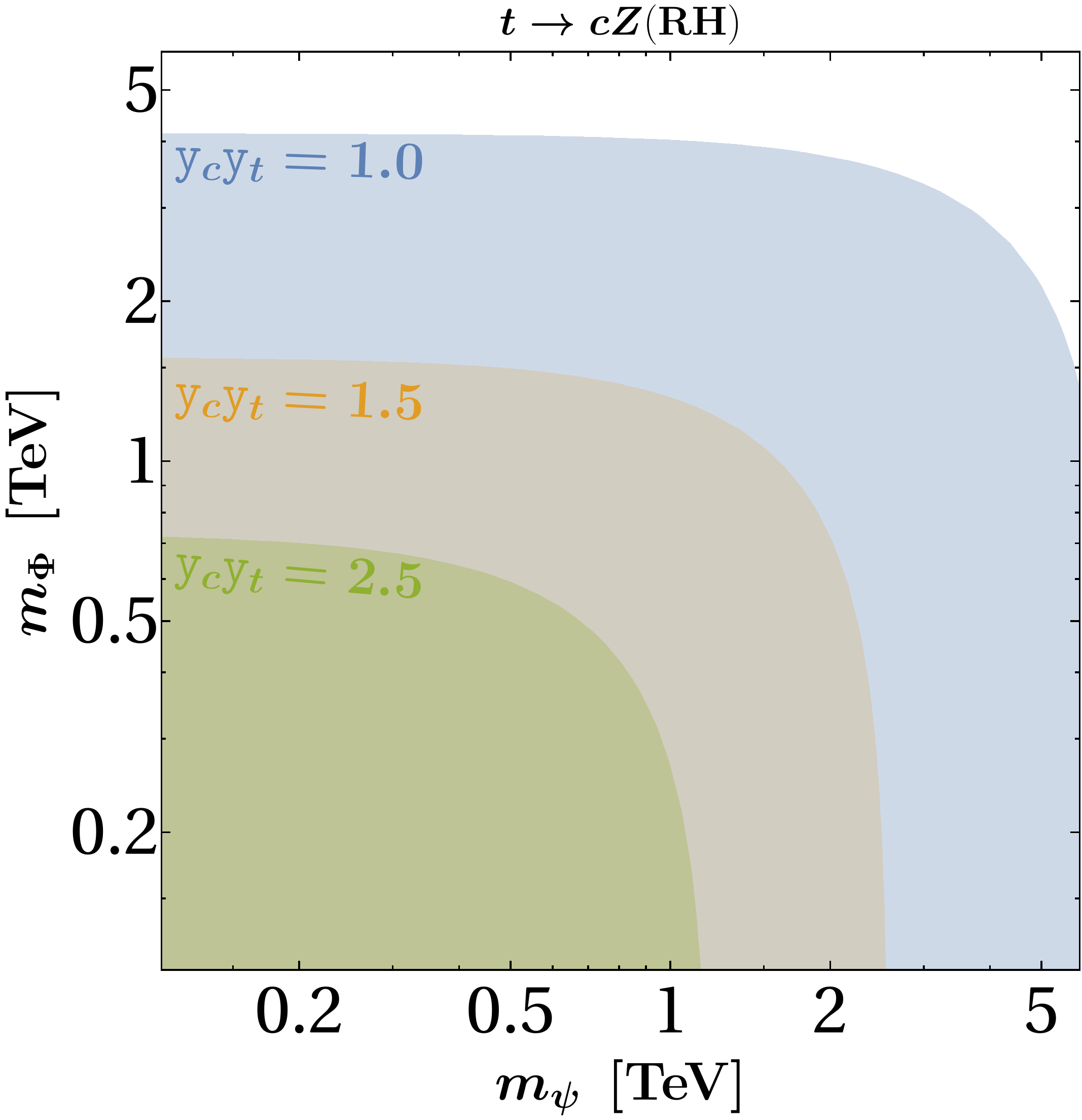}\quad
\includegraphics[width=0.475\linewidth]{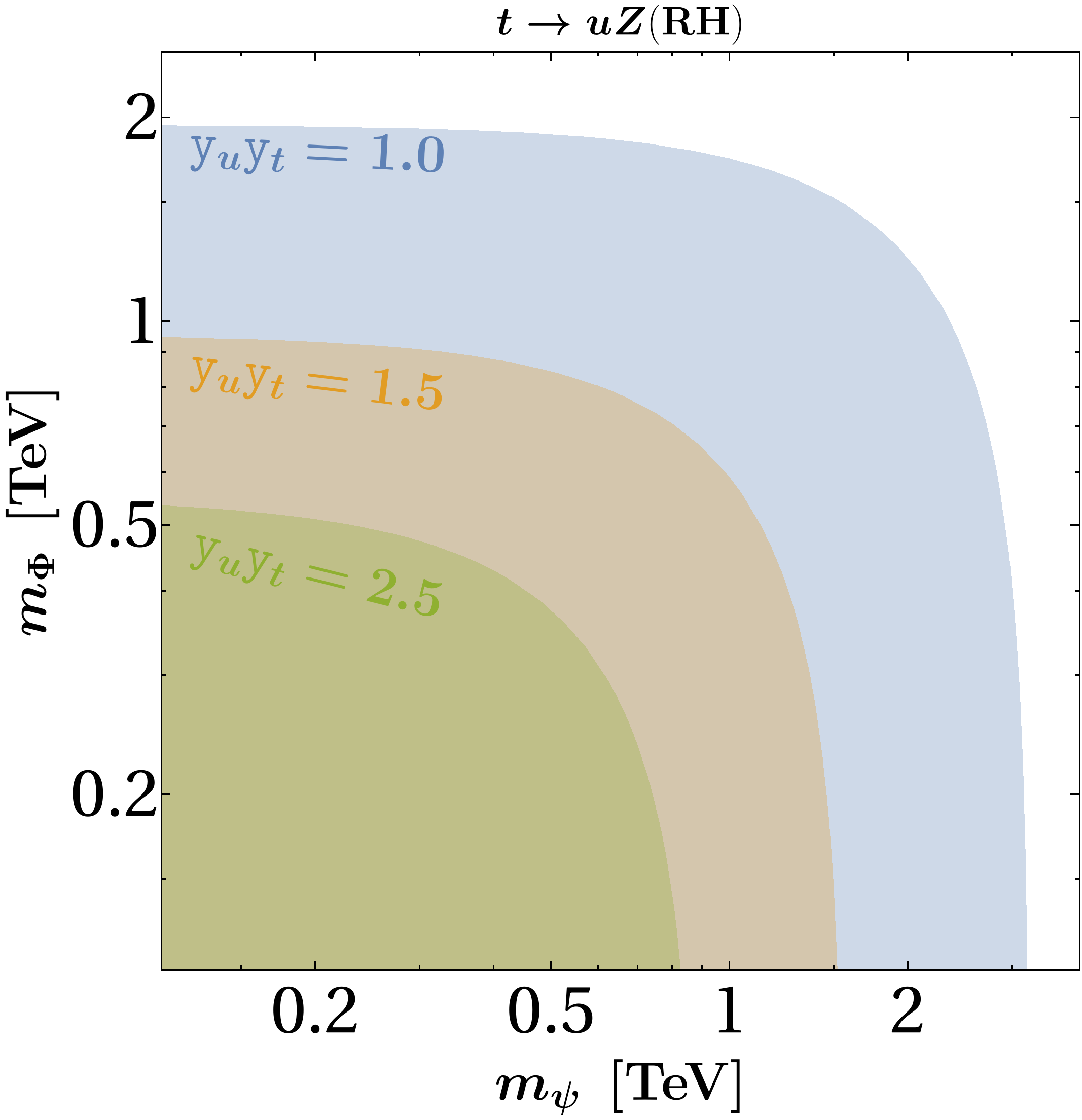}
\caption{In the left (right) panel, the shaded regions depict the allowed parameter space constrained by the decays $t\to c ~(u)Z$, corresponding to different combinations of coupling constants.}
\label{fig:t2cuZ}
\end{figure}
\section*{Top-FCNC decays:} 
This model also induces top-FCNC decays, specifically $t \to c(u)X$ with $X = \gamma, g, Z, h$, via one-loop processes. Among these channels, the most significant contributions from our model arise in the $t \to c(u)Z$, $t \to c(u)\gamma$, and $t \to c(u)g$ decay modes. Since these processes involve two up-type quarks, the corresponding decay widths scale as:
$$
\Gamma_{t \to c(u)X} \propto |\mathtt{y}_t\, \mathtt{y}_{c(u)}|^2\,.
$$
The explicit expressions for the branching ratios are provided in \Eq\eqref{eq:Br_top_FCNC}. These are evaluated in the limit $m_{q_j} \to 0$, and we have verified that including the mass of the final-state quark leads to no significant numerical differences.
In contrast to the case of neutral meson mixing, only upper limits on the branching ratios are available for these processes. Consequently, the experimental constraints translate into upper bounds on the allowed parameter space, rather than defining a band. For the decay involving the \( Z \) boson, the experimental upper limits are provided separately for scenarios in which the effective couplings are purely left-handed (LH) or right-handed (RH). The experimental branching ratios for the decays $t\to c(u)X $ with $X=(\gamma, g, Z, h)$, as reported by ATLAS \cite{ATLAS:2023qzr} and CMS \cite{CMS:2024ubt}, summarized in \Cref{tab:top_FCNC_prediction}. It shows that the decay involving the up quark is more suppressed than the corresponding decay involving the charm quark. We have performed the analysis for both cases and, in what follows, we present the results corresponding to the case that imposes the most stringent constraint. As mentioned earlier, the dependence of the branching ratio on the mass of the final-state light quark is negligible and does not affect the parameter space. Therefore, the branching ratios for these two channels differ only by the coupling of the DM-mediator to the \( u \) and \( c \) quarks. Also, for a particular decay process, the branching ratio will have a significant dependency on the mass of the particles in the loop, as will be discussed following. 

\Fig\ref{fig:t2cuZ} shows the allowed parameter space in the mediator-DM mass plane (\( \mpsi - \mphi \)). The left panel corresponds to the decay \( \tcZ \), while the right panel shows results for \( \tuZ \). The parameter space shown is allowed by the experimental upper bounds on the branching ratios, assuming purely right-handed effective couplings, as indicated by the label ``RH" in both plots.
Since the branching ratio depends on the product of couplings \( \mathtt{y}_{c(u)} \, \yt \), and not on the individual values, the plots are shown for different values of the parameters \( \yt \, \yc \) and \( \yt \, \yu \), corresponding to the decays \( \tcZ \) and \( \tuZ \), respectively. As only upper bounds are available, the allowed parameter space becomes more relaxed for smaller coupling values. We have illustrated this for three different coupling combinations: \( \yt \, \mathtt{y}_{c(u)} = 1.0, 1.5 \), and \( 2.5 \). For the \( t \to uZ \) decay, the region \( \mphi \leq 2 \) TeV and \( \mpsi \leq 3 \) TeV is allowed for  $\yt \yu = 1.0 $, which shrinks with increasing coupling values. For example, when \( \yt \, \yu = 2.5 \), the allowed region reduces to \( \mphi \leq 0.5 \) TeV and \( \mpsi \lesssim 0.8 \) TeV. The bounds are slightly more relaxed for the \( \tcZ \) decay for the same coupling combinations. For instance, with \( \yt \, \yc = 2.5 \), we obtain \( \mphi \leq 0.7 \) TeV and \( \mpsi \leq 1.2 \) TeV.

For both decay modes, we present only the allowed region corresponding to the ``RH" case. In our scenario, the left-handed contribution to the loop is highly suppressed. We have verified that even with a large coupling, $\yc \, \yt = 3.5$, the entire parameter space remains allowed. Consequently, for the same coupling value, the region allowed in the ``RH" case is already encompassed by that of the ``LH" case. Therefore, we do not show the corresponding plots explicitly.

\begin{figure}[htb!]
\centering
\includegraphics[width=0.45\linewidth]{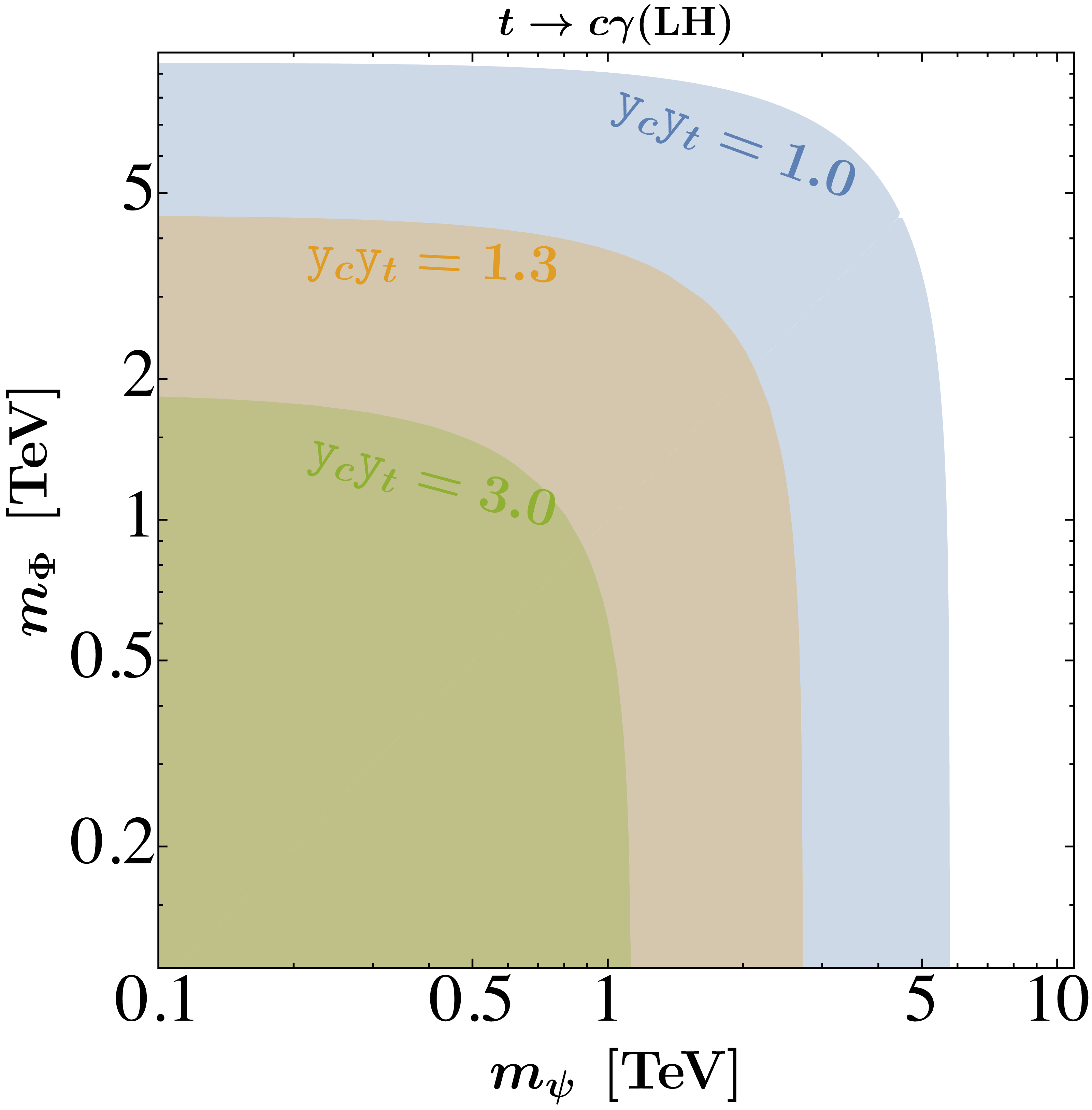}\hspace{0.5cm}
\includegraphics[width=0.45\linewidth]{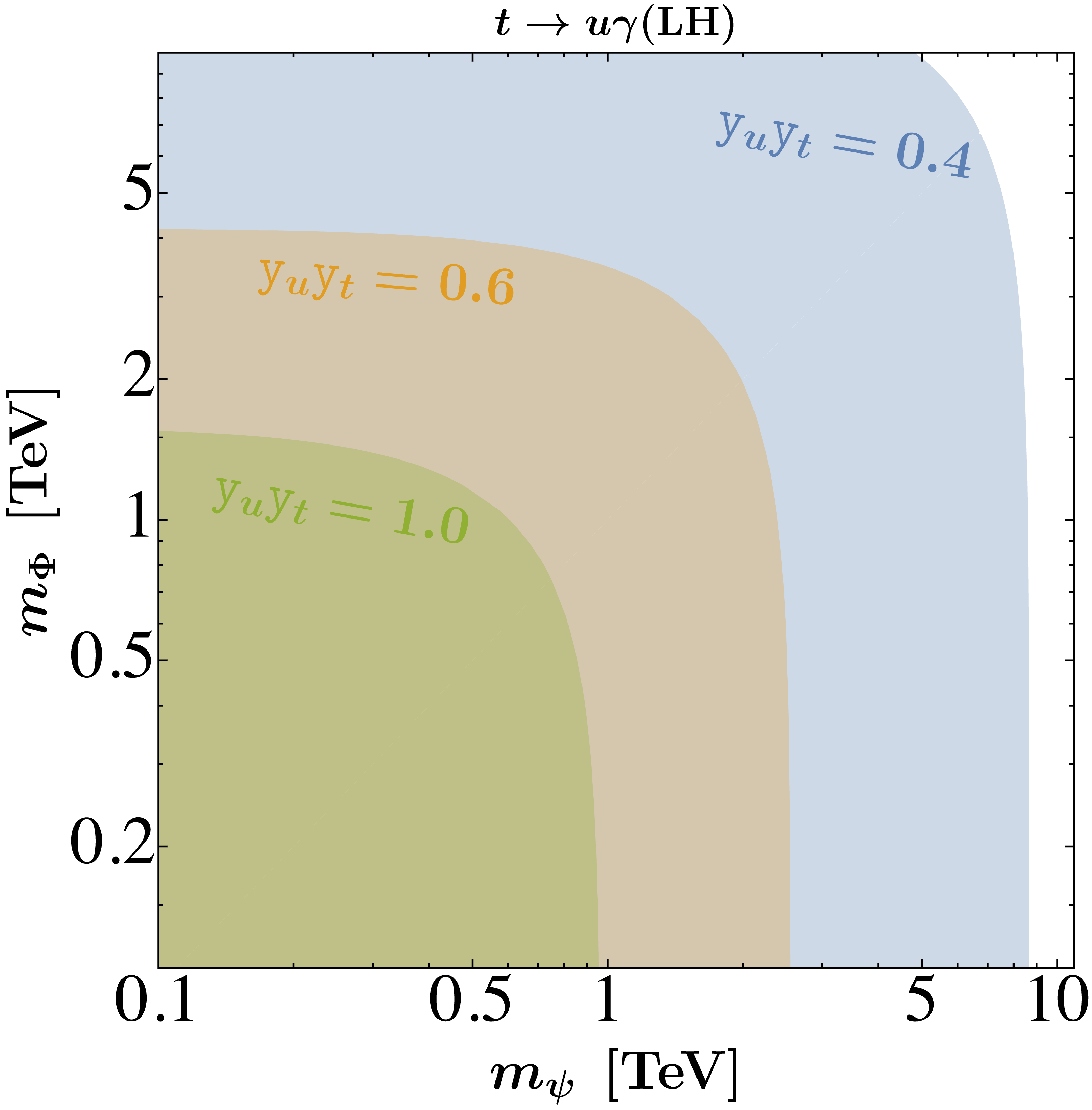}
\caption{In the left (right) panel, the shaded regions represent the allowable parameter space constrained by the decays $t \to c~(u)\gamma$, corresponding to various combinations of coupling constants $\yc\yt$ and $\yu\yt$, respectively.}
\label{fig:t2cugamma}
\end{figure}

\Fig\ref{fig:t2cugamma} shows the allowed parameter space from the decays $t \to c(u) \gamma$. The experimental bounds on these decays are given in \Cref{tab:top_FCNC_prediction}, for considering purely left and right-handed couplings. In this work, the dominant contribution arises from the effective left-handed coupling, while the entire parameter space remains allowed for the right-handed coupling. The left panel plot shows the decay $t \to c \gamma (\rm LH)$, whereas the right panel corresponds to $t \to u \gamma \, \rm (LH)$. The different colored regions represent the allowed parameter space for various coupling combinations. Since the decay width for $t \to c(u) \gamma$ depends on the couplings as $\Gamma_{t \to c(u) \gamma } \propto \mathtt{y}_{c(u)}^2 \mathtt{y}_t^2$, similar to $t \to c(u) Z$, the plots are presented as a function of the product coupling $\mathtt{y}_{c(u)} \mathtt{y}_t$. From \Cref{tab:top_FCNC_prediction}, the branching ratio for top decay to an up quark is seen to be an order of magnitude smaller than that for a charm quark. As a result, for equal values of the couplings to charm and up quarks, the $t \to u \gamma$ process imposes stronger constraints on the parameter space. This distinction is clearly illustrated in \Fig\ref{fig:t2cugamma}, where the left panel shows constraints from $t \to c \gamma$, and the right panel from $t \to u \gamma$. The allowed parameter space from $t \to c \gamma \, \rm (LH) $ has shown for the coupling combinations $\yc\yt = \{ 1.0, 1.3 , 3.0\}$, whereas for the decay $ t\to u \gamma \, \rm (LH)$, the coupling combinations are: $ \yu\yt = \{ 0.4, 0.6 , 1.0\} $. For $\yc \yt = 1.0$, the $\mphi$ is allowed in the region $\mphi \gtrsim 4 \TeV$, whereas for $\mpsi$, the region $\mpsi \gtrsim 6~\TeV$ is disallowed by $t \to c \gamma $ decay. Similarly, for $ t \to u \gamma$ decay, for $ \yu \yt = 0.4, $ the region with $\mpsi \gtrsim 9~\TeV$ and the whole region of $\mphi$ is allowed. With increasing coupling constants, the allowed region decreases. If we take $\yc \yt $ as high as $3.0$, the allowed region becomes very much constrained, i.e., $\mphi \lesssim 2~\TeV$ and $\mpsi \lesssim 1~\TeV$.

\begin{figure}[htb!]
\centering
\includegraphics[width=0.45\linewidth]{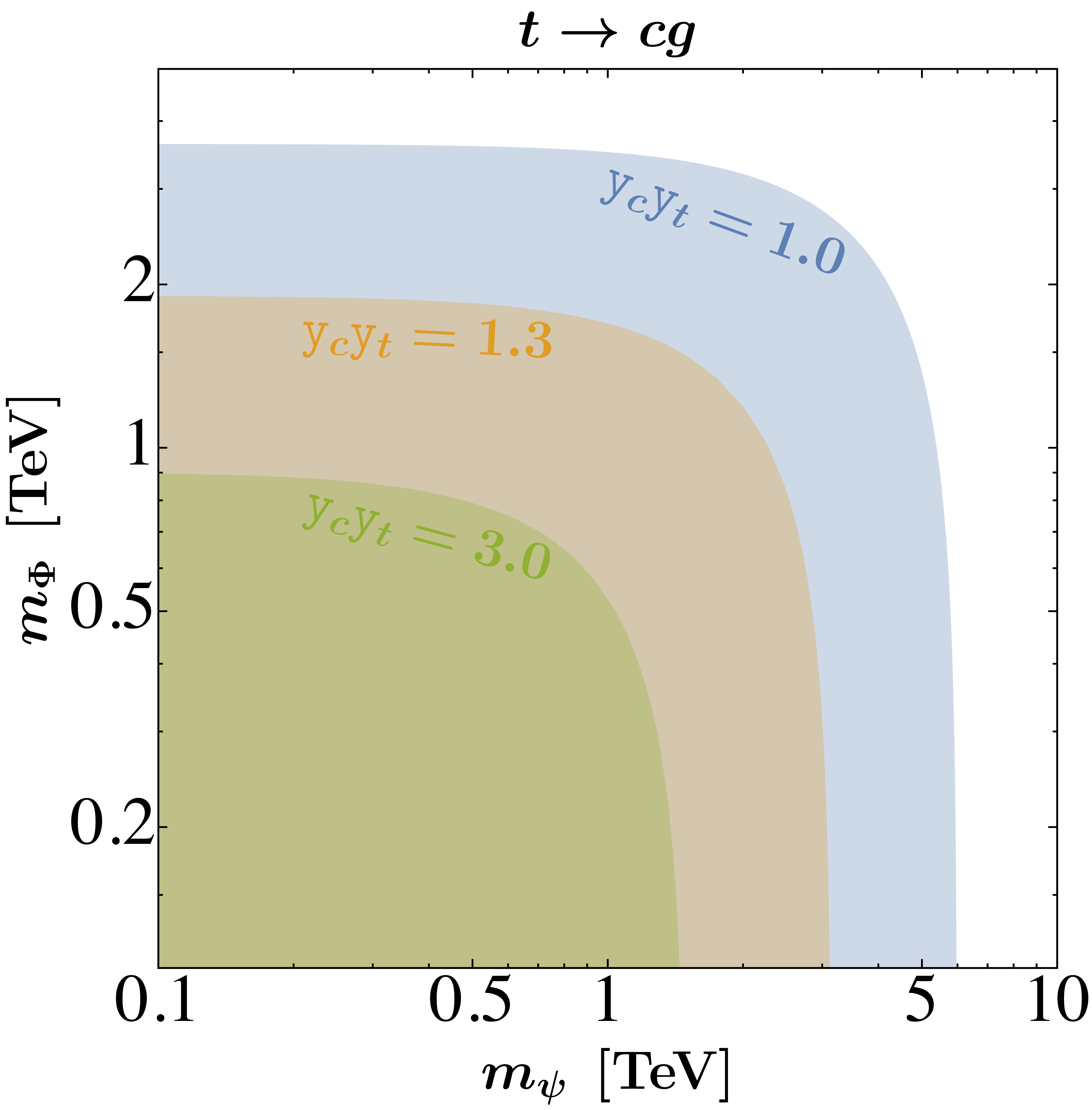}\hspace{0.5cm}
\includegraphics[width=0.45\linewidth]{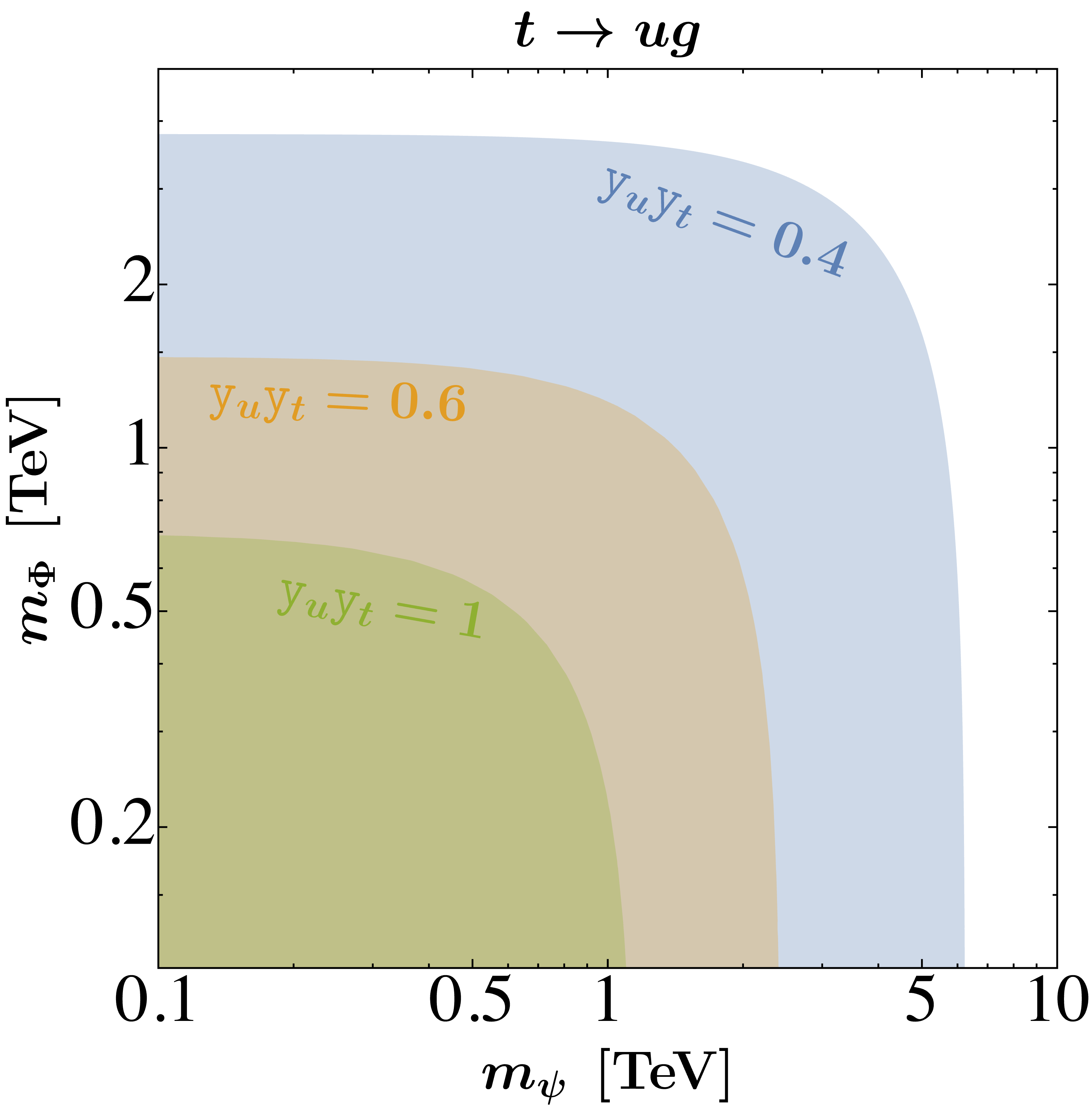}
\caption{In the left (right) panel, the shaded regions represent the allowed parameter space constrained by the decays $t \to c~(u)g$, corresponding to various combinations of the coupling constants $\yc\yt$ and $\yu\yt$, respectively.}
\label{fig:t2cugluon}
\end{figure} 

 \Fig\ref{fig:t2cugluon} shows the allowed parameter space arising from the decays $t \to c(u)g$. The experimental limits on these processes are summarized in \Cref{tab:top_FCNC_prediction}. Unlike previous decay channels, the branching ratios in this case are not distinguished by the chirality of the couplings. Accordingly, both left- and right-handed dipole operators are included in our analysis. However, the contributions from right-handed couplings are significantly suppressed relative to their left-handed ones. As a result, the dominant effects in our study arise from the effective left-handed couplings.
The left panel of \Fig\ref{fig:t2cugluon} shows the allowed parameter space in the $\mpsi-\mphi$ plane corresponding to the $t \to c g$ decay, while the right panel depicts the $t \to u g$ decay. The decay width for $t \to c(u) g$ scales as $\Gamma_{t \to c(u) g} \propto \mathtt{y}_{c(u)}^2 \mathtt{y}*t^2$, similar to the $t \to c(u) Z$ decay. Therefore, we express our results in terms of the coupling product $\mathtt{y}_{c(u)} \mathtt{y}_t$. The different colored regions represent the allowed parameter space for various combinations of couplings. 
From \Cref{tab:top_FCNC_prediction}, we observe that the branching ratio for decays into an up quark is approximately one order of magnitude smaller than that for charm quark final states. As a result, for identical coupling values to both charm and up quarks, the $\tug$ process imposes more stringent constraints on the parameter space. This difference is clearly reflected in \Fig\ref{fig:t2cugluon}, where the left panel shows the constraints from $t \to c g$, and the right panel shows those from $t \to u g$.
The allowed parameter space from the decay process $t \to c g$ is shown for coupling combinations $\yc \yt = \{1.0, 1.3, 3.0\}$, while for $t \to u g$, the corresponding values are $\yu \yt = \{0.4, 0.6, 1.0\}$. For $\yc \yt = 1.0$, the region with $\mphi \gtrsim 4~\TeV$ remains allowed, whereas $\mpsi \gtrsim 6~\TeV$ is excluded by constraints from $t \to c g$. In the case of $t \to u g$ with $\yu \yt = 0.4$, the parameter space with $\mphi \gtrsim 4~\TeV$ and $\mpsi \gtrsim 6~\TeV$ remains viable. As the coupling values increase, the allowed regions become progressively more constrained. For example, at $\yc \yt = 3.0$, the viable parameter space shrinks significantly to $\mphi \lesssim 1 ~\TeV$ and $\mpsi \lesssim 1.5~\TeV$.

To summarise the parameter space constraints from top-FCNC decays, we perform the analysis separately using different top-FCNC decay modes. Significant constraints arise from decays $t \to c(u)X$, where $X = \gamma,\, g,\, Z$. In contrast, the top decay into the Higgs boson does not impose any substantial constraint on the parameter space. Among these, the radiative decays $t \to c(u)\gamma$ and $t \to c(u)g$ yield comparatively stronger bounds. Furthermore, the branching ratios of these decays exhibit a pronounced dependence on the masses of both the DM and VLQ particles, as illustrated in the figures and discussed above. The loop factor grows with the mass of particles in the loop; thus, the branching ratio increases with $m_{\psi}$ and $m_{\Phi}$. As a result, the high-mass region of the parameter space is subject to more stringent constraints from top-FCNC decays.

Additionally, for the same values of \( \yc \yt \) and \( \yu \yt \), the decay to the up quark yields stricter bounds on the mass plane, as the branching ratio to the up quark is more suppressed. In the previous section, we observed that satisfying the $D^0\!-\!\bar{D}^0$ mixing bound requires the product $\yc \yu$ to be very small, of the order $\lesssim \mathcal{O}(10^{-2})$. Therefore, to simultaneously satisfy the bounds from top-FCNC decays and mixing, we require $\yc \yu \sim \mathcal{O}(10^{-2})$ and $\yt \mathtt{y_{c(u)}} \lesssim \mathcal{O}(1)$. This can be achieved either by taking all three couplings to be small, $\lesssim \mathcal{O}(10^{-1})$, or by taking $\yu \sim \mathcal{O}(10^{-2})$ and $\yc, \yt \lesssim \mathcal{O}(1)$. After discussing other constraints from the dark sector and colliders, we will use a few suitable benchmarks for our analysis.

\subsection*{Unified parameter space from all Flavor-Changing processes:}

\begin{figure}[htb!]
\centering
\subfloat[]{\includegraphics[width=0.475\linewidth]{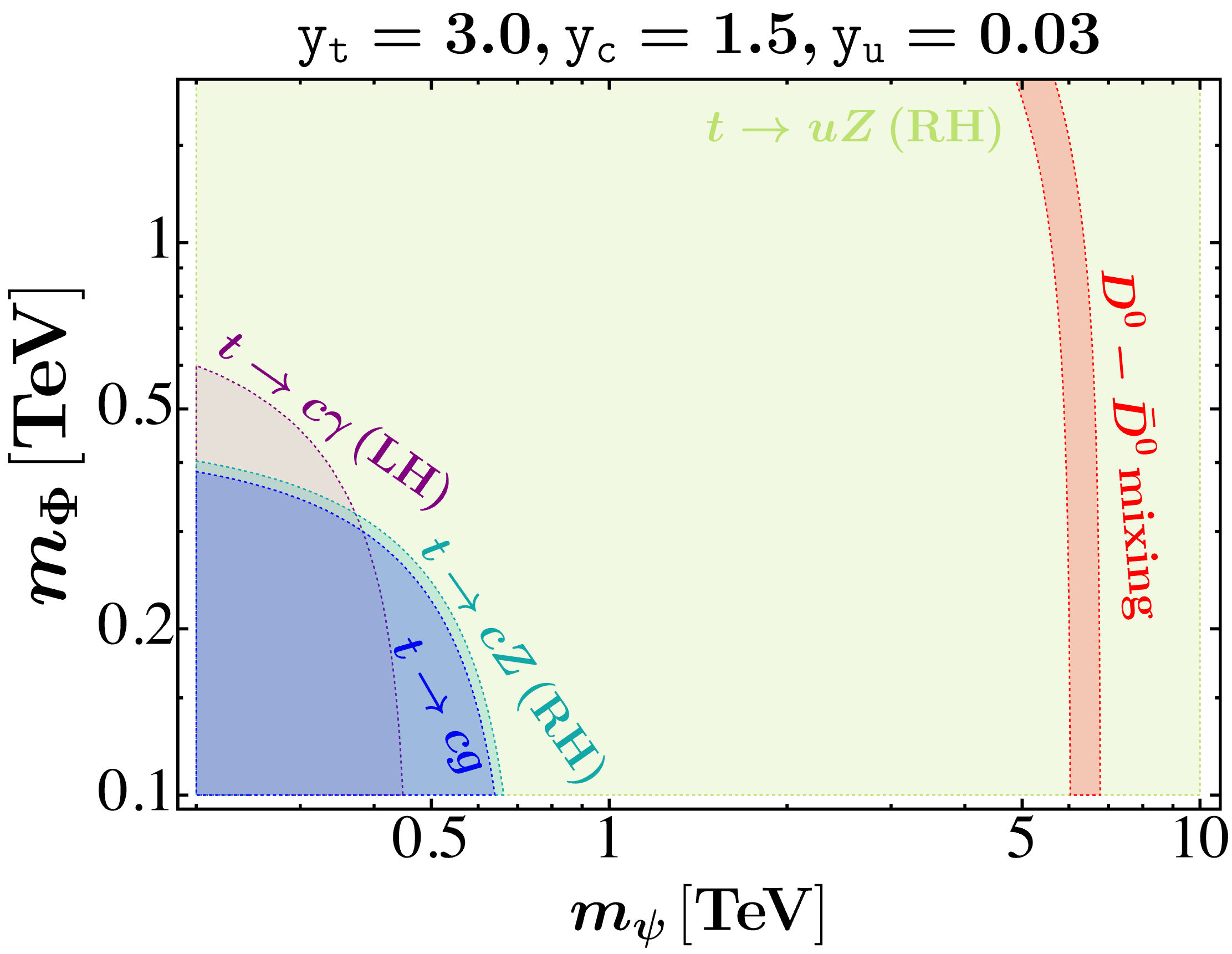}\label{fig:flavor_common_1}}\hspace{0.5cm}
\subfloat[]{\includegraphics[width=0.475\linewidth]{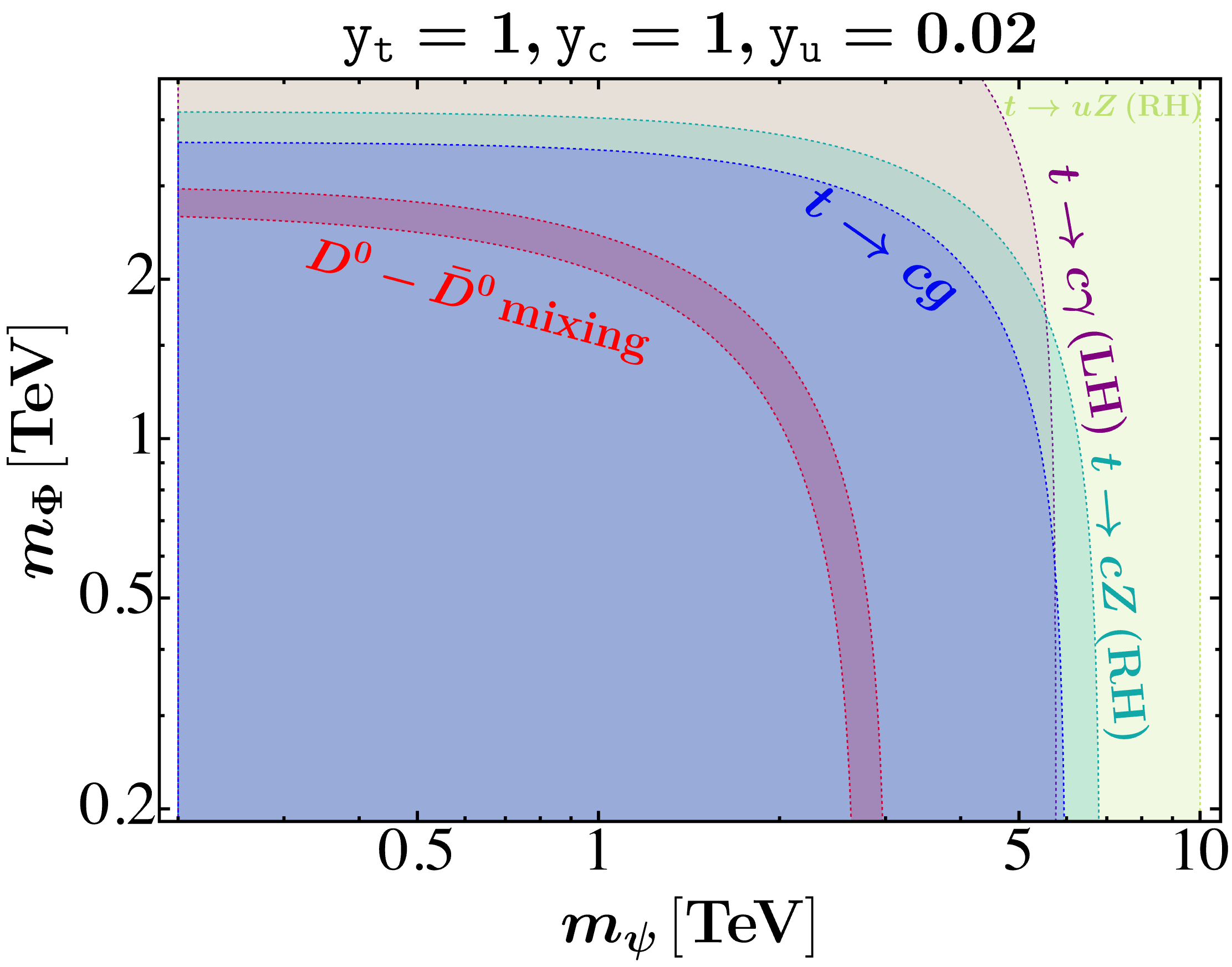}\label{fig:flavor_common_2}}
\caption{The shaded region represents the allowed parameter space constrained by observables from $\ddbar$ mixing and top-FCNC decays.}
\label{fig:Flavor_region}
\end{figure}
In the preceding sections, we analyzed the constraints imposed by various FCNC processes individually. Among these, $\ddbar$ mixing provides the most stringent bounds on the parameter space, while the constraints from top-FCNC decays are comparatively weaker, resulting in a more relaxed allowed region due to the current availability of only upper limits. Here, we perform a combined analysis of all relevant processes to delineate the region of the coupling parameter space that remains viable under the simultaneous application of all constraints. Processes found to be non-constraining in the individual analyses are omitted here, as they are already consistent with the parameter space of interest.

\Fig\ref{fig:Flavor_region} displays the allowed regions in the DM–mediator mass plane (i.e., the $\mphi-\mpsi$ plane), where the colored areas correspond to parameter regions permitted by the respective decay processes, as indicated by labels of the same color. The figure includes constraints from the decays $t \to c\gamma$, $t \to cg$, $\tcZ$, $\tuZ$, and from $\ddbar$ mixing. Other processes—such as $D^0 \to \ell^+ \ell^-$, $t \to c(u)Z$ (LH), and $\tch$, $\tuh$—are not shown, as they impose significantly weaker constraints, as discussed earlier. The plots are generated for two benchmark scenarios corresponding to different coupling choices. In both cases, the coupling $\yu$ is taken to be very small, as required by our DM analysis, in order to evade direct detection bounds.

\Fig\ref{fig:flavor_common_1} show the parameter space for a set of Yukawa couplings: $\yu = 0.03$, $\yc = 1.5$, and $\yt = 3$. The yellow region corresponds to the constraint from $t \to uZ$ (\text{RH}), while the sea-green, purple, blue, and red regions represent the allowed spaces for the processes $\tcZ$ (\text{RH}), $t \to c \gamma$ (\text{LH}), $t \to c g$, and $\ddbar$ mixing, respectively. The region with $\mphi \leq 0.40$ TeV and $\mpsi \leq 0.45$ TeV is allowed by the top-FCNC decays but excluded by the $\ddbar$ mixing process. It is possible to reduce the coupling $\yu$ sufficiently to satisfy all the constraints simultaneously. However, in \Sec\ref{sec:collider-constraints}, we discussed that the smaller mass region is already excluded by the LHC bounds on VLQ mass, which are $\gtrsim 1.5~\TeV$, making this benchmark unsuitable for our analysis.
In contrast, \Fig\ref{fig:flavor_common_2} corresponds to the coupling values $\yu = 0.02$, $\yc = \yt = 1.0$. With $\yc \, \yt = 1$ in this case, a large portion of the parameter space is allowed, extending up to $\mphi \leq 2.5~\TeV$ and $\mpsi \leq 4~\TeV$ from top FCNC decays. The red band represents the region allowed within the $1\sigma$ experimental bound from the $\ddbar$ mixing observable. The region consistent with all processes lies within $\mphi \leq 3.0~\TeV$ and $1.0~\TeV \leq \mpsi \lesssim 3.0~\TeV$, which corresponds to the region of interest for both DM and collider analysis.

\subsection{Dark Matter analysis}
\label{sec:dm-result}
\begin{figure}[htb!]
\centering
\subfloat[]{\includegraphics[width=0.45\linewidth]{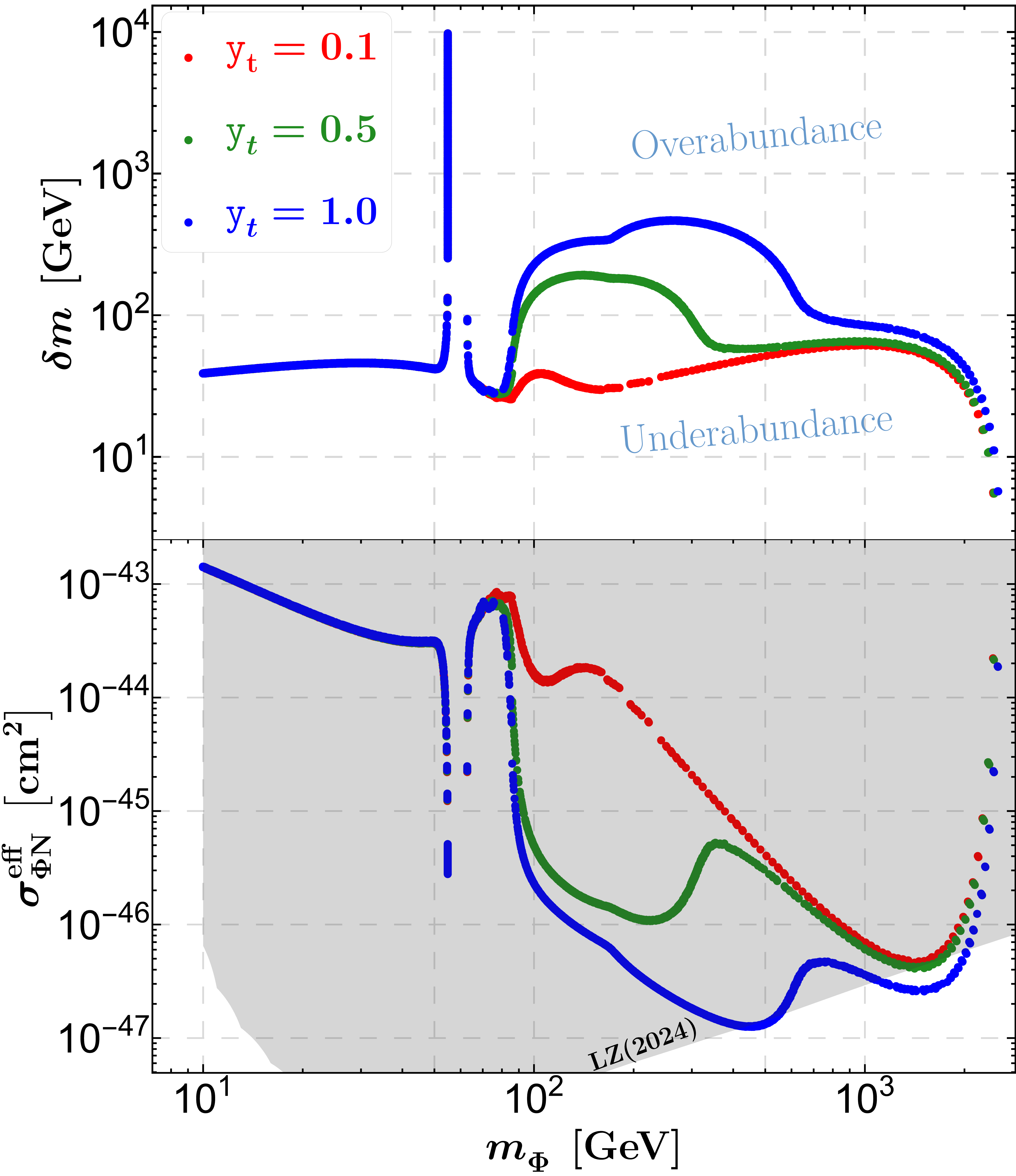}\label{fig:relic1}}\hspace{0.5cm}
\subfloat[]{\includegraphics[width=0.45\linewidth]{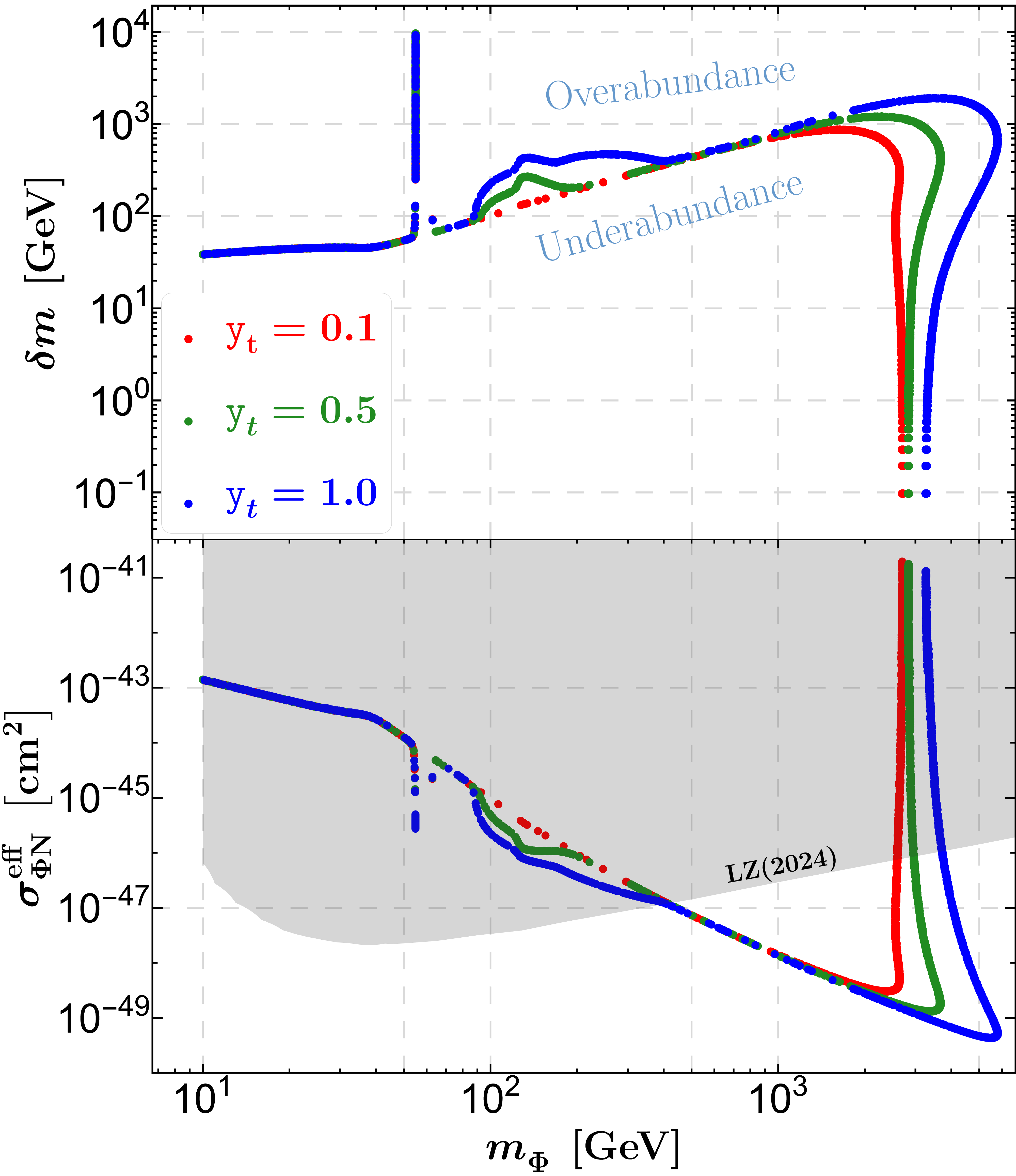}\label{fig:relic2}}
\caption{The left~(right) figure represents the relic allowed parameter space where $\deltam=\mpsi-\mphi,~\yu=10^{-2},~\yc=0.5,~\lphiH=10^{-2},~\mut=\mphi/3~(2\mphi)$.}
\label{fig:relic}
\end{figure}

The relic density of DM is calculated numerically by solving the Boltzmann Equation (BEQ) corresponding to the CSDM using \texttt{micrOMEGAs} \cite{Alguero:2023zol}. The annihilation and co-annihilation channels that dominantly contribute to the DM number density evaluation are shown in \Fig\ref{fig:feynman-relic}. The relic density of DM is inversely related to the annihilation cross-section. It depends on the initial and final state particle masses, which determine whether the process will be contributing or not, and the annihilation rate is directly proportional to the couplings. To analyze the dependence of the model parameters (see \Eq\eqref{eq:free}) on the dark matter relic density and its direct detection prospects, it is necessary to constrain certain parameters by making specific choices.
For the analysis presented in \Fig\ref{fig:relic}, all parameters have been fixed except for $\mphi$ and $\mpsi$ (written in terms of $\deltam = \mpsi - \mphi$), which are varied within the range of approximately $100~\GeV$ to $10~\TeV$, subject to the condition $\mpsi > \mphi$. The different colored lines illustrate the variation of $\yt$. The grey-shaded region in the lower plane represents the parameter space excluded by the latest constraints from the LUX-ZEPLIN experiment on the spin-independent dark matter-nucleon scattering cross-section.

\Fig\ref{fig:relic} represents the parameter space allowed by relic density constraints in the $\mphi-\deltam$ (top) and $\mphi-\sphiN$ (bottom) planes. The left and right panels correspond to different coupling choices: $\mut=\mphi/3$ and $\mut=2\mphi$, respectively.
To systematically investigate the contributions of annihilation and co-annihilation processes to the dark matter relic density, the dark matter mass range is categorized into three regimes based on its mass splitting with the top quark ($\mt$).
The effects of two couplings ($\yu$ and $\lphiH$) are minimized, while $\yt$ is varied across three distinct values along with a fixed $\yc$, ensuring that dark matter remains consistent with direct detection constraints.

\begin{itemize}
\item \texttt{First} ($\mphi< \mt$): If the DM particle mass is significantly smaller than $\mt$, the relic density dynamics in this regime are predominantly governed by the Higgs portal s-channel and $\psi$-mediated t-channel interactions, $\Phi\Phi^* \to c\bar{c}$. However, in the limit of a very small $\yc$, and effectively in the absence of $\psi$, the model reduces to a minimal CSDM scenario.
Here, the choice of $\lphiH$ is very small, rendering its effect subdominant in \Fig\ref{fig:feynman-relic11}. However, if $\yc$ is sufficiently larger than the other couplings, then $\Phi\Phi^* \to c\bar{c}$ (\Fig\ref{fig:feynman-relic13}) becomes the dominant process when $\mphi>\mpsi/2$, while $\Phi\Phi \to c\bar{\psi}$ (\Fig\ref{fig:feynman-relic51}) dominates when $\mphi<\mpsi/2$. As $\deltam$ decreases, \ie as $\mpsi$ becomes smaller, the relic density gradually decreases. Consequently, for $10~\GeV<\mphi<50~\GeV$, the process $\Phi\Phi^* \to c\bar{c}$ predominantly contributes to the observed DM relic density.
For these two benchmark scenarios, around the Higgs resonance (with $\mphi > 50~\GeV$), the only viable channels to achieve the correct relic density are $\Phi\Phi^*\to b\bar{b}$ below $\mh/2$ and $\Phi\Phi^*\to c\bar{\psi}$ above $\mh/2$. A similar reasoning applies to a more underabundant regime for smaller $\deltam$. However, this regime is excluded by the spin-independent direct detection limit on DM-nucleon scattering from the LZ-2024 experiment (see the gray-shaded region in the bottom panel of each figure) for this choice of parameters.

\item \texttt{Second} ($\mphi\sim \mt$): In this regime, the process $\Phi\Phi \to c\bar{\psi}$ dominantly contributes to the relic density. As $\yt$ increases, the process $\Phi\Phi^* \to c\bar{t}$ (\Fig\ref{fig:feynman-relic13}) begins to contribute, leading to a decrease in the relic density, which is compensated by increasing $\mpsi$, \ie $\deltam$. However, when $\mphi$ exceeds $\mt$, the contribution from $\Phi\Phi^* \to c\bar{t}$ becomes subdominant, and subsequently, decreasing $\deltam$ yields the correct relic density.

\item \texttt{Third} ($\mphi> \mt$): If \(\mphi\) is much larger than \(\mt\), the processes \(\Phi\Phi \to c\bar{\psi}\) (dominant) and \(\Phi\Phi \to t\bar{\psi}\) (sub-dominant) contribute to the relic density. Consequently, with the enhancement of \(\yt\), a slight increase in \(\deltam\) is required to adjust the relic density. However, if the DM mass is sufficiently large (\(\mphi > 1~\TeV\)), co-annihilation channels such as \(\Phi\bar{\psi} \to \SM~\SM\) and \(\psi\bar{\psi} \to \SM~\SM\) begin to dominate over DM self-annihilation. In this regime, \(\deltam\) plays a crucial role in determining the relic density, as reducing the mass splitting between co-annihilating partners enhances the annihilation cross-section. At very high masses (\(\mphi > 6~\TeV\)), both self-annihilation and co-annihilation channels become inefficient in achieving the correct relic density, resulting in an overabundant DM scenario. Furthermore, the enhancement of \(\mut\) leads to an increase in the annihilation cross-section, which can be compensated by an increase in either \(\mpsi\) or \(\deltam\), as clearly illustrated in both plots of \Fig\ref{fig:relic}.
\end{itemize}

\begin{figure}[htb!]
\centering
\subfloat[]{\includegraphics[width=0.45\linewidth]{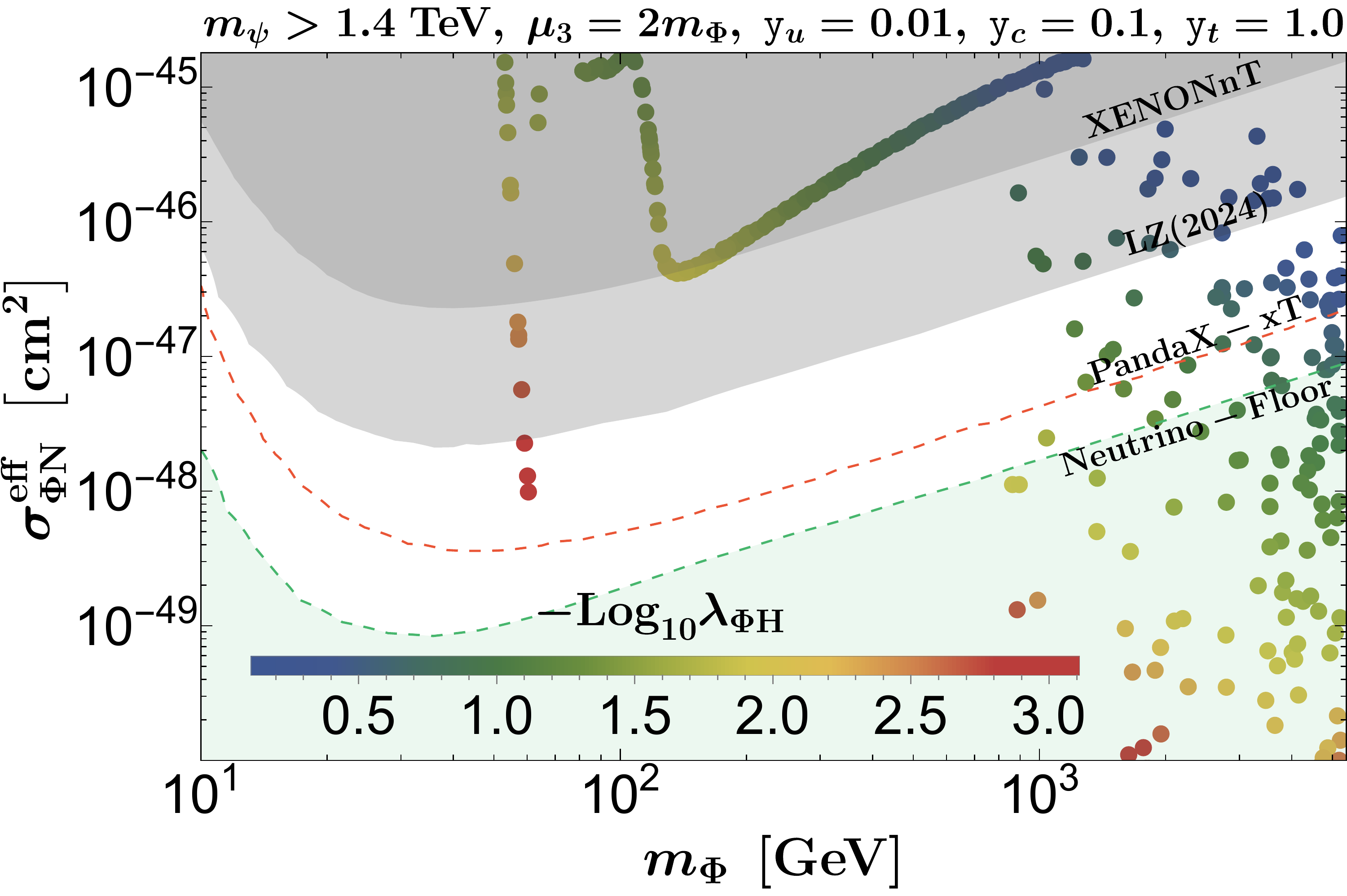}\label{fig:dd1}}\hspace{0.5cm}
\subfloat[]{\includegraphics[width=0.45\linewidth]{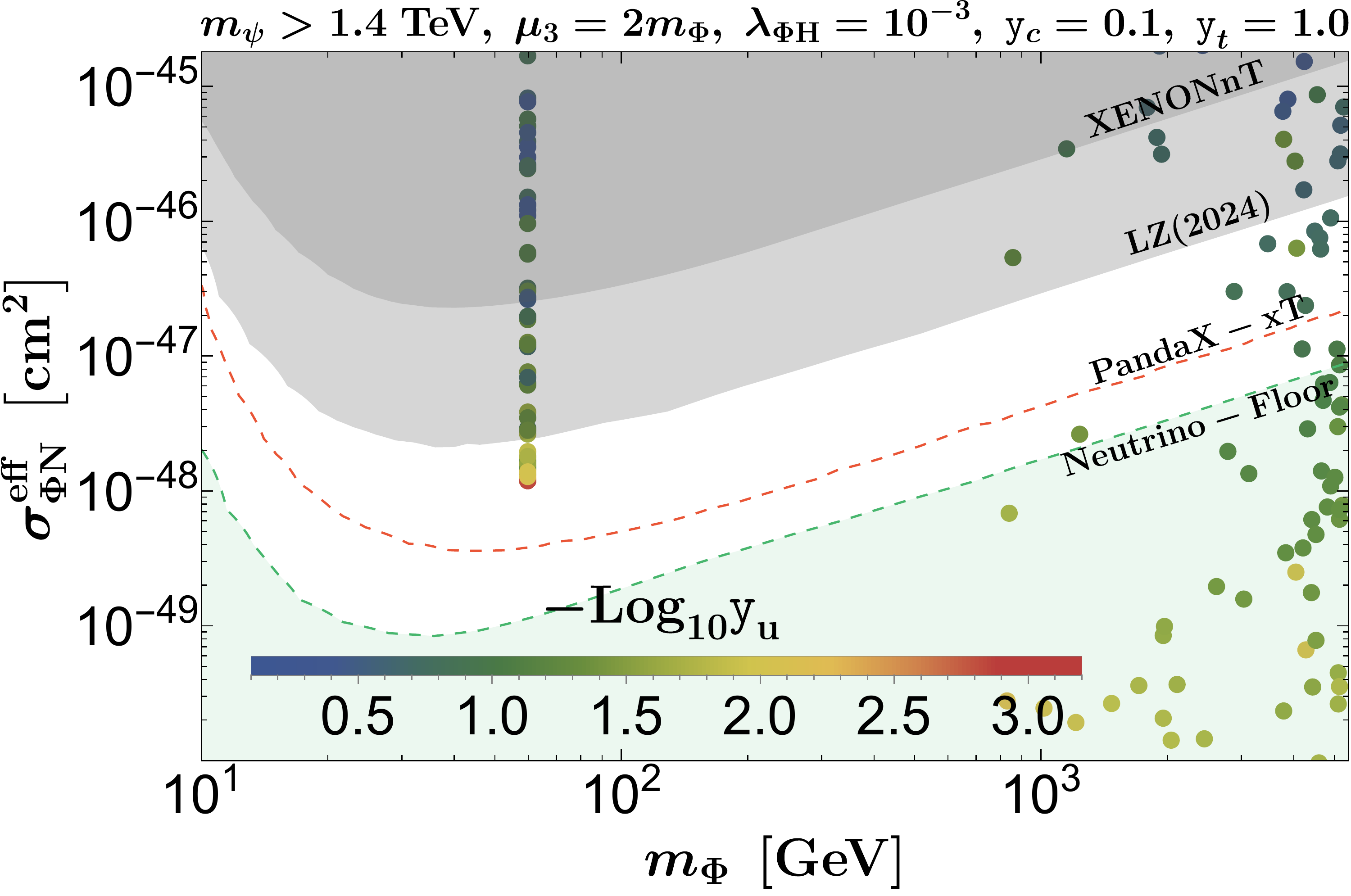}\label{fig:dd2}}
\caption{Figures represent the relic allowed parameter space in the $\mphi-\sphiN$ plane while the rainbow colorbar show the variation of $\lphiH$ (left) and $\yu$ (right). The gray-shaded regions are excluded by the respective DD experiments mentioned in the figure inset.}
\label{fig:dd}
\end{figure}

In \Fig\ref{fig:dd}, we show the relic-allowed parameter space in the $\mphi-\sphiN$ plane. The rainbow color bar represents the variation of $\lphiH$ (left) and $\yu$ (right), while the remaining parameter values are specified at the top of the figure. The gray-shaded regions are excluded by the corresponding experimental constraints, while the green-shaded regions denote the neutrino floor, where distinguishing DM signals from neutrino backgrounds becomes highly challenging in DD experiments. The relevant Feynman diagrams for SI DM-nucleon scattering are shown in \Fig\ref{fig:feynman-dd}.
The cross-section \(\sphiN\) has an implicit dependency on the parameters \(\{\mphi,~\mpsi,~\lphiH,~\yu\}\). In earlier discussions, this dependence was analyzed in the context of the relic density, assuming fixed values of \(\lphiH\) and \(\yu\). In contrast, the present analysis allows these couplings to vary and is represented by the color bar. Given that $\sphiN$ scales directly with $\lphiH$ and $\yu$, smaller values ($\lphiH < 1$ and $\yu < 0.1$) remain consistent with the $\rm LZ-2024$ constraints while still satisfying the DM relic density requirement. In the regime where $\mphi > 1~\TeV$, co-annihilation channels become more effective, enabling further relaxation of the coupling values. Consequently, this leads to an enhanced relic density and a broader region of parameter space allowed by direct detection constraints at higher dark matter masses.

The non-observation of DM may be strongly constrained by indirect searches such as those from Fermi-LAT, H.E.S.S., and Planck. In this work, we consider only the Fermi-LAT data, which impose severe limits on DM annihilation into bottom and up quark pairs, as the corresponding annihilation rates are not sufficiently suppressed to account for the observed DM relic density.
\begin{figure}[htb!]
\centering
\subfloat[]{\includegraphics[width=0.45\linewidth]{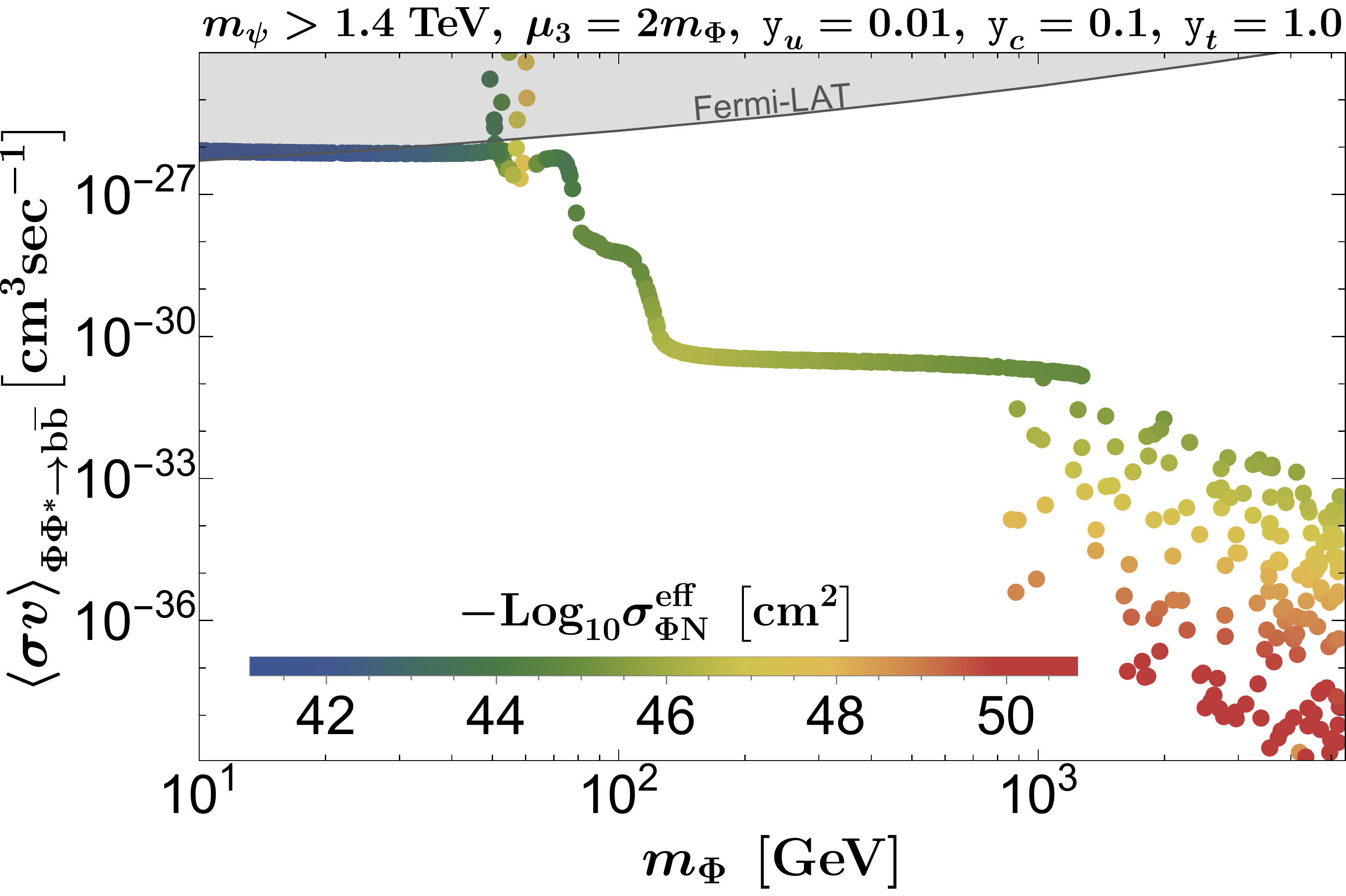}\label{fig:id1}}\hspace{0.5cm}
\subfloat[]{\includegraphics[width=0.45\linewidth]{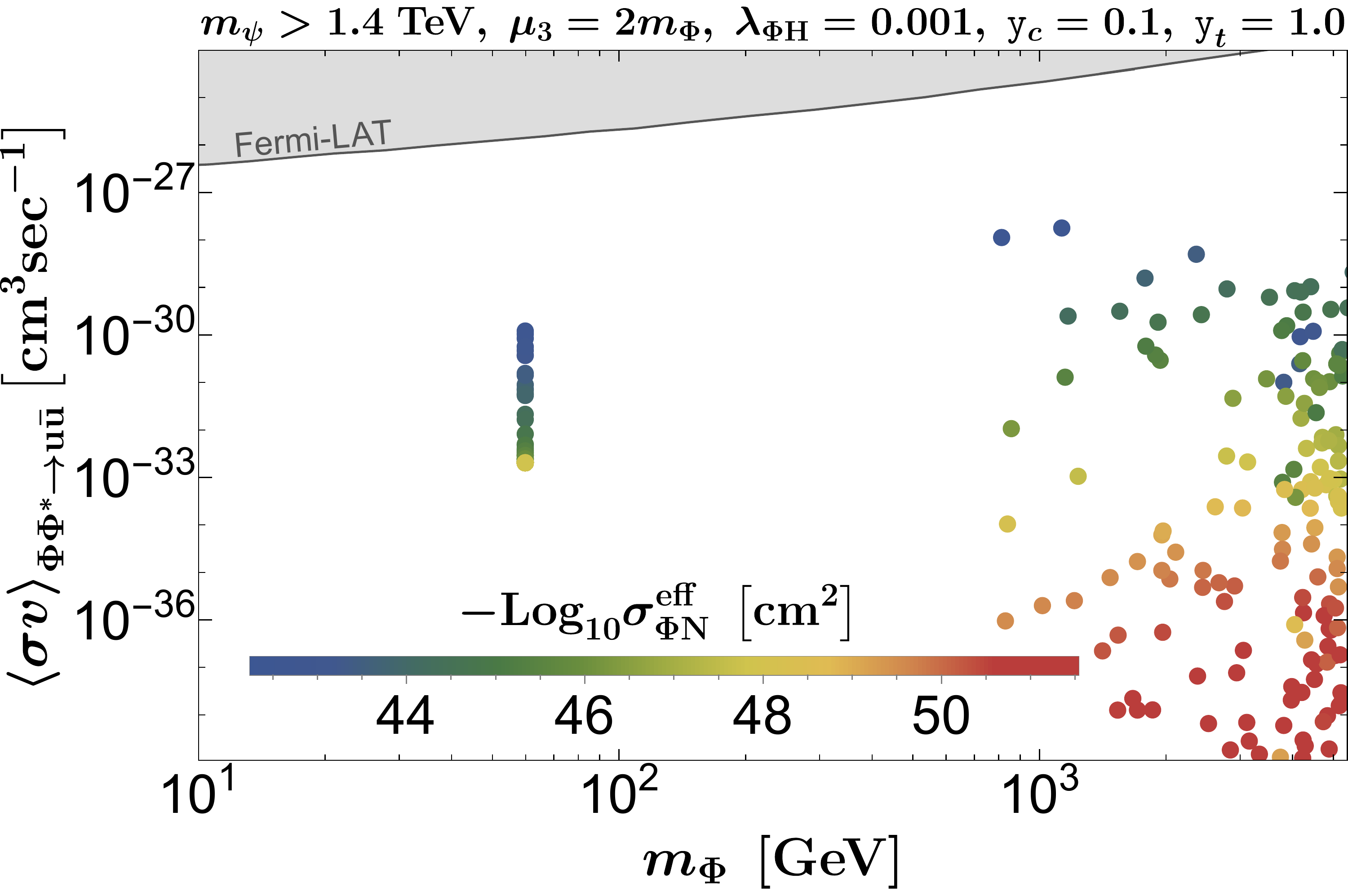}\label{fig:id2}}
\caption{The Fermi-LAT observed bounds are illustrated on the relic allowed parameter-space in $\mphi-\langle\sigma v\rangle_{\Phi\Phi^*\to b\bar{b}}$ (\textit{left}) and $\mphi-\langle\sigma v\rangle_{\Phi\Phi^*\to u\bar{u}}$ (\textit{right}) plane, by the thick gray line.}
\label{fig:id}
\end{figure}
In \Fig\ref{fig:id}, we have represented the DM relic-allowed parameter space in the $\mphi$–$\langle\sigma v\rangle_{\Phi\Phi^*\to u\bar{u}~(b\bar{b})}$ plane. The DM annihilation cross-sections $\mphi$–$\langle\sigma v\rangle_{\Phi\Phi^*\to b\bar{b}}$ and $\mphi$–$\langle\sigma v\rangle_{\Phi\Phi^*\to u\bar{u}}$ are proportionally related to the couplings $\lphiH$ and $\yu$, respectively, similar to the behavior observed in the DD cross-section. As a result, both the DD cross-section ($\sphiN$) and the ID annihilation cross-section decrease with the decrement of these couplings. This trend is depicted using the rainbow color bar, which represents $\rm -Log_{10}\sphiN~[cm^2]$. In \Fig\ref{fig:id1} and \Fig\ref{fig:id2}, we show the variation of the $\lphiH~(\yu)$ coupling as functions of $\sphiN$, represented through the rainbow colorbar, since the DD cross-section depends on these couplings as $\propto\lambda_{\rm \Phi H}^2~(\mathtt{y}_u^4)$. Beyond the Higgs resonance, the DM annihilation cross-section decreases gradually with increasing $\mphi$, and also diminishes with smaller values of $\yu$ and $\lphiH$, as shown in both panels of \Fig\ref{fig:id}. Around the Higgs resonance, some points in \Fig\ref{fig:id1} are excluded by the Fermi-LAT constraint on the DM annihilation cross-section into bottom quark pairs, $\langle\sigma v\rangle_{\Phi\Phi^*\to b\bar{b}}$. In \Fig\ref{fig:id2}, we vary $\yu \geq 0.1$ to ensure that the entire parameter space remains consistent with the Fermi-LAT limit; otherwise, some regions, although allowed by other constraints, would be excluded by this limit.

Therefore, a more suitable choice of couplings should be adopted for analyzing the described flavor data and predicting the collider signal, particularly in the regime of very small $\lphiH$ and $\yu$ couplings. This approach ensures that the parameter space remains consistent with DM relic density constraints while also satisfying the bounds from DD and ID experiments.
\subsection{VLQ search at future muon colliders}
\label{sec:collider-result}
Traditionally, hadron colliders have been the primary facility for searching for VLQs (discussed in \Cref{sec:collider-constraints}), given its connection to the quark sector and large dataset from proton-proton collisions. However, searching for VLQs connected to the dark sector at hadron colliders faces significant challenges, particularly in reconstructing final states involving MET in the presence of complex QCD backgrounds. The overwhelming hadronic environment at the hadron leads to substantial background contamination, making it difficult to isolate new physics signals from SM processes. In contrast, muon colliders present a cleaner environment with distinct advantages for such searches. The reduced QCD background in lepton collisions allows for better signal discrimination with distinct observables. Secondly, muon colliders offer a significantly enhanced ability to probe NP scenarios and exotic decays due to more precise energy control and suppressed initial-state radiation compared to hadron colliders, significantly improving searches involving invisible particles in the final states. Moreover, muon colliders can achieve TeV-scale partonic collision energies with high luminosities, making them well-suited for direct searches of heavy VLQs that may be difficult to detect at the LHC.

\begin{figure}[htb!]
\centering
\begin{tikzpicture}
\begin{feynman}
\vertex(a);
\vertex[above left = 1cm and 1cm  of a] (a1){\(\mu^-\)};
\vertex[below left = 1cm and 1cm  of a] (a2){\(\mu^+\)};
\vertex[right =1.5cm  of a] (b);
\vertex[above right = 0.5cm and 0.5cm  of b] (b1);
\vertex[below right = 0.5cm and 0.5cm  of b] (b2);
\vertex[above right = 0.5cm and 0.5cm  of b1] (b3){\(\Phi\)};
\vertex[below right = 0.5cm and 0.5cm  of b2] (b4){\(\Phi^*\)};
\vertex[right = 0.5cm  of b1] (b11){\(t/c\)};
\vertex[right = 0.5cm  of b2] (b22){\(\overline{c}/\overline{t}\)};
\diagram*{
(a1) -- [line width=0.25mm,fermion, arrow size=0.7pt,style=black] (a),
(a) -- [line width=0.25mm,fermion, arrow size=0.7pt,style=black] (a2),
(a) -- [line width=0.25mm,boson, arrow size=0.7pt,style=black!75,edge label'={\({\color{black}\rm\gamma/Z} \)}] (b),
(b) -- [line width=0.25mm,fermion, arrow size=0.7pt,style=black!50,edge label={\({\color{black}\rm\psi}\)}] (b1),
(b1) -- [line width=0.25mm,charged scalar, arrow size=0.7pt,style=black] (b3),
(b1) -- [line width=0.25mm,fermion, arrow size=0.7pt,style=black] (b11),
(b2) -- [line width=0.25mm,fermion, arrow size=0.7pt,style=black!50,edge label={\({\color{black}\rm\psi} \)}] (b),
(b22) -- [line width=0.25mm,fermion, arrow size=0.7pt,style=black] (b2),
(b4) -- [line width=0.25mm,charged scalar, arrow size=0.7pt,style=black] (b2)};
\end{feynman}
\end{tikzpicture}
\caption{Feynman diagrams correspond to VLQ production ($t\overline{c}/\overline{t}c + \slashed{E}$ signal) at lepton colliders.}
\label{fig:ee}
\end{figure}
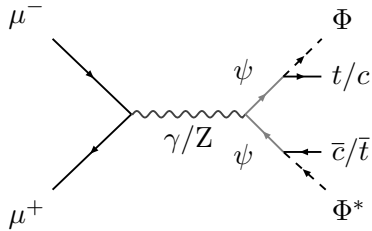

In this section, we study VLQ pair production at the muon collider with CM energy, $\sqrt{s} = 10$ TeV with an integrated luminosity of 10 ab$^{-1}$. We focus on the $\tc$ + missing energy final state, where one VLQ decays to a top quark and DM, and the other VLQ decays to a charm quark and DM. The Feynman diagram for the signal process is shown in \Fig\ref{fig:ee}. At multi-TeV muon colliders, collisions take place at a CM energy surpassing that achieved by the current LHC (parton level) or potential future $e^{+}e^{-}$ colliders. Consequently, it is anticipated that jets which are closely clustered together in these collisions, will exhibit collimation, effectively coalescing into a singular, consolidated jet. A notable illustration of this behavior can be observed in the jets stemming from the hadronic decay processes of top quarks, Higgs or $W/Z$ bosons, where they converge to form a single ``top", ``Higgs" or ``W/Z" jet \cite{Sun:2023cuf, Bhattacharya:2023beo}. Consequently, our signal process reduces to a dijet plus missing energy signature, with one `heavy' jet (resulting from top decay). We further impose the selection criteria of no detected photons or leptons in the final state. The event generation follows the same approach as described in \Cref{sec:collider-constraints}. The detector simulation is done using the default \texttt{Delphes} card in \texttt{MG5\_aMC}, which takes care of detector resolution and tagging efficiencies. The jet clustering is done using the \texttt{anti-kt} algorithm \cite{Cacciari:2005hq, Cacciari:2008gp, Cacciari:2011ma}. The jet radius is chosen to be 0.5, and the minimum $p_{T}$ of the jet is 20 GeV. For the analysis, we choose the benchmark point,
\begin{equation}
    {\rm BP}:\;\;\; \mphi = 800\; {\rm GeV},\;\; \mpsi = 1575\; {\rm GeV}, \;\; (\yu,\yc,\yt) = (0.02,0.75,1.5),
\end{equation}
which satisfy the observed DM relic density and abide by the different constraints discussed in previous sections. 

\begin{figure}[htb!]
\centering
\includegraphics[width=0.475\linewidth]{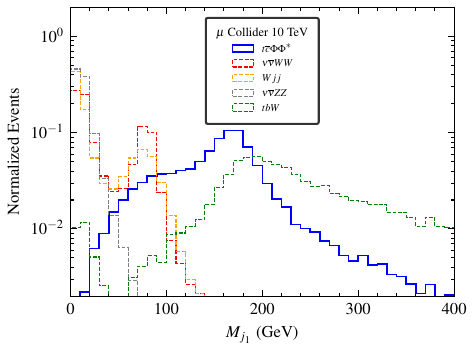}~~
\includegraphics[width=0.475\linewidth]{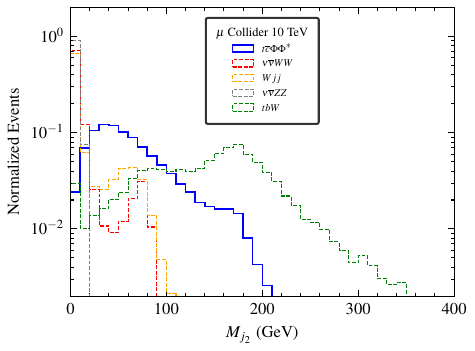} \\
\includegraphics[width=0.475\linewidth]{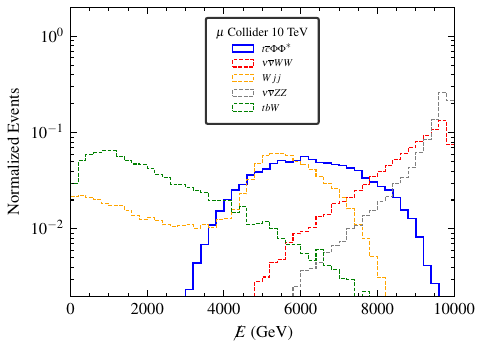}~~
\includegraphics[width=0.475\linewidth]{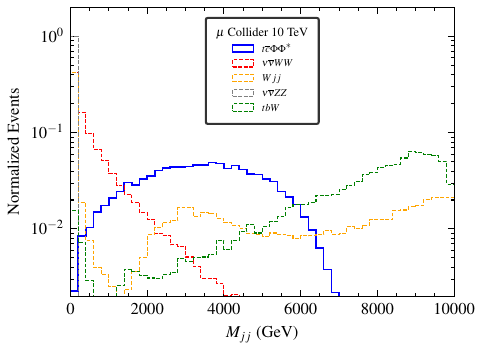}
\caption{Kinematic distributions: \textit{Top left:} Mass of heavier jet, $M_{j_1}$, \textit{Top right:} Mass of lighter jet, $M_{j_2}$, \textit{Bottom left:} Missing energy, $\slashed{E}$, \textit{Bottom right:} Invariant mass of jet pair, $M_{jj}$, corresponding to signal and major background processes at the $10~\TeV$ muon collider.}
\label{fig:muc}
\end{figure}

The major SM background processes are $tbW$, $\nu \overline{\nu} WW$, $Wjj$, $\nu \overline{\nu} ZZ$, $t \overline{t} Z$ and $Zjj$. In addition to the NP signal, $t \overline{c}\Phi\Phi^*$, there are NP backgrounds, $t \overline{t}\Phi\Phi^*$ and $c \overline{c}\Phi\Phi^*$, which should also be taken into account. Some kinematic variables for the NP signal process and the SM backgrounds are plotted in \Fig\ref{fig:muc}. The variables are defined as follows:
\begin{itemize}
    \item $M_{j_1}$ and $M_{j_2}$ are masses of the heavier and lighter jet respectively. Jet mass is defined as the invariant mass of the reconstructed particle constituents of the jet.
    \item The missing energy, $\slashed{E}$ of an event is defined as $\slashed{E} = \sqrt{s} - \sum_{i} E_{i},$ where $\sum_{i} E_{i}$ is the sum of the energy of the visible final state particles.
    \item The invariant mass of the jets, $M_{jj}$ is defined as $M_{jj} = \sqrt{(p_{j_1} + p_{j_2})^{2}}$, where $p_{j_1}$ and $p_{j_2}$ are the 4-momenta of the heavier and lighter jet respectively.
\end{itemize}

\begin{figure}[htb!]
\centering
\includegraphics[width=0.6\linewidth]{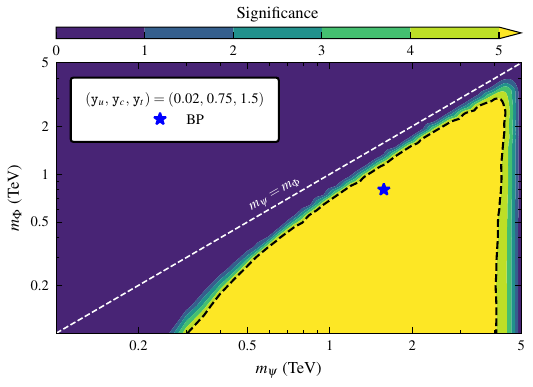}
\caption{Signal significance for $\tc \Phi \Phi^*$ signal process at the muon collider 10 TeV 10 ab$^{-1}$ on the $\mpsi-\mphi$ plane. The black dashed line corresponds to the 5$\sigma$ significance limit. The blue star corresponds to the benchmark point, BP.}
\label{fig:sig}
\end{figure}

In order to segregate the NP signal from the background processes, we apply sequential kinematic cuts (on top of selection and detector cuts discussed earlier):
\begin{itemize}
\itemsep-0em
\item $\mathcal{C}_1: 125$ GeV $< M_{j_1} < 200$ GeV,
\item $\mathcal{C}_2: M_{j_1} < 150$ GeV,
\item $\mathcal{C}_3: 3000$ GeV $< \slashed{E} < 8000$ GeV,
\item $\mathcal{C}_4: 2000$ GeV $< M_{jj} < 7000$ GeV.
\end{itemize}
The $M_{j_1}$ distribution for the signal is expected to peak at the top quark mass, $m_t \sim 175$ GeV. We impose a wide cut, $\mathcal{C}_1$ owing to the difficulty in the reconstruction of constituents of a boosted jet. This effectively reduces the background without top quarks in the final state. The next cut, $\mathcal{C}_2$ is imposed to backgrounds containing more that a single top quark in the final state like the background, $tbW$, where the dominant contribution comes from on-shell $t \overline{t}$ process. The cut windows, $\mathcal{C}_3$ and $\mathcal{C}_4$ are chosen to reduce the remaining background while losing minimum signal in this process. These cuts are selected in order the optimize the signal-background discrimination for our chosen benchmark. The detailed cutflow is shown in \Cref{tab:cuts} for the signal benchmarks and the backgrounds. The backgrounds $t\overline{t}Z$ and $Zjj$ do not survive the cutflow and hence are not mentioned in the table. The signal significance is defined as:
\begin{align}
{\rm Significance} = \frac{\sigma_{s} \times \mathfrak{L}_{\rm int}}{\sqrt{\sigma_{b}}}\,,
\end{align}
where, $\sigma_{s}$ and $\sigma_{b}$ are total cross sections of the signal and backgrounds respectively, after applying the sequential cuts. $\mathfrak{L}_{\rm int}$ is the integrated luminosity of the collider, chosen to be $10~\text{ab}^{-1}$ here. We observe that both the signal benchmark has signal significances higher than 5$\sigma$, the commonly accepted discovery limit for NP signals \cite{ParticleDataGroup:2024cfk}. The signal significance for the $\tc\Phi\Phi^*$ signal is plotted in the $\mpsi-\mphi$ plane, for the two different benchmark cases, in \Fig\ref{fig:sig}.
\begin{table}[htb!]
\centering
\begin{tabular}{|c|c|c|c|c|c|}
\hline
\rowcolor{gray!25} \multicolumn{2}{|c|}{Processes} & $\sigma(\mathcal{C}_1)$ (fb) & $\sigma(\mathcal{C}_2)$ (fb) & $\sigma(\mathcal{C}_3)$ (fb) & $\sigma(\mathcal{C}_4)$ (fb)  \\ \hline \hline
\rowcolor{cyan!10}  & $\nu \overline{\nu} WW$ & $7.955 \times 10^{-1}$ & $7.914 \times 10^{-1}$ & $2.867 \times 10^{-1}$ & $5.249 \times 10^{-2}$ \\
\rowcolor{cyan!10} & $Wjj$ & $5.702 \times 10^{-2}$ & $5.604 \times 10^{-2}$ & $1.640 \times 10^{-2}$ & $1.468 \times 10^{-2}$ \\
\rowcolor{cyan!10} & $tbW$ & $2.205 \times 10^{0}$ & $1.513 \times 10^{0}$ & $7.587 \times 10^{-1}$ & $7.112 \times 10^{-1}$ \\
\rowcolor{cyan!10} \multirow{-4}*{SM} & $\nu \overline{\nu} ZZ$ & $1.251 \times 10^{-1}$ & $1.251 \times 10^{-1}$ & $5.041 \times 10^{-2}$ & $2.054 \times 10^{-2}$ \\  \hline \hline
\rowcolor{magenta!10} & {$\tc \Phi \Phi^*$} & $8.425 \times 10^{-2}$ & $8.058 \times 10^{-2}$ & $7.423 \times 10^{-2}$ & $6.882 \times 10^{-2}$ \\ \cline{2-6}
\rowcolor{magenta!10} & $t \overline{t} \Phi \Phi^*$ & $7.883 \times 10^{-2}$ & $5.771 \times 10^{-2}$ & $5.246 \times 10^{-2}$ & $4.818 \times 10^{-2}$ \\ 
\rowcolor{magenta!10} & $c \overline{c} \Phi \Phi^*$ & $9.310 \times 10^{-3}$ & $9.172 \times 10^{-3}$ & $8.574 \times 10^{-3}$ & $7.933 \times 10^{-3}$ \\ \cline{2-6}
\rowcolor{magenta!10} \multirow{-4}*{BP} & {Significance} & 4.658 & 5.044 & 6.853 & 7.443 \\  \hline
\end{tabular}
\caption{Cutflow and signal significance for $\tc \Phi \Phi^*$ signal process along with NP ($t\bar{t} \Phi \Phi^*$ and $c\bar{c} \Phi \Phi^*$) and SM ($\nu \bar{\nu} WW$, $Wjj$, $tbW$ and $\nu \bar{\nu} ZZ$) backgrounds at the muon collider $10~\TeV~10~\text{ab}^{-1}$. Here, $\sigma(\mathcal{C}_{i})$ refers to as the cross section of the corresponding process following sequential cut, $\mathcal{C}_{i}$.}
\label{tab:cuts}
\end{table}

\subsection{Combined interpretation}
\label{sec:combined-result}

\begin{figure}[htb!]
\centering
\includegraphics[width=0.75\linewidth]{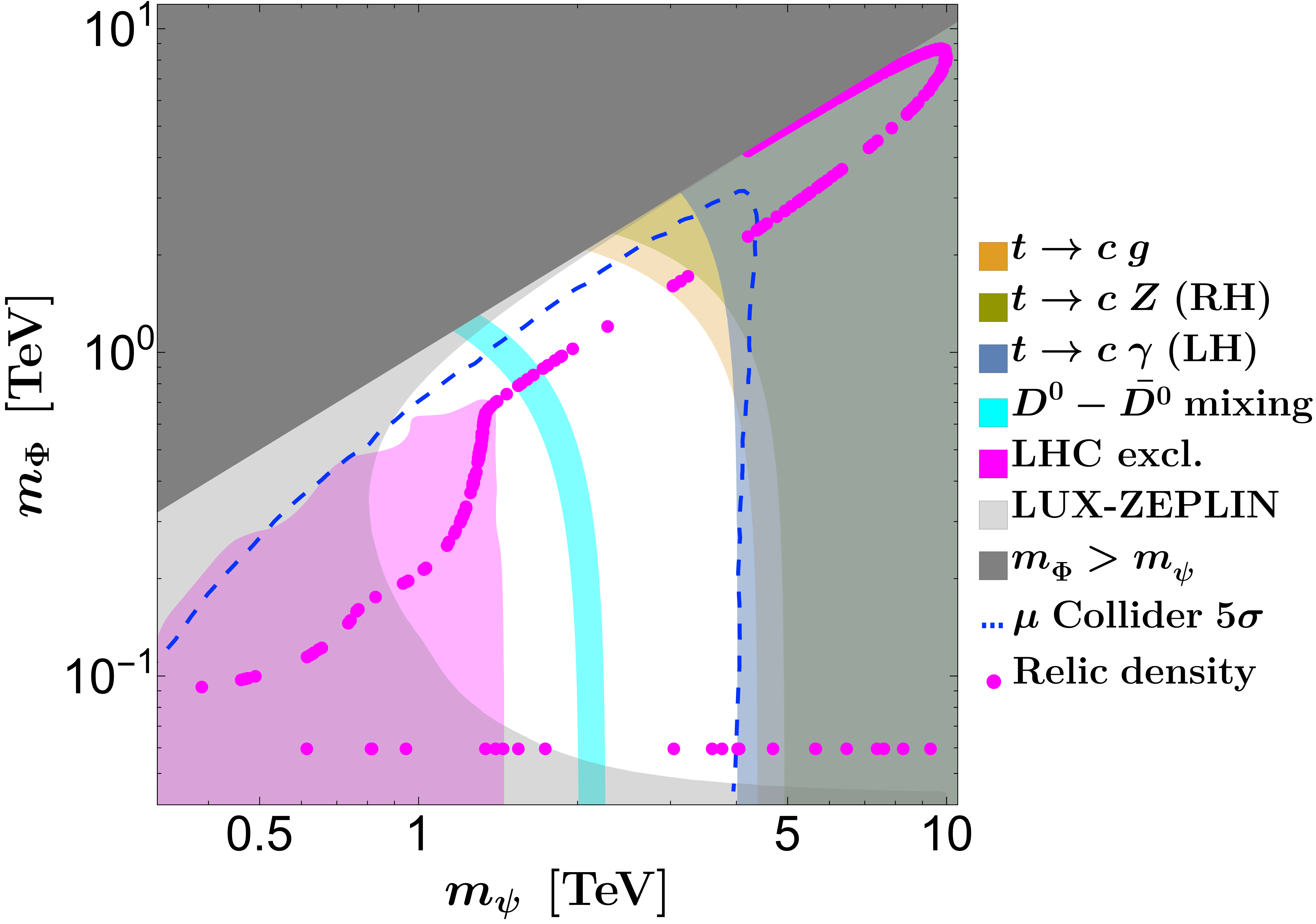}
\caption{The magenta color points in $\mpsi-\mphi$ plane, satisfy the DM relic density and also respect the indirect detection limit on DM annihilation. The other model parameters are fixed as: $\mathtt{y}_u = 0.02,~\mathtt{y}_c = 0.75,~\mathtt{y}_t = 1.5,~\mut = 2\mphi$, and $\lphiH = 10^{-3}$. The cyan-colored band represents the $\ddbar$ mixing. The grey, magenta, light-yellow, light-olive, and light-blue shaded regions are excluded by the LUX-ZEPLIN (2024), LHC, $\tcg$, $\tcZ~\rm (RH)$, and $t\to~c\rm \gamma\,(LH)$ decay, respectively. The dark gray shaded region, corresponding to $\mphi > \mpsi$, is excluded based on the known properties of DM. The blue dashed line represents the $5\sigma$ signal significance at the $10~\TeV$ muon collider with an integrated luminosity of $10~\text{ab}^{-1}$.}
\label{fig:comb}
\end{figure}

In \Fig\ref{fig:comb}, we show a combined interpretation of the exclusion limits and future sensitivities in the $\mpsi - \mphi$ parameter space for a benchmark value: 
\begin{equation}\label{eq:benchmark}
    \mathtt{y}_u = 0.02,~\mathtt{y}_c = 0.75,~\mathtt{y}_t = 1.5,~\mut = 2\mphi,~\text{and} ~\lphiH = 10^{-3}.
\end{equation}
The cyan band represents the region allowed by $\ddbar$ mixing constraints (see \Cref{sec:flavor-constraints} and \ref{sec:flavor-result}). The light-blue shaded region represents the excluded parameter space from the $\tcZ ~(\rm RH)$ decay, discussed in \Cref{sec:flavor-result}.
The grey shaded area is excluded by the SI DM-nucleon scattering limits from LUX-ZEPLIN (2024), discussed in \Cref{sec:dm-result}.
The Magenta shaded region shows exclusion limits from a recast of the $t\bar{t} +$ MET signal at the LHC ($13~\TeV,~139$ fb$^{-1}$) for this benchmark (see \Cref{sec:collider-constraints}).
The blue dashed curve indicates the $5\sigma$ discovery reach for the $\tc + \slashed{E}$ signal at a future $10~\TeV$ muon collider with an integrated luminosity of $10~\text{ab}^{-1}$ (\Cref{sec:collider-result}).
The Magenta colored points indicate the parameters are consistent with the observed DM relic density, and indirect detection limits from Fermi-LAT and Planck observations (\Cref{sec:dm-result}).

It is important to note that the CSDM is only consistent with the observed relic density, DD, and ID constraints in the regions ($\mphi \sim \mh/2$, $\mpsi \gtrsim 2~\TeV$) and ($\mphi \gtrsim \mt$, $\mpsi \gtrsim 1~\TeV$). This compatibility arises due to the small values chosen for $\lphiH$ and $\yu$, which are effective for the spin-independent DD and CSDM annihilation cross-sections relevant to indirect detection. In contrast, the relic density is mostly influenced by other model parameters such as $\yc$, $\yt$, $\mut$, $\mphi$, and $\mpsi$, which are less relevant for DM detection apart from collider studies, where $\yt$ and $\yc$ are significant along the masses.
In the high-mass regime ($\mphi \gtrsim 2~\TeV$ and $\mpsi \gtrsim 4~\TeV$), co-annihilation channels become dominant (as discussed in \Cref{sec:dm-result}), resulting in nearly degenerate CSDM masses for a fixed $\mpsi$ could achieve the correct relic.
This bending kind of behavior arises only in the higher mass regime ($\mphi\gtrsim 1~\TeV$) due to the following reasons: if the CSDM mass is $\sim \mpsi/2$, then the $\Phi\Phi^* \to \qu\bar{\psi}$ channel (\Fig\ref{fig:feynman-relic51}) dominantly contributes to the DM freeze-out and gives a relic-allowed points.
However, if we gradually increase $\mphi$ for a fixed $\mpsi$, other co-annihilation processes (\Fig\ref{fig:feynman-relic}) start to contribute and yield under-abundant relic points.
On the contrary, for $\mpsi/2\ll\mphi\lesssim \mpsi$, the effect of the co-annihilation channels starts to decrease, leading to another correct relic point before entering into the overabundance regime.

Finally, in \Fig\ref{fig:comb}, we see that the points around $\mpsi \sim 1.5~\TeV$ and $\mphi \sim 1~\TeV$ are consistent with all current experimental limits for this benchmark.
However, other benchmark scenarios may also satisfy these constraints and could be probed by future experiments. But the model parameters simultaneously influence dark matter, flavor, and collider phenomenology. Therefore, the viable parameter space is highly constrained and typically requires fine-tuning to simultaneously accommodate all observables.

\subsection{Prediction of $t \to c (u) \ell \ell$ branching ratios}
\label{sec:prediction}
In \Cref{sec:constraints}, we discussed how the VLQ and DM can modify the top-FCNC processes, allowing us to derive constraints on the three couplings $\yu$, $\yc$, and $\yt$. \Fig\ref{fig:comb} represents the combined allowed parameter space for a choice of benchmark point. Since significant constraints arise from top-FCNC processes such as $t \to c \gamma$, $t \to c g$, and $\tcZ$, it is also interesting to investigate the impact of these couplings on the branching ratios of the rare decays $t \to c \ell \ell$ and $t \to u \ell \ell$. Furthermore, we analyze how closely the benchmark point approaches current experimental limits and evaluate its viability in the context of more stringent future constraints.

\begin{table}[htb!]
\centering
\renewcommand{\arraystretch}{1.3}
\setlength{\tabcolsep}{5pt}
\begin{tabular}{|c|c|}
\hline
\rowcolor{cyan!5} 
\multicolumn{2}{|c|}{ }  \\
\rowcolor{cyan!5} 
\multicolumn{2}{|c|}{\multirow{-2}{*}{$\boldsymbol{\yu = 0.02 \,, \yc = 0.75 \,, \yt = 1.5\,, \mpsi = 1.5  ~\TeV$  and $\mphi = 0.7 ~\TeV}$}}
 \\ \hline \hline
\rowcolor{gray!10}\textbf{Process}& \textbf{Our Prediction} \\  \hline
\rowcolor{magenta!5} $\mathcal{B}(t \to c \ell \ell )$ &   $1.56 \times 10^{-8}$  \\
\rowcolor{magenta!5} $\mathcal{B}(t \to u \ell \ell )$  &   $ 1.10 \times 10^{-11}$     \\ 
\rowcolor{magenta!5} $\mathcal{B}(t \to c \nu \bar{\nu} )$ &   $ 1.17 \times 10^{-12}$   \\ \hline

\rowcolor{lime!5}  $\mathcal{B}(t \to c g )$    &  $1.84 \times 10^{-4} $\\ 
\rowcolor{lime!5}  $\mathcal{B}(t \to u g )$    & $1.31 \times 10^{-7} $\\

\rowcolor{brown!5} $\mathcal{B}(t \to c \gamma \,(\rm LH) \, )$   & $ 1.99  \times 10^{-5}$  \\
\rowcolor{brown!5} $\mathcal{B}(t \to u \gamma\, (\rm LH) \, )$   & $ 1.41  \times 10^{-8}$  \\ 

\rowcolor{green!5} $\mathcal{B}(t \to c Z \, \rm (RH) )$    & $5.45 \times 10^{-5} $ \\ 
\rowcolor{green!5} $\mathcal{B}(t \to u Z  \, \rm (RH) )$   & $3.86 \times 10^{-8} $  \\ 

\rowcolor{olive!5} $\mathcal{B}(t \to c Z \, \rm (LH) )$  &  $2.93 \times 10^{-6} $  \\ 
\rowcolor{olive!5} $\mathcal{B}(t \to u Z  \, \rm (LH) )$ & $2.08 \times 10^{-9}$ \\ 

\rowcolor{blue!5}  $\mathcal{B}(t \to c h )$       & $1.13 \times 10^{-9} $ \\ 
\rowcolor{blue!5}  $\mathcal{B}(t \to u h ) $      & $8.07 \times 10^{-13} $ \\   \hline
\end{tabular}
\caption{The predictions for benchmark model parameters, considering left (right) handed effective couplings denoted by $\rm LH~(RH)$, are presented, while the SM predictions and experimental bounds on top-FCNC decays have already been discussed in \Cref{tab:top_FCNC_prediction}.}
\label{tab:top-FCNC_prediction2}
\end{table}
\Cref{tab:top-FCNC_prediction2} shows the predictions of rare top-FCNC decays like $t \to c (u) \ell \ell$ along with the processes that we have used to constrain our parameter space. The branching ratio of the semi-leptonic top decay is very suppressed in the SM. In SM, their branching ratio, calculated at one-loop, is given by \cite{Frank:2006ku}:
\begin{align}
\label{eq:t2cll_SM}
&\mathcal{B}( t \to c e^{+} e^{-} )_{\rm SM} = 8.48 \times 10^{-15} \,, \\
&\mathcal{B}( t \to c \mu^{+} \mu^{-} )_{\rm SM} = 9.55 \times 10^{-15} \,,  \\
&\mathcal{B}( t \to c \tau^{+} \tau^{-} )_{\rm SM} = 1.91 \times 10^{-15} \,, \\
&\mathcal{B}( t \to c \nu \bar{\nu} )_{\rm SM} = 2.99 \times 10^{-14} \,.
\end{align}
As can be seen from \Eq \eqref{eq:t2cll_SM}, the branching ratio of the semi-leptonic decays of top-quark in SM is $\sim \mathcal{O} (10^{-15})$. However, no experimental bounds are available for these processes.

For the final benchmark point, we conducted a combined analysis considering constraints from $\ddbar$, top-FCNC decays, dark sector constraints such as relic density and DD cross-section bounds, along with the LHC bound, as given in \Eq\eqref{eq:benchmark}. Here, we choose a benchmark point for the mass of the DM and VLQ as, $m_{\Phi} = 700~\GeV$ and $m_{\psi} = 1.5~\TeV$. 
\Cref{tab:top-FCNC_prediction2} presents the predictions for all FCNC processes involving the top quark. The effects of the final state lepton and quark mass are negligible in our case, as the dominant contribution arises from photon and $Z$-mediated decays. Here, $\ell = (e, \mu)$, as these two leptons yield the same branching ratio in the final state. The branching ratio obtained for $t \to c \ell \ell$ is $\sim \mathcal{O}(10^{-8})$, which is approximately $\mathcal{O}(10^{7})$ times larger than the SM prediction but several orders smaller than the current existing bound. In our benchmark point, $\mathtt{y_{c}}/ \mathtt{y_{u}} = 2.67 \times 10^{-2}$. Therefore, the difference in the branching ratio of processes involving the $u$ quark is approximately $\mathcal{O}(10^{-3})$ orders smaller, as shown in Table~\ref{tab:top-FCNC_prediction2}.

We have also predicted the branching ratios for the processes $t \to q X$, with $X = (\gamma, g, h, Z)$, to assess their compatibility with current experimental bounds. Except for $t \to c h$, all processes with a charm quark in the final state lie very close to the present experimental limits. In contrast, the branching ratios for processes involving an up quark in the final state are suppressed, as expected from the earlier discussion.

\paragraph{\underline{Prediction of $Z \to c u $}:} Along with the top FCNC processes, we will also have a contribution to the $Z \to c u$ decay. 
The SM prediction for this decay at NLO is given by \cite{dEnterria:2023wjq}:
\begin{equation}
\mathcal{B}(Z \to \bar{c} u ) = 1.7 \times 10^{-18}\,.
\end{equation}
However, we will use the total branching ratio: 
\begin{equation}
    \mathcal{B}(Z \to c u) = \mathcal{B}(Z \to c \bar{u}) +\mathcal{B}(Z \to \bar{c} u)\,.  
\end{equation}
There are currently no experimental upper limits available for this decay. Our predicted branching ratio at the final benchmark point, as defined in \Eq\eqref{eq:benchmark}, is given below:
\begin{equation}
\mathcal{B}(Z \to c u) = 1.33 \times 10^{-12}\,.
\end{equation}
So, our prediction is $\mathcal{O}(10^{6})$ order higher than the value of SM.

\section{Summary and Conclusion}
\label{sec:summary}
In this article, we present a simple model that addresses two key issues in particle physics: Dark Matter (DM) and the observables related to FCNC processes. 
This is accomplished by extending the Standard Model (SM) particle contents with a complex scalar field and a singlet Dirac vector-like quark (VLQ), which couples 
with the SM up-type quarks due to chosen $\rm U(1)_{\mathtt{Y}}$ hypercharge.
In this minimal extension, the complex scalar field acts as the DM candidate, with its stability ensured by an unbroken $\mathcal{Z}_3$ symmetry. 
The VLQ transforms similarly under this symmetry but is chosen to be heavier than the complex scalar, allowing it to decay into the DM and visible particles. 

The study elaborates upon the correct DM relic density allowed parameter space consistent with the current direct detection (DD) and indirect detection (ID) limits. Due to the existence of heavy dark sector VLQ, the depletion is aided by co-annihilation processes whenever the mass splitting remains small. 
Apart, the $\mathcal{Z}_3$ symmetry allows $\Phi^3$ interaction, which plays a crucial role in achieving the correct relic density via semi-annihilation processes 
of the type $\Phi\Phi \to \bar{\psi} u$, whenever $2\mphi > \mpsi$. Apart, the interaction between the dark sector and visible sector particles is also mediated by 
the VLQ and Higgs portal interaction, while the latter is suppressed to remain consistent with current DD and ID constraints. The Yukawa couplings between 
VLQ, DM, and SM particles play a crucial role in DM relic as well as FCNC and collider observables. 

Existing LHC searches for DM via missing energy impose strong bounds, pushing both the VLQ and DM particles into the high-mass regime. 
The most stringent limits come from the observation of $t\bar{t} + \rm MET$ at ATLAS, which yields $\mpsi \gtrsim 1.4~\TeV$ for $(\mphi + \mt) < \mpsi$.

As the model offers interaction between VLQ, DM, and right-handed up-type quarks, it contributes to FCNC processes; we have considered $\ddbar$ mixing, 
rare decays of $D^{0}$ meson, and top-FCNC decays. The processes related to $D^{0}$ depend on the couplings $\yu$ and $\yc$, while top-FCNCs are impacted by all three Yukawa couplings. The most relevant flavor observables, $\ddbar$ mixing, $\tcZ$ (RH), $\tcg$, and $t\to c\gamma$ (LH), significantly impact DM phenomenology and collider search prospects.
Similar extensions can be constructed for left-handed quark doublets or right-handed down-type quarks, with appropriate charge assignments for the VLQ and DM. Such variations would introduce additional flavor constraints from the other low-energy observables and open up a direct connection between the DM and the full quark sector. A more comprehensive analysis can be taken up in future in this direction. 

We explore the viable parameter space of the model consistent with the constraints outlined in \Cref{sec:constraints}, and provide predictions 
in future muon collider experiments operating at $\sqrt{s} = 10~\TeV$ via $t\bar{c}$ plus missing energy. Missing energy and invariant mass of the jet pair help in signal-background segregation. Due to high VLQ mass, the prospects at the LHC seems difficult, and so is at future electron positron machines. 
The summary plot (\Fig\ref{fig:comb}) represents the relic-allowed parameter space for a benchmark scenario consistent with direct, indirect DM search constraints, 
collider sensitivity, observed $\ddbar$ mixing data, and the exclusion limits from $\tcg$, $t\to c\gamma$ (LH), $\tcZ$ (RH) decays. 

In summary, our model presents a minimal yet compelling framework linking DM and flavor physics, where there exists a correlation between both sectors. 
It remains consistent with current experimental searches, while the detection seems to be possible at future muon colliders, providing a compelling reason to build such machines to explore BSM physics. 

%

\newpage
\appendix
\section{Feynman diagrams related to DM relic}
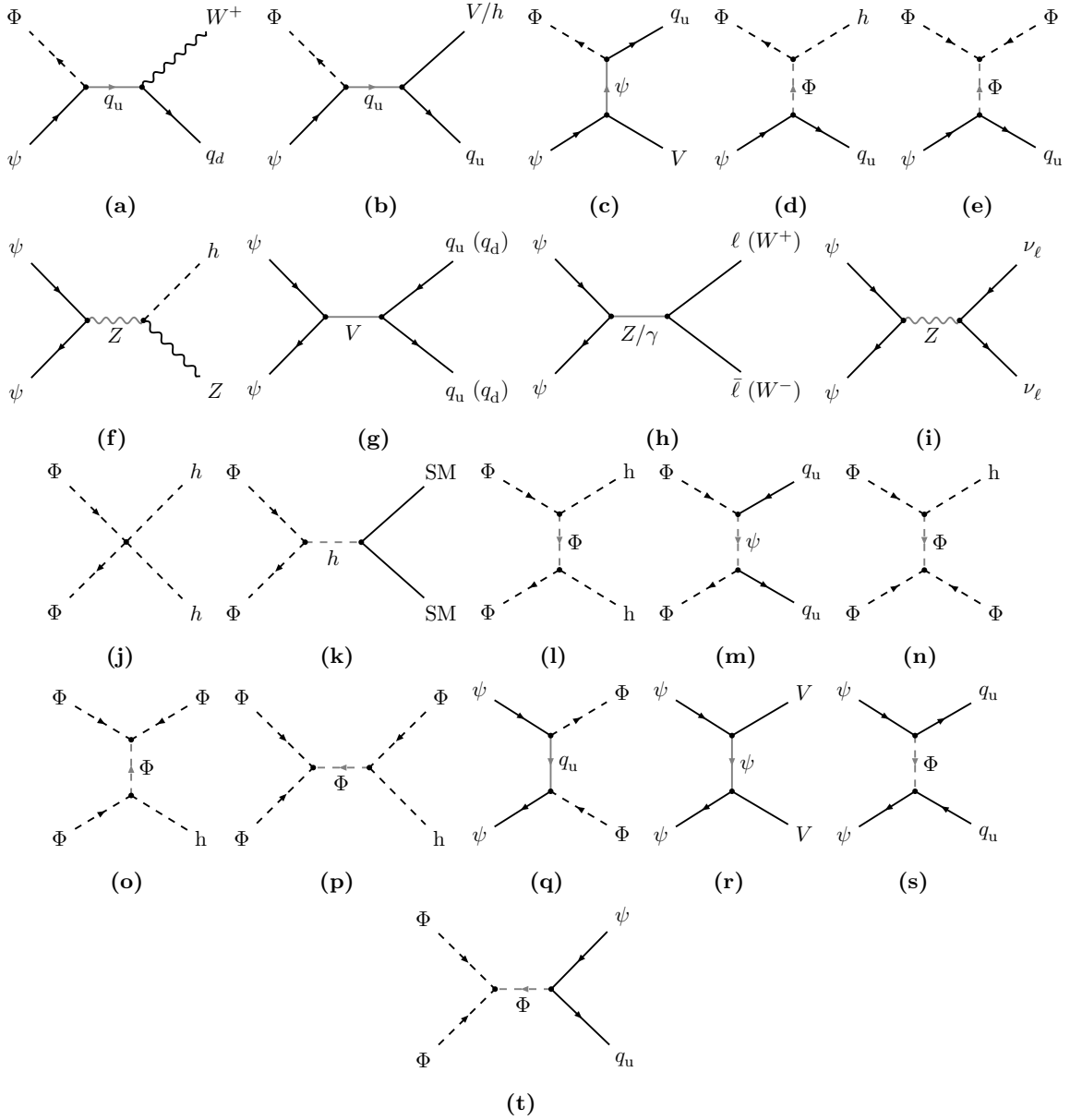
\begin{figure}[htb!]
\centering
\subfloat[]{\begin{tikzpicture}[baseline={(current bounding box.center)},style={scale=0.80, transform shape}]
\begin{feynman}
\vertex (a);
\vertex[above left = 1cm and 1cm  of a] (a1){\(\Phi\)};
\vertex[below left = 1cm and 1cm  of a] (a2){\(\psi\)}; 
\vertex[right = 1cm of a] (b); 
\vertex[above right = 1cm and 1cm of b] (c1){\(W^+\)};
\vertex[below right = 1cm and 1cm of b] (c2){\(q_d\)};
\diagram*{
(a) -- [line width=0.25mm, charged scalar, arrow size=0.7pt,style=black] (a1),
(a2) -- [line width=0.25mm,fermion, arrow size=0.7pt,style=black] (a),
(a) -- [line width=0.25mm,fermion,arrow size=0.7pt, edge label'={\(\color{black}{\qu}\)}, style=black!50] (b),
(b) -- [line width=0.25mm, boson, arrow size=0.7pt] (c1),
(b) -- [line width=0.25mm,fermion, arrow size=0.7pt] (c2)};
\node at (a)[circle,fill,style=black,inner sep=1pt]{};
\node at (b)[circle,fill,style=black,inner sep=1pt]{};
\end{feynman}
\end{tikzpicture}
\label{fig:feynman-relic1}}
\subfloat[]{\begin{tikzpicture}[baseline={(current bounding box.center)},style={scale=0.80, transform shape}]
\begin{feynman}
\vertex (a);
\vertex[above left=1cm and 1cm  of a] (a1){\(\Phi\)};
\vertex[below left=1cm and 1cm  of a] (a2){\(\psi\)}; 
\vertex[right=1cm of a] (b); 
\vertex[above right=1cm and 1cm of b] (c1){\(V/h\)};
\vertex[below right=1cm and 1cm of b] (c2){\(\qu\)};
\diagram*{
(a) -- [line width=0.25mm, charged scalar, arrow size=0.7pt,style=black] (a1),
(a2) -- [line width=0.25mm,fermion, arrow size=0.7pt,style=black] (a),
(a) -- [line width=0.25mm,fermion,arrow size=0.7pt, edge label'={\(\color{black}{\qu}\)}, style=black!50] (b),
(b) -- [line width=0.25mm, plain, arrow size=0.7pt] (c1),
(b) -- [line width=0.25mm,fermion, arrow size=0.7pt] (c2)};
\node at (a)[circle,fill,style=black,inner sep=1pt]{};
\node at (b)[circle,fill,style=black,inner sep=1pt]{};
\end{feynman}
\end{tikzpicture}
\label{fig:feynman-relic2}}
\subfloat[]{\begin{tikzpicture}[baseline={(current bounding box.center)},style={scale=0.80, transform shape}]
\begin{feynman}
\vertex (a);
\vertex[above left=0.5cm and 1cm of a] (a1){\(\Phi\)};
\vertex[above right=0.5cm and 1cm  of a] (a2){\(\qu \)}; 
\vertex[below = 1.0cm of a] (b); 
\vertex[below left=0.5cm and 1cm of b] (c1){\(\psi\)};
\vertex[below right=0.5cm and 1cm of b] (c2){\(V\)};
\diagram*{
(a) -- [line width=0.25mm, charged scalar, arrow size=0.7pt,style=black] (a1),
(a) -- [line width=0.25mm, fermion, arrow size=0.7pt,style=black] (a2),
(b) -- [line width=0.25mm, fermion, arrow size=0.7pt, edge label'={\(\color{black}{\psi}\)}, style=black!50] (a) ,
(c1) -- [line width=0.25mm, fermion, arrow size=0.7pt] (b),
(b) -- [line width=0.25mm, plain, arrow size=0.7pt] (c2)};
\node at (a)[circle,fill,style=black,inner sep=1pt]{};
\node at (b)[circle,fill,style=black,inner sep=1pt]{};
\end{feynman}
\end{tikzpicture}
\label{fig:feynman-relic3}}
\subfloat[]{\begin{tikzpicture}[baseline={(current bounding box.center)},style={scale=0.80, transform shape}]
\begin{feynman}
\vertex (a);
\vertex[above left=0.5cm and 1cm of a] (a1){\(\Phi\)};
\vertex[above right=0.5cm and 1cm  of a] (a2){\(h \)}; 
\vertex[below = 1.0cm of a] (b); 
\vertex[below left=0.5cm and 1cm of b] (c1){\(\psi\)};
\vertex[below right=0.5cm and 1cm of b] (c2){\(\qu\)};
\diagram*{
(a) -- [line width=0.25mm,charged scalar, arrow size=0.7pt,style=black] (a1),
(a2) -- [line width=0.25mm, scalar, arrow size=0.7pt,style=black] (a),
(b) -- [line width=0.25mm, charged scalar, arrow size=0.7pt, edge label'={\(\color{black}{\Phi}\)}, style=black!50] (a) ,
(c1) -- [line width=0.25mm,fermion, arrow size=0.7pt] (b),
(b) -- [line width=0.25mm,fermion, arrow size=0.7pt] (c2)};
\node at (a)[circle,fill,style=black,inner sep=1pt]{};
\node at (b)[circle,fill,style=black,inner sep=1pt]{};
\end{feynman}
\end{tikzpicture}
\label{fig:feynman-relic4}}
\subfloat[]{\begin{tikzpicture}[baseline={(current bounding box.center)},style={scale=0.80, transform shape}]
\begin{feynman}
\vertex (a);
\vertex[above left=0.5cm and 1cm of a] (a1){\(\Phi\)};
\vertex[above right=0.5cm and 1cm  of a] (a2){\(\Phi \)}; 
\vertex[below = 1.0cm of a] (b); 
\vertex[below left=0.5cm and 1cm of b] (c1){\(\psi\)};
\vertex[below right=0.5cm and 1cm of b] (c2){\(\qu\)};
\diagram*{
(a1) -- [line width=0.25mm,charged scalar, arrow size=0.7pt,style=black] (a),
(a2) -- [line width=0.25mm,charged scalar, arrow size=0.7pt,style=black] (a),
(b) -- [line width=0.25mm, charged scalar, arrow size=0.7pt, edge label'={\(\color{black}{\Phi}\)}, style=black!50] (a) ,
(c1) -- [line width=0.25mm,fermion, arrow size=0.7pt] (b),
(b) -- [line width=0.25mm,fermion, arrow size=0.7pt] (c2)};
\node at (a)[circle,fill,style=black,inner sep=1pt]{};
\node at (b)[circle,fill,style=black,inner sep=1pt]{};
\end{feynman}
\end{tikzpicture}
\label{fig:feynman-relic5}}

\subfloat[]{\begin{tikzpicture}[baseline={(current bounding box.center)},style={scale=0.80, transform shape}]
\begin{feynman}
\vertex (a);
\vertex[above left=1cm and 1cm  of a] (a1){\(\psi\)};
\vertex[below left=1cm and 1cm  of a] (a2){\(\psi\)}; 
\vertex[right=1cm of a] (b);
\vertex[above right=1cm and 1cm of b] (c1){\(h\)};
\vertex[below right=1cm and 1cm of b] (c2){\(Z\)};
\diagram*{
(a1) -- [line width=0.25mm, fermion, arrow size=0.7pt,style=black] (a),
(a)  -- [line width=0.25mm, fermion, arrow size=0.7pt,style=black] (a2),
(a)  -- [line width=0.25mm, boson,arrow size=0.7pt, edge label'={\(\color{black}{Z}\)}, style=black!50] (b),
(b) -- [line width=0.25mm, scalar, arrow size=0.7pt] (c1),
(b) -- [line width=0.25mm, boson, arrow size=0.7pt] (c2)};
\node at (a)[circle,fill,style=black,inner sep=1pt]{};
\node at (b)[circle,fill,style=black,inner sep=1pt]{};
\end{feynman}
\end{tikzpicture}
\label{fig:feynman-relic6}}
\subfloat[]{\begin{tikzpicture}[baseline={(current bounding box.center)},style={scale=0.80, transform shape}]
\begin{feynman}
\vertex (a);
\vertex[above left=1cm and 1cm  of a] (a1){\(\psi\)};
\vertex[below left=1cm and 1cm  of a] (a2){\(\psi\)}; 
\vertex[right=1cm of a] (b);
\vertex[above right=1cm and 1cm of b] (c1){\(\qu~(\qd)\)};
\vertex[below right=1cm and 1cm of b] (c2){\(\qu~(\qd)\)};
\diagram*{
(a1) -- [line width=0.25mm, fermion, arrow size=0.7pt,style=black] (a),
(a)  -- [line width=0.25mm, fermion, arrow size=0.7pt,style=black] (a2),
(a)  -- [line width=0.25mm, plain,arrow size=0.7pt, edge label'={\(\color{black}{V}\)}, style=black!50] (b),
(c1) -- [line width=0.25mm, fermion, arrow size=0.7pt] (b),
(b) -- [line width=0.25mm, fermion, arrow size=0.7pt] (c2)};
\node at (a)[circle,fill,style=black,inner sep=1pt]{};
\node at (b)[circle,fill,style=black,inner sep=1pt]{};
\end{feynman}
\end{tikzpicture}
\label{fig:feynman-relic7}}
\subfloat[]{\begin{tikzpicture}[baseline={(current bounding box.center)},style={scale=0.80, transform shape}]
\begin{feynman}
\vertex (a);
\vertex[above left=1cm and 1cm  of a] (a1){\(\psi\)};
\vertex[below left=1cm and 1cm  of a] (a2){\(\psi\)}; 
\vertex[right=1cm of a] (b);
\vertex[above right=1cm and 1cm of b] (c1){\(\ell~(W^+)\)};
\vertex[below right=1cm and 1cm of b] (c2){\(\bar{\ell}~(W^-)\)};
\diagram*{
(a1) -- [line width=0.25mm, fermion, arrow size=0.7pt,style=black] (a),
(a)  -- [line width=0.25mm, fermion, arrow size=0.7pt,style=black] (a2),
(a)  -- [line width=0.25mm, plain,arrow size=0.7pt, edge label'={\(\color{black}{Z/\gamma}\)}, style=black!50] (b),
(c1) -- [line width=0.25mm, plain, arrow size=0.7pt] (b),
(b) -- [line width=0.25mm, plain, arrow size=0.7pt] (c2)};
\node at (a)[circle,fill,style=black,inner sep=1pt]{};
\node at (b)[circle,fill,style=black,inner sep=1pt]{};
\end{feynman}
\end{tikzpicture}
\label{fig:feynman-relic8}}
\subfloat[]{\begin{tikzpicture}[baseline={(current bounding box.center)},style={scale=0.80, transform shape}]
\begin{feynman}
\vertex (a);
\vertex[above left=1cm and 1cm  of a] (a1){\(\psi\)};
\vertex[below left=1cm and 1cm  of a] (a2){\(\psi\)}; 
\vertex[right=1cm of a] (b);
\vertex[above right=1cm and 1cm of b] (c1){\(\nu^{}_\ell\)};
\vertex[below right=1cm and 1cm of b] (c2){\(\nu^{}_{\ell}\)};
\diagram*{
(a1) -- [line width=0.25mm, fermion, arrow size=0.7pt,style=black] (a),
(a)  -- [line width=0.25mm, fermion, arrow size=0.7pt,style=black] (a2),
(a)  -- [line width=0.25mm, boson,arrow size=0.7pt, edge label'={\(\color{black}{Z}\)}, style=black!50] (b),
(c1) -- [line width=0.25mm, fermion, arrow size=0.7pt] (b),
(b) -- [line width=0.25mm, fermion, arrow size=0.7pt] (c2)};
\node at (a)[circle,fill,style=black,inner sep=1pt]{};
\node at (b)[circle,fill,style=black,inner sep=1pt]{};
\end{feynman}
\end{tikzpicture}
\label{fig:feynman-relic9}}

\subfloat[]{\begin{tikzpicture}[baseline={(current bounding box.center)},style={scale=0.80, transform shape}]
\begin{feynman}
\vertex (a);
\vertex[above left=1cm and 1cm  of a] (a1){\(\Phi\)};
\vertex[below left=1cm and 1cm  of a] (a2){\(\Phi\)}; 
\vertex[right=1cm of a] (b);
\vertex[above right=1cm and 1cm of a] (c1){\(h\)};
\vertex[below right=1cm and 1cm of a] (c2){\(h\)};
\diagram*{
(a1) -- [line width=0.25mm, charged scalar, arrow size=0.7pt,style=black] (a),
(a)  -- [line width=0.25mm, charged scalar, arrow size=0.7pt,style=black] (a2),
(a) -- [line width=0.25mm, scalar, arrow size=0.7pt] (c1),
(a) -- [line width=0.25mm, scalar, arrow size=0.7pt] (c2)};
\node at (a)[circle,fill,style=black,inner sep=1pt]{};
\end{feynman}
\end{tikzpicture}
\label{fig:feynman-relic10}}
\subfloat[]{\begin{tikzpicture}[baseline={(current bounding box.center)},style={scale=0.80, transform shape}]
\begin{feynman}
\vertex (a);
\vertex[above left=1cm and 1cm  of a] (a1){\(\Phi\)};
\vertex[below left=1cm and 1cm  of a] (a2){\(\Phi\)}; 
\vertex[right=1cm of a] (b);
\vertex[above right=1cm and 1cm of b] (c1){\(\rm SM\)};
\vertex[below right=1cm and 1cm of b] (c2){\(\rm SM\)};
\diagram*{
(a1) -- [line width=0.25mm, charged scalar, arrow size=0.7pt,style=black] (a),
(a)  -- [line width=0.25mm, charged scalar, arrow size=0.7pt,style=black] (a2),
(a)  -- [line width=0.25mm, scalar, arrow size=0.7pt, edge label'={\(\color{black}{h}\)}, style=black!50] (b),
(b) -- [line width=0.25mm, plain, arrow size=0.7pt] (c1),
(b) -- [line width=0.25mm, plain, arrow size=0.7pt] (c2)};
\node at (a)[circle,fill,style=black,inner sep=1pt]{};
\node at (b)[circle,fill,style=black,inner sep=1pt]{};
\end{feynman}
\end{tikzpicture}
\label{fig:feynman-relic11}}
\subfloat[]{\begin{tikzpicture}[baseline={(current bounding box.center)},style={scale=0.80, transform shape}]
\begin{feynman}
\vertex (a);
\vertex[above left=0.5cm and 1cm of a] (a1){\(\Phi\)};
\vertex[above right=0.5cm and 1cm  of a] (a2){\(\rm h \)}; 
\vertex[below = 1.0cm of a] (b); 
\vertex[below left=0.5cm and 1cm of b] (c1){\(\Phi\)};
\vertex[below right=0.5cm and 1cm of b] (c2){\(\rm h\)};
\diagram*{
(a1) -- [line width=0.25mm, charged scalar, arrow size=0.7pt,style=black] (a),
(a2) -- [line width=0.25mm, scalar, arrow size=0.7pt,style=black] (a),
(a) -- [line width=0.25mm, charged scalar, arrow size=0.7pt, edge label={\(\color{black}{\Phi}\)}, style=black!50] (b) ,
(b) -- [line width=0.25mm, charged scalar, arrow size=0.7pt] (c1),
(b) -- [line width=0.25mm, scalar, arrow size=0.7pt] (c2)};
\node at (a)[circle,fill,style=black,inner sep=1pt]{};
\node at (b)[circle,fill,style=black,inner sep=1pt]{};
\end{feynman}
\end{tikzpicture}
\label{fig:feynman-relic12}}
\subfloat[]{\begin{tikzpicture}[baseline={(current bounding box.center)},style={scale=0.80, transform shape}]
\begin{feynman}
\vertex (a);
\vertex[above left=0.5cm and 1cm of a] (a1){\(\Phi\)};
\vertex[above right=0.5cm and 1cm  of a] (a2){\(\qu \)}; 
\vertex[below = 1.0cm of a] (b); 
\vertex[below left=0.5cm and 1cm of b] (c1){\(\Phi\)};
\vertex[below right=0.5cm and 1cm of b] (c2){\(\qu\)};
\diagram*{
(a1) -- [line width=0.25mm, charged scalar, arrow size=0.7pt,style=black] (a),
(a2) -- [line width=0.25mm, fermion, arrow size=0.7pt,style=black] (a),
(a) -- [line width=0.25mm, charged scalar, arrow size=0.7pt, edge label={\(\color{black}{\psi}\)}, style=black!50] (b) ,
(b) -- [line width=0.25mm, charged scalar, arrow size=0.7pt] (c1),
(b) -- [line width=0.25mm, fermion, arrow size=0.7pt] (c2)};
\node at (a)[circle,fill,style=black,inner sep=1pt]{};
\node at (b)[circle,fill,style=black,inner sep=1pt]{};
\end{feynman}
\end{tikzpicture}
\label{fig:feynman-relic13}}
\subfloat[]{\begin{tikzpicture}[baseline={(current bounding box.center)},style={scale=0.80, transform shape}]
\begin{feynman}
\vertex (a);
\vertex[above left=0.5cm and 1cm of a] (a1){\(\Phi\)};
\vertex[above right=0.5cm and 1cm  of a] (a2){\(\rm h \)}; 
\vertex[below = 1.0cm of a] (b); 
\vertex[below left=0.5cm and 1cm of b] (c1){\(\Phi\)};
\vertex[below right=0.5cm and 1cm of b] (c2){\(\Phi\)};
\diagram*{
(a1) -- [line width=0.25mm, charged scalar, arrow size=0.7pt,style=black] (a),
(a2) -- [line width=0.25mm, scalar, arrow size=0.7pt,style=black] (a),
(a) -- [line width=0.25mm, charged scalar, arrow size=0.7pt, edge label={\(\color{black}{\Phi}\)}, style=black!50] (b) ,
(c1) -- [line width=0.25mm, charged scalar, arrow size=0.7pt] (b),
(c2) -- [line width=0.25mm, charged scalar, arrow size=0.7pt] (b)};
\node at (a)[circle,fill,style=black,inner sep=1pt]{};
\node at (b)[circle,fill,style=black,inner sep=1pt]{};
\end{feynman}
\end{tikzpicture}
\label{fig:feynman-relic14}}

\subfloat[]{\begin{tikzpicture}[baseline={(current bounding box.center)},style={scale=0.80, transform shape}]
\begin{feynman}
\vertex (a);
\vertex[above left=0.5cm and 1cm of a] (a1){\(\Phi\)};
\vertex[above right=0.5cm and 1cm  of a] (a2){\(\Phi \)}; 
\vertex[below = 1.0cm of a] (b); 
\vertex[below left=0.5cm and 1cm of b] (c1){\(\Phi\)};
\vertex[below right=0.5cm and 1cm of b] (c2){\(\rm h\)};
\diagram*{
(a1) -- [line width=0.25mm, charged scalar, arrow size=0.7pt,style=black] (a),
(a2) -- [line width=0.25mm, charged scalar, arrow size=0.7pt,style=black] (a),
(b) -- [line width=0.25mm, charged scalar, arrow size=0.7pt, edge label'={\(\color{black}{\Phi}\)}, style=black!50] (a) ,
(c1) -- [line width=0.25mm, charged scalar, arrow size=0.7pt] (b),
(c2) -- [line width=0.25mm, scalar, arrow size=0.7pt] (b)};
\node at (a)[circle,fill,style=black,inner sep=1pt]{};
\node at (b)[circle,fill,style=black,inner sep=1pt]{};
\end{feynman}
\end{tikzpicture}
\label{fig:feynman-relic15}}
\subfloat[]{\begin{tikzpicture}[baseline={(current bounding box.center)},style={scale=0.80, transform shape}]
\begin{feynman}
\vertex (a);
\vertex[above left=1cm and 1cm  of a] (a1){\(\Phi\)};
\vertex[below left=1cm and 1cm  of a] (a2){\(\Phi\)}; 
\vertex[right=1cm of a] (b); 
\vertex[above right=1cm and 1cm of b] (c1){\(\Phi\)};
\vertex[below right=1cm and 1cm of b] (c2){\(\rm h\)};
\diagram*{
(a1) -- [line width=0.25mm, charged scalar, arrow size=0.7pt,style=black] (a),
(a2) -- [line width=0.25mm, charged scalar, arrow size=0.7pt,style=black] (a),
(b) -- [line width=0.25mm, charged scalar,arrow size=0.7pt, edge label={\(\color{black}{\Phi}\)}, style=black!50] (a),
(c1) -- [line width=0.25mm, charged scalar, arrow size=0.7pt] (b),
(b) -- [line width=0.25mm, scalar, arrow size=0.7pt] (c2)};
\node at (a)[circle,fill,style=black,inner sep=1pt]{};
\node at (b)[circle,fill,style=black,inner sep=1pt]{};
\end{feynman}
\end{tikzpicture}
\label{fig:feynman-relic16}}
\subfloat[]{\begin{tikzpicture}[baseline={(current bounding box.center)},style={scale=0.80, transform shape}]
\begin{feynman}
\vertex (a);
\vertex[above left=0.5cm and 1cm of a] (a1){\(\psi\)};
\vertex[above right=0.5cm and 1cm  of a] (a2){\(\Phi \)}; 
\vertex[below = 1.0cm of a] (b); 
\vertex[below left=0.5cm and 1cm of b] (c1){\(\psi\)};
\vertex[below right=0.5cm and 1cm of b] (c2){\(\Phi\)};
\diagram*{
(a1) -- [line width=0.25mm, fermion, arrow size=0.7pt,style=black] (a),
(a) -- [line width=0.25mm, charged scalar, arrow size=0.7pt,style=black] (a2),
(a) -- [line width=0.25mm, fermion, arrow size=0.7pt, edge label={\(\color{black}{\qu}\)}, style=black!50] (b),
(c2) -- [line width=0.25mm, charged scalar, arrow size=0.7pt] (b),
(b) -- [line width=0.25mm, fermion, arrow size=0.7pt] (c1)};
\node at (a)[circle,fill,style=black,inner sep=1pt]{};
\node at (b)[circle,fill,style=black,inner sep=1pt]{};
\end{feynman}
\end{tikzpicture}
\label{fig:feynman-relic17}}
\subfloat[]{\begin{tikzpicture}[baseline={(current bounding box.center)},style={scale=0.80, transform shape}]
\begin{feynman}
\vertex (a);
\vertex[above left=0.5cm and 1cm of a] (a1){\(\psi\)};
\vertex[above right=0.5cm and 1cm  of a] (a2){\(V \)}; 
\vertex[below = 1.0cm of a] (b); 
\vertex[below left=0.5cm and 1cm of b] (c1){\(\psi\)};
\vertex[below right=0.5cm and 1cm of b] (c2){\(V\)};
\diagram*{
(a1) -- [line width=0.25mm, fermion, arrow size=0.7pt,style=black] (a),
(a) -- [line width=0.25mm, plain, arrow size=0.7pt,style=black] (a2),
(a) -- [line width=0.25mm, fermion, arrow size=0.7pt, edge label={\(\color{black}{\psi}\)}, style=black!50] (b),
(c2) -- [line width=0.25mm, plain, arrow size=0.7pt] (b),
(b) -- [line width=0.25mm, fermion, arrow size=0.7pt] (c1)};
\node at (a)[circle,fill,style=black,inner sep=1pt]{};
\node at (b)[circle,fill,style=black,inner sep=1pt]{};
\end{feynman}
\end{tikzpicture}
\label{fig:feynman-relic18}}
\subfloat[]{\begin{tikzpicture}[baseline={(current bounding box.center)},style={scale=0.80, transform shape}]
\begin{feynman}
\vertex (a);
\vertex[above left=0.5cm and 1cm of a] (a1){\(\psi\)};
\vertex[above right=0.5cm and 1cm  of a] (a2){\(\qu \)}; 
\vertex[below = 1.0cm of a] (b); 
\vertex[below left=0.5cm and 1cm of b] (c1){\(\psi\)};
\vertex[below right=0.5cm and 1cm of b] (c2){\(\qu\)};
\diagram*{
(a1) -- [line width=0.25mm, fermion, arrow size=0.7pt,style=black] (a),
(a) -- [line width=0.25mm, fermion, arrow size=0.7pt,style=black] (a2),
(a) -- [line width=0.25mm, charged scalar, arrow size=0.7pt, edge label={\(\color{black}{\Phi}\)}, style=black!50] (b),
(c2) -- [line width=0.25mm, fermion, arrow size=0.7pt] (b),
(b) -- [line width=0.25mm, fermion, arrow size=0.7pt] (c1)};
\node at (a)[circle,fill,style=black,inner sep=1pt]{};
\node at (b)[circle,fill,style=black,inner sep=1pt]{};
\end{feynman}
\end{tikzpicture}
\label{fig:feynman-relic19}}

\subfloat[]{\begin{tikzpicture}[baseline={(current bounding box.center)},style={scale=0.80, transform shape}]
\begin{feynman}
\vertex (a);
\vertex[above left=1cm and 1cm  of a] (a1){\(\Phi\)};
\vertex[below left=1cm and 1cm  of a] (a2){\(\Phi\)}; 
\vertex[right=1cm of a] (b); 
\vertex[above right=1cm and 1cm of b] (c1){\(\psi\)};
\vertex[below right=1cm and 1cm of b] (c2){\(\qu\)};
\diagram*{
(a1) -- [line width=0.25mm, charged scalar, arrow size=0.7pt,style=black] (a),
(a2) -- [line width=0.25mm, charged scalar, arrow size=0.7pt,style=black] (a),
(b) -- [line width=0.25mm, charged scalar,arrow size=0.7pt, edge label={\(\color{black}{\Phi}\)}, style=black!50] (a),
(c1) -- [line width=0.25mm, fermion, arrow size=0.7pt] (b),
(b) -- [line width=0.25mm, fermion, arrow size=0.7pt] (c2)};
\node at (a)[circle,fill,style=black,inner sep=1pt]{};
\node at (b)[circle,fill,style=black,inner sep=1pt]{};
\end{feynman}
\end{tikzpicture}
\label{fig:feynman-relic51}}
\caption{The Feynman diagrams are related to the annihilation and co-annihilation of WIMP ($\psi~\Phi^*\to \qu~V/h$, $\psi~\Phi\to \qu~\Phi$,~$\Phi~\Phi^*\to\rm SM~SM$,~$\Phi~\Phi\to\rm \Phi^*~h$,~$\rm\psi~\bar{\psi}\to\Phi~\Phi^*/h~Z/q~\bar{q}/\qu~\bar{q}_u/\ell~\bar{\ell}/\nu_{\ell}~\bar{\nu_{\ell}}/V~V/W^+~W^-$) where $V\in \{\gamma,~ Z,~g\}$, $\qu\in\{u,~c,~t\}$, $q_d\in\{d,~s,~b\}$, $\ell\in\{e,\mu,\tau\}$, $\nu^{}_\ell\in\{\nu_e,\nu_\mu,\nu_\tau\}$ and $\rm SM=\{h,~\qu,~\qd,~\ell,~Z,~W^{\pm}\}$.}
\label{fig:feynman-relic}
\end{figure}


\newpage
\bibliographystyle{JHEP}
\bibliography{biblio.bib}
\end{document}